\documentclass{aa}

\usepackage[varg]{txfonts}
\usepackage{subfig}
\usepackage{epsfig} 
\usepackage{amssymb}
\usepackage{graphicx}
\usepackage{natbib}
\usepackage{longtable}
\usepackage{longtable,lscape}
\usepackage{lscape}
\usepackage{ifpdf}
\bibpunct{(}{)}{;}{a}{}{,}

\newcommand{\ch}{{\it Chandra}}
\newcommand{\spz}{{\it Spitzer}}
\newcommand{\her}{{\it Herschel}} 

\begin{document}


\title{GOODS--\her: radio-excess signature of hidden AGN activity in distant star-forming galaxies}
\author{A.~Del Moro
\inst{1}
\and D.~M.~Alexander
\inst{1}
\and J.~R.~Mullaney
\inst{1,2}
\and E.~Daddi
\inst{2}
\and M.~Pannella
\inst{2}
\and F.~E.~Bauer
\inst{3,4}
\and A.~Pope
\inst{5}
\and M.~Dickinson
\inst{6}
\and D.~Elbaz
\inst{2}
\and P.~D.~Barthel
\inst{7}
\and M.~A.~Garrett
\inst{8,9,10}
\and W.~N.~Brandt
\inst{11}
\and V.~Charmandaris
\inst{12}
\and R.~R.~Chary
\inst{13}
\and K.~Dasyra
\inst{2,14}
\and R.~Gilli
\inst{15}
\and R.~C.~Hickox
\inst{16}
\and H.~S.~Hwang
\inst{17}
\and R.~J.~Ivison
\inst{18}
\and S.~Juneau
\inst{19}
\and E.~Le Floc'h
\inst{2}
\and B.~Luo
\inst{11}
\and G.~E.~Morrison
\inst{20}
\and E.~Rovilos
\inst{1,15}
\and M.~T.~Sargent
\inst{2}
\and Y.~Q.~Xue
\inst{11,21}
}

\institute{Department of Physics, Durham University, South Road, Durham, DH1 3LE, UK
\and Laboratoire AIM, CEA/DSM-CNRS-Universit\'{e} Paris Diderot, IRFU/Service d'Astrophysique, B\^{a}t. 709, CEA-Saclay, 
91191 Gif-sur-Yvette Cedex, France 
\and Pontificia Universidad Cat\'{o}lica de Chile, Departamento de Astronom\'{\i}a y Astrof\'{\i}sica, Casilla 306, Santiago 22, Chile
\and Space Science Institute, 4750 Walnut Street, Suite 205, Boulder, Colorado 80301, USA
\and Department of Astronomy University of Massachusetts, LGRT-B618, 710 North Pleasant Street, Amherst, MA 01003, USA
\and National Optical Astronomy Observatory, 950 North Cherry Avenue, Tucson, AZ 85719, USA
\and Kapteyn Astronomical Institute, University of Groningen, Groningen, The Netherlands
\and ASTRON, Netherlands Institute for Radio Astronomy, Post box 2, 7990AA, Dwingeloo, The Netherlands  
\and Leiden Observatory, Leiden University, Post box 9513, 2300RA Leiden, The Netherlands 
\and Centre for Astrophysics and Supercomputing, Swinburne University of Technology, Australia. 
\and Department of Astronomy and Astrophysics, 525 Davey Lab, Pennsylvania State University, University Park, PA 16802, USA
\and Department of Physics \& ITCP, University of Crete, GR-71003, Heraklion, Greece
\and U.S. Planck Data Center, MS220-6 Caltech, Pasadena, CA 91125, USA
\and Observatoire de Paris, LERMA (CNRS:UMR8112), 61 Av. de l' Observatoire, F-75014, Paris, France 
\and INAF - Osservatorio Astronomico di Bologna, Via Ranzani, 1, 40127 Bologna, Italy
\and Department of Physics and Astronomy, Dartmouth College, 6127 Wilder Laboratory, Hanover, NH 03755, USA
\and Smithsonian Astrophysical Observatory, 60 Garden St., Cambridge, MA 02138, USA
\and UK Astronomy Technology Centre, Science and Technology Facilities Council, Royal Observatory, Blackford Hill, Edinburgh EH9 3HJ, UK
\and Steward Observatory, University of Arizona, Tucson, AZ 85721, USA
\and Institute for Astronomy, Universityof Hawaii, Manoa, HI 96822, USA; Canada-France-Hawaii Telescope Corp., Kamuela, HI 96743, USA 
\and Key Laboratory for Research in Galaxies and Cosmology, Department of Astronomy, University of Science and Technology of China, Chinese Academy of Sciences, Hefei, Anhui 230026, China
}
\date{Accepted 28 September 2012}
\abstract{We present here a new spectral energy distribution (SED) fitting approach that we adopt to select radio-excess sources amongst distant star-forming galaxies in the GOODS-\her\ (North) field and to reveal the presence of hidden, highly obscured AGN. Through extensive SED analysis of 458 galaxies with radio 1.4 GHz and mid-IR 24 $\mu$m detections using some of the deepest \ch\ X-ray, \spz\ and \her\ infrared, and VLA radio data available to date, we have robustly identified a sample of 51 radio-excess AGN ($\sim$1300 deg$^{-2}$) out to redshift $z\approx3$. These radio-excess AGN have a significantly lower far-IR/radio ratio ($q<1.68$, 3$\sigma$) than the typical relation observed for star-forming galaxies ($q\approx2.2$). We find that $\approx$45\% of these radio-excess sources have a dominant AGN component in the mid-IR band, while for the remainders the excess radio emission is the only indicator of AGN activity. The presence of an AGN is also confirmed by the detection of a compact radio core in deep VLBI 1.4 GHz observations for eight of our radio-excess sources ($\approx$16\%; $\approx$66\% of the VLBI detected sources in this field), with the excess radio flux measured from our SED analysis agreeing, to within a factor of two, with the radio core emission measured by VLBI. We find that the fraction of radio-excess AGN increases with X-ray luminosity reaching $\sim$60\% at $L_{\rm X}\approx10^{44}-10^{45}$ erg s$^{-1}$, making these sources an important part of the total AGN population. However, almost half (24/51) of these radio-excess AGN are not detected in the deep \ch\ X-ray data, suggesting that some of these sources might be heavily obscured. Amongst the radio-excess AGN we can distinguish three groups of objects: i) AGN clearly identified in infrared (and often in X-rays), a fraction of which are likely to be distant Compton-thick AGN, ii) moderate luminosity AGN ($L_{\rm X}\lesssim10^{43}$ erg s$^{-1}$) hosted in strong star-forming galaxies, and iii) a small fraction of low accretion-rate AGN hosted in passive (i.e. weak or no star-forming) galaxies. We also find that the specific star formation rates (sSFRs) of the radio-excess AGN are on average lower that those observed for X-ray selected AGN hosts, indicating that our sources are forming stars more slowly than typical AGN hosts, and possibly their star formation is progressively quenching.}

\titlerunning{Radio-excess sources in GOODS--\her.}
\maketitle

\section{Introduction}\label{intro}
The discovery of a strong correlation between the properties of galaxies and those of the supermassive black holes (SMBH) hosted in their centres, such as the $M_{\rm BH}-M_{\rm bulge}$ or $M_{\rm BH}-\sigma$ relations \citep{magorrian1998, ferrarese2000,gebhardt2000,marconi2003}, has pointed out that SMBHs must play an important role in the growth and evolution of galaxies (see \citealt{alexander2012}, for a general review). In the past decades many studies have focussed on understanding the relation between nuclear activity (AGN) and host galaxies and have revealed a common history, where both star formation and black hole accretion were much more common in the past, with a peak at redshift $z\approx2$ (e.g. \citealt{fiore2003, merloni2004, marconi2004, hopkins2006b, hopkins2007, merloni2007}). Completing the census of AGN activity, especially at redshifts where most of the accretion occurred, is therefore essential in order to understand the nature of the link between SMBH and galaxies and their cosmic co-evolution. This paper aims to expand our knowledge of the AGN population, by selecting objects with radio emission in excess of that expected from star formation. As discussed below, this method selects many AGN that cannot be identified using other established techniques, and so moves us closer to a complete census of growing SMBHs in the Universe.

Deep X-ray surveys have proved to be a very powerful tool in detecting obscured and unobscured AGN down to faint fluxes and to high redshifts ($z\sim5$; e.g. \citealt{alexander2001, fiore2003, hasinger2008, brusa2009, xue2011, lehmer2012}). However, it is now evident that even the deepest X-ray surveys are not complete (e.g. \citealt{tozzi2006,hasinger2008}) and miss a significant part of the AGN population, in particular the most obscured, Compton-thick (CT) AGNs, where the X-ray emission below 10 keV is strongly suppressed by large column density gas ($N_{\rm H}>10^{24}$~cm$^{-2}$). A large population of heavily obscured AGN is indeed predicted by synthesis models of the X-ray background (XRB; \citealt{comastri1995, gilli2001, gilli2007,treister2009,ballantyne2011}) in order to reproduce the high energy peak of the observed X-ray background emission ($E\approx30$ keV), which has not yet been directly resolved by current X-ray surveys. 

Since large amounts of gas and dust are responsible for the suppression of the radiation in the UV, optical and soft X-ray bands, perhaps the most obvious waveband to search for these heavily obscured objects is where the dust emission peaks, i.e. the infrared (IR) band. In fact, the dust surrounding the SMBH is heated by the nuclear radiation, reaching temperatures $T\sim200-1000$ K, and re-emits the radiation predominantly in the mid-infrared (MIR; $\lambda\approx5-40\ \mu$m) band, peaking at $\sim20-30\ \mu$m \citep{netzer2007,mullaney2011}. Moreover, at these wavelengths the effects of extinction are small, making it easier, in theory, to find even the most obscured AGN \citep[e.g.][]{gandi2009,goulding2012}. The downside of using the IR band to search for AGN activity is that dust is present not only in the circumnuclear region of AGN, but also in the host galaxy, in particular in star-forming regions. The dust in these regions is heated on average to lower temperatures, and therefore its emission peaks at longer wavelengths, than that around the black hole (typically at $\lambda\approx$100 $\mu$m, $T_{dust}\sim20-50$ K; e.g. \citealt{calzetti2000, chary2001}). However star formation often dominates the spectral energy distribution (SED) over the entire IR band (e.g. \citealt{elvis1994, richards2006, netzer2007, mullaney2011}) and it is often not trivial to separate it from the AGN emission. 

Potentially a very powerful approach to obtaining an unbiased look at obscured and unobscured AGN is through radio observations. In fact, at radio frequencies, where the emission is mainly due to non-thermal processes, such as synchrotron radiation, the effects of extinction are negligible. Historically, AGN detected in radio surveys have been divided into two main classes: i) radio-loud (RL) AGN, which are the strongest radio emitters (typically $L_{rad}\gtrsim10^{24-25}$ W Hz$^{-1}$; \citealt{miller1990, yun2001}) and show strong extended radio emission, such as kpc-scale relativistic jets and lobes, and ii) radio-quiet (RQ) AGN, the weaker radio emitters, whose radio emission is confined in a small, unresolved region ($\le0.1$ pc; ``core''); the latter group constitutes the majority of the population ($\sim$90\%; e.g. \citealt{miller1990, stocke1992}). The separation between radio-loud and radio-quiet AGN has typically been set at $R=10$, where $R$, the radio-loudness parameter, is defined as the ratio between the monochromatic flux density in the radio and optical bands\footnote{Most recently, other definitions of the radio-loudness parameter have been used, e.g.  $R_{X}=\rm log\ (\nu L_{\rm rad}/L_{\rm 2-10\ keV})$, which uses the monochromatic radio luminosity and hard X-ray luminosity (e.g. \citealt{ballantyne2009, lafranca2010}).} $R=S_{\rm rad}/S_{\rm opt}$ (e.g. \citealt{kellermann1989,laor2000}). However, more recent studies based on deep radio surveys have shown that while the radio-loudness parameter spans a very wide range of values for AGN, there is no clear evidence of bimodality in the population (e.g. \citealt{white2000,brinkmann2000,cirasuolo2003a,lafranca2010}). AGN can therefore be identified in the radio band with a wide distribution of radio power.

While at bright fluxes the radio population is almost entirely composed of AGN, at low radio fluxes (sub-mJy regime) star-forming galaxies (SFGs) constitute a significant fraction of the radio source population and become dominant at $\mu$Jy fluxes \citep[e.g.][]{seymour2008}. The non-thermal radio continuum observed in star-forming galaxies is produced by synchrotron radiation from cosmic ray electrons and positrons, accelerated by supernova remnants, which mainly occur in young stellar populations in star-forming regions (see \citealt{condon1992}; for a review). The radio emission observed in star-forming galaxies tightly correlates with the emission in the far-infrared (FIR; $\lambda\approx40-120\ \mu$m) band, since they both originate from star formation processes (e.g. \citealt{helou1985, condon1992, yun2001, appleton2004,ivison2010}). This correlation, observed primarily in local star-forming galaxies and starbursts, is found to hold out to high redshifts ($z\approx2$; e.g. \citealt{ibar2008,sargent2010b,ivison2010, mao2011, bourne2011}).

Joint analyses in the FIR and radio bands, therefore, allow us to separate star-forming galaxies from the AGN population. Although the weakest radio AGN (RQ) have been found to follow the same FIR/radio correlation of star-forming galaxies (e.g. \citealt{moric2010,padovani2011}), given their wide range in radio-loudness, it is possible to identify AGN via their deviation from the expected FIR/radio relation, the so-called ``radio-excess'' sources (e.g. \citealt{roy1997,donley2005}). Since the radio emission is not affected (or only lightly affected) by extinction, the radio-excess source selection can potentially identify AGN that are often missed in optical or even deep X-ray surveys (i.e. the most obscured Compton-thick AGN; e.g. \citealt{donley2005}).



In this work, we combine two methods to identify the presence of AGN in star-forming galaxies out to high redshift ($z\approx3$): i) detailed IR SED decomposition, which allows us to measure the AGN contribution to the total SED, typically dominated by star formation emission; ii) radio-excess signature compared to the typical FIR--radio relation observed for SFGs, which is most likely due to the presence of nuclear activity. We investigate here the radio-excess sources in the GOODS-North field, using deep infrared, radio and X-ray data, which are some of the deepest data available to date. The paper is organised as follows: the data and catalogues used in our investigation are presented in Section \ref{data}. In Section \ref{rexc}, our SED fitting approach is described in details as well as the definition of the FIR--radio flux ratio ($q$) and the radio-excess sample selection, together with a comparison with other selection criteria used in previous studies. In Section \ref{res}, we investigate the X-ray, radio and IR properties of our radio-excess AGN sample and their SEDs. In Section \ref{discus} we discuss the mixed population found amongst our radio-excess AGN, attempting to constrain the fraction of candidate Compton-thick AGN; we also examine the star formation properties of the radio-excess AGN hosts through their specific star-formation rate (sSFR) in comparison with those of X-ray selected AGN hosts. In Section \ref{conc} we summarise our results and give our conclusions. In Appendix A, the tests performed to refine our SED fitting approach are explained and the best-fit SEDs for the entire radio-excess sample are shown in Appendix B.

Throughout the paper we assume a cosmological model with $H_0=70\ \rm km\ s^{-1}\ Mpc^{-1}$, $\Omega_M=0.27$ and $\Omega_{\Lambda}=0.73$ \citep{spergel2003}.

\section{Observations and catalogs}\label{data}
The Great Observatories Origins Deep Survey--North field (GOODS-N; \citealt{giavalisco2004}) is one of the deepest multi-wavelength surveys currently available and it constitutes an unprecedented resource in terms of its broad-band coverage and sensitivity. It covers $\approx$160 arcmin$^2$ centred on the Hubble Deep Field North (HDF-N, $12^h\ 36^m,\ +62^{\circ}\ 14'$; \citealt{williams1996}) and it includes very deep X-ray {\it Chandra} data (2 Ms; \citealt{alexander2003}), optical {\it Hubble Space Telescope} (HST; \citealt{giavalisco2004}) and mid-infrared (MIR) \spz\ observations (PI: M. Dickinson); the GOODS-N field has also been the target of several deep optical imaging and spectroscopic campaigns from 8--10 m ground-based telescopes. Recently, new deep observations of this field in the far-infrared (FIR) band with \her\ \citep{elbaz2011} and the radio band with VLA (\citealt{morrison2010}) have usefully increased the potential of the GOODS-N data set.

\subsection{\spz\ MIR data}\label{mir}

The GOODS-N field has been observed at MIR wavelengths by \spz\ at 3.6, 4.5, 5.8 and 8.0 $\mu$m with IRAC \citep{fazio2004}, with a mean exposure time per position of $\approx$23 hours per band, and at 24 $\mu$m with MIPS \citep{rieke2004}, as part of the GOODS \spz\ Legacy program (PI: M. Dickinson). 
The source catalogue was produced using the {\it SExtractor} source detection routine \citep{bertin1996} on a combined 3.6 $\mu$m $+$ 4.5 $\mu$m image, with matched aperture photometry performed in the four IRAC bands individually (M. Dickinson et al., in preparation). The resulting IRAC catalogue includes 19437 objects detected at 3.6 $\mu$m with a $\sim$50\% completeness limit of 0.5 $\mu$Jy. 

The 24 $\mu$m observations consist of a final mosaic image of 1.2$''$ pixel scale and a $5\sigma$ sensitivity limit of $\sim$30 $\mu$Jy. The source extraction was performed with a PSF fitting technique using the positions of the IRAC 3.6 $\mu$m sources detected at $>5\sigma$ as priors (see \citealt{magnelli2011}, for details). The IRAC 3.6 $\mu$m data is used to define the source priors because it is $\sim$30 times deeper than the 24 $\mu$m observations, and therefore all real 24 $\mu$m detected sources should also be detected at 3.6 $\mu$m. The resulting 24 $\mu$m catalogue includes 2552 sources detected with signal-to-noise ratio ($\rm S/N$) $>3$ in the GOODS-N field. However, we note that in the outer regions of the GOODS-N field the MIPS data is shallower and the uncertainties on the source fluxes are typically larger; we therefore limit our catalogues to a smaller area ($\sim$135 arcmin$^2$) within the GOODS-N field where the MIPS data is deeper (1943 sources detected at 24 $\mu$m, $\sim$76\%). We require at least a detection ($\rm S/N>3$) at 24 $\mu$m for the sources in our sample to be able to constrain the source SEDs in the MIR band (see Sect. \ref{sed}).


An area of $\sim$150 arcmin$^{2}$ of the GOODS-N field has also been surveyed at 16 $\mu$m using the {\it Infrared Spectrograph} (IRS) peak-up Imaging (PUI) with pointings of $\sim$10 min each. The observations and data reduction are described in detail by \citet{teplitz2011}; the resulting mosaic image is characterised by 0.9$''$ pixel scale and has an average 5$\sigma$ depth of $\sim$40 $\mu$Jy (\citealt{teplitz2011}). The source catalogue was constructed using \spz-MIPS 24 $\mu$m priors ($>5\sigma$ sources) and the 16 $\mu$m fluxes were calculated through PSF-fitting, similarly to the procedure used for the 24 $\mu$m data; the 16 $\mu$m catalogue contains 770 sources (Daddi et al. 2012, in prep.). 

\subsection{GOODS--\her\ FIR data}
\label{fir}
The GOODS-N field has been observed by the \her\ Space Observatory as part of the GOODS-\her\ survey (PI: D. Elbaz), which consists of deep FIR observations of the GOODS-North and GOODS-South fields for a total exposure of 361.3 hours. Imaging of the full northern field (GOODS-N; $10'\times16'$) was performed using PACS \citep{poglitsh2008} at 100 $\mu$m and 160 $\mu$m (124 hours of observations) and SPIRE \citep{griffin2010} at 250 $\mu$m, 350 $\mu$m and 500 $\mu$m (31.2 hours in total); see \citet{elbaz2011} for details on the PACS and SPIRE observations. 

The data reduction was performed following the procedure described in \citet{berta2010} and the resulting images have pixel scales of 1.2$''$ and 2.4$''$ at PACS 100 $\mu$m and 160 $\mu$m, respectively and 3.6$''$, 5.0$''$ and 7.2$''$ at SPIRE 250 $\mu$m, 350 $\mu$m and 500 $\mu$m. Since the 350 $\mu$m and 500 $\mu$m data suffer from strong source blending due to the large pixel scales, we did not include data at these wavelengths in our SED fitting procedure (Sect. \ref{sed}) and we only used the PACS 100 $\mu$m and 160 $\mu$m and SPIRE 250 $\mu$m data. The 350 $\mu$m and 500 $\mu$m flux densities and/or upper limits ($S_{350}\approx20.0$ mJy and $S_{500}\approx30.0$ mJy, 5$\sigma$) from the catalogue described in \citet{elbaz2011} were only used in some of the plots (see e.g. Fig. \ref{fig.sedall}) to verify the accuracy of our SED fitting at these longer wavelengths. 

The \her\ 100 $\mu$m, 160 $\mu$m and 250 $\mu$m fluxes were calculated using PSF fitting at the positions of the \spz-MIPS 24 $\mu$m sources, which are used as priors (Daddi et al. 2012, in prep.). In the resulting catalogue 819 sources ($\sim$42\%) have $\rm S/N> 3$ in at least one of the \her\ bands: 633 sources at 100 $\mu$m, 537 sources at 160 $\mu$m and 435 at 250 $\mu$m, with 5$\sigma$ (3$\sigma$) sensitivity limits of $\sim$1.7 ($\sim$1.2) mJy, $\sim$4.5 ($\sim$2.3) mJy and $\sim$6.5 ($\sim$4.0) mJy, respectively. It is important to note, however, that the sensitivity of the 250 $\mu$m data strongly varies across the field, depending on the local source density of the 24 $\mu$m priors.
We included in our sample all of the 1943 24 $\mu$m detected sources within the GOODS--\her\ field (restricted to the area with deeper MIPS data), with or without a significant \her\ detection; this is to avoid biases against faint FIR sources, which are more likely to be AGN dominated. 


\subsection{VLA radio data}
\label{rad}
Deep, high-resolution radio observations of the GOODS-N field were taken at 1.4 GHz using the National Radio Astronomy Observatory's (NRAO) Very Large Array (VLA) in the A, B, C and D configurations (165 hours). The combined radio image reaches a rms noise level of $\sim3.9\ \mu$Jy beam$^{-1}$ near the centre with a beam size of $\sim1.7''$. These are amongst the deepest radio data taken so far. From the VLA image the radio flux density measurements have been obtained through PSF fitting at each 3.6 $\mu$m source position (coinciding with the 24 $\mu$m positions; see Sect. \ref{mir}; Daddi et al. 2012, in prep.): 1.4 GHz flux measurements were obtained for all 1943 sources detected at 24 $\mu$m within the GOODS--\her\ area, with 489 sources having $\rm S/N>3$ and $S_{\rm \nu}>13\ \mu$Jy. The remaining 1454 sources have ${\rm S/N}<3$ and they are considered as radio upper limits. Within this field, we estimated that $\sim$20 radio detected sources (5$\sigma$) are undetected at 24 $\mu$m down to a flux limit of $S_{\rm \nu}\approx21\ \mu$Jy; this gives an estimate of the completeness of our radio and 24 $\mu$m detected sample (hereafter VLA/24 $\mu$m sample) of $\sim$93\%, as compared to a pure radio-selected sample. We note that the detection limit adopted for our VLA radio catalogue is lower than that used in \citet{morrison2010} (5$\sigma$ detection threshold) and therefore the number of radio detections found here (489 sources) is much larger than that found by \citet{morrison2010} over the same area (256 sources).  

 
\subsection{{\it Chandra} X-ray data}
\label{xx}
The \ch\ X-ray observations of GOODS--N field cover an area of $\approx448$ arcmin$^2$ in the 0.5--8.0 keV energy band, with an exposure of $\approx$2 Ms (\ch\ Deep Field North, CDF-N; \citealt{alexander2003}), reaching a sensitivity (on-axis) of $\approx2.5\times10^{-17}\rm\ erg~cm^{-2}~s^{-1}$ (0.5--2.0 keV) and $\approx1.4\times10^{-16}\rm\ erg~cm^{-2}~s^{-1}$ (2--8 keV). 
The main source catalogue of the CDF-N includes 503 X-ray detected sources (\citealt{alexander2003}). A supplementary catalogue containing 430 X-ray sources is also available in this field. These catalogues where constructed using a source detection algorithm with false-positive probability threshold of 10$^{-7}$ for the main catalogue, and a more relaxed threshold of 10$^{-5}$ for the supplementary catalogue (see \citealt{alexander2003}, for details). Although this second catalogue is likely to include many spurious X-ray sources, it can be used to robustly identify fainter X-ray counterparts associated to known sources (see e.g., Sect. 3.4.2. of \citealt{alexander2003}).

These catalogues were used to identify the X-ray counterparts of the 489 VLA/24 $\mu$m sources. The 24 $\mu$m positions were matched to the X-ray positions using a small search radius of $1.5''$; taking into account the high positional resolution of the \ch\ data (median positional uncertainties $\approx$0.3$''$) and the small pixel size of the MIPS-24 $\mu$m images, the majority of the true counterparts are expected to lie within this radius. In fact, calculating the probability $P$ to find a random object within 1.5$''$ from the X-ray positions following the prescription of \citet{downes1986}, we obtained a maximum probability of random association $P=0.03$ (considering a space density of 24 $\mu$m sources of $n=5.2\times10^{4}$ deg$^{-2}$). 

From the main catalogue we found X-ray counterparts (in the 0.5--8 keV energy band) 
for 137 of the 489 VLA/24 $\mu$m sources ($\approx$28\%), with a median positional separation of $\approx$0.2$''$. Amongst the matched sources, we found that in none of the cases was there more than one counterpart within the search radius, with the closest neighbours being at separations $\gtrsim2''$. From the supplementary catalogue we identified a further 
22 X-ray counterparts to the VLA/24 $\mu$m sources, yielding a total of 159 X-ray detected sources (i.e. $\approx$33\% of the VLA/24 $\mu$m sample). We note that since the source catalogues in all of the MIR and FIR bands considered here, as well as the VLA radio catalogue, are based on the 3.6 $\mu$m positions there was no need to cross-match the X-ray source positions with any of the other bands. For the X-ray undetected sources, 3$\sigma$ upper limits were derived from aperture-corrected photometry in the \ch\ images at the 24 $\mu$m source positions, assuming a power-law model with $\Gamma=1.4$ (see e.g., Sect. 3.4.1 of \citealt{alexander2003,bauer2010}).  


The X-ray luminosities (2--10 keV; rest-frame) of the sources were extrapolated from the observed $2-8$ keV fluxes calculated from detailed X-ray spectral analysis (Bauer et al., in preparation) and from \citet{alexander2003} for the sources in the supplementary \ch\ catalogue. We used the redshifts described in Sect. \ref{reds} and assumed a constant photon index $\Gamma=1.9$ (which gives a band conversion factor $L_{\rm2-10\ keV}=1.08~L_{\rm2-8\ keV}$), including appropriate $k-$correction. We note that the X-ray luminosities have not been corrected for absorption because the column density estimates ($N_{\rm H}$, available from Bauer et al., in prep.) often have large uncertainties; moreover, $N_{\rm H}$ values were not available for the sources in the supplementary catalogue (\citealt{alexander2003}). Therefore, to avoid adding further uncertainties to the X-ray luminosities and to keep consistency in the $L_{\rm2-10\ keV}$ measurements between the main and the supplementary \ch\ catalogues, we did not apply any absorption corrections. 


\subsection{Redshifts}\label{reds}
Thanks to the large spectroscopic follow-up observations performed in the GOODS-N field, $\approx$3000 redshift identifications are available for the objects in this field. 
A compilation of spectroscopic redshifts ($z_{spec}$) were obtained from the major publicly available spectroscopic redshift surveys of the GOODS-N field (i.e., \citealt{wirth2004,cowie2004,chapman2005,barger2008,chapin2009}), as well as some unpublished spectroscopic redshift identifications (courtesy of M. Dickinson). The optical positions of the sources in these catalogues were matched with the 24 $\mu$m positions of our detected sources using a search radius of 1.0$''$. Since the errors on the optical positions are typically very small, we used a smaller search radius than that used for the X-ray catalogues (see Sect. \ref{xx}). 
With this search radius and considering the sky density of the 24 $\mu$m detected sources (Sect. \ref{xx}), we estimated the spurious detections to be $\sim$3\%.

Spectroscopic redshift measurements were found for 1225 sources amongst the 24 $\mu$m detected sample ($\sim$63\%), with the large majority coming from the \citet{barger2008} spectroscopic redshift catalogue (1030/1225). Two more redshift identifications were obtained from {\it Spitzer} IRS MIR spectra \citep{murphy2009}, yielding a total of 1227 $z_{spec}$ measurements for our sources. 
\begin{figure}[!t]
\centering{
\includegraphics[scale=0.6]{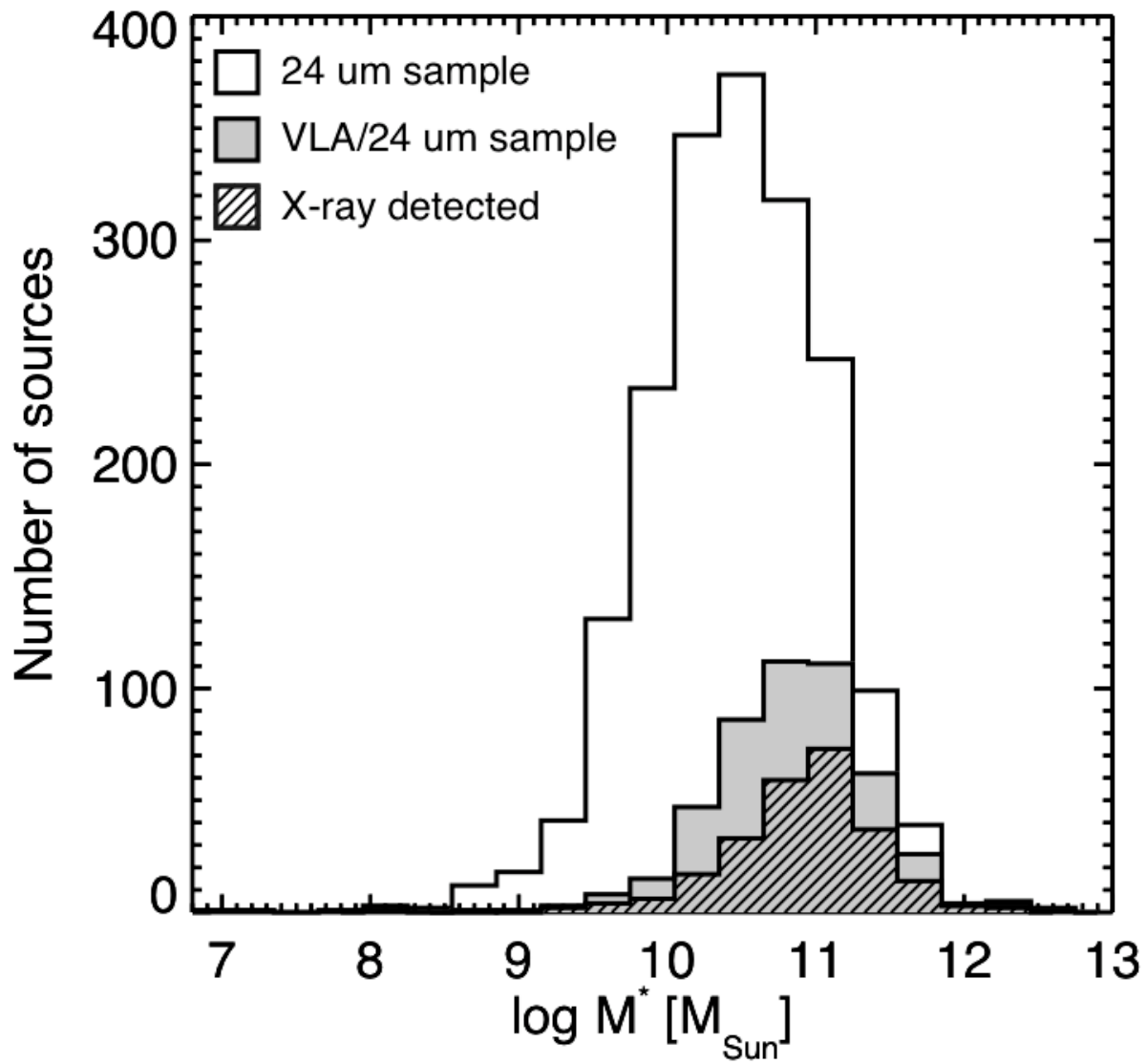}}
\caption{Galaxy stellar masses in units of M$_{\odot}$ for the entire 24 $\mu$m detected sample with redshift identification (spectroscopic or photometric). The stellar masses for the sources detected in the VLA radio band are shown as grey histogram and the X-ray detected sources are shown as shaded black histogram.}
\label{fig.mass}
\end{figure}

In order to increase the redshift identification completeness of the sample, we also included photometric redshifts from a catalogue built following the procedure described in \citet{pannella2009b} and \citet{strazzullo2010}. The photometric redshifts ($z_{phot}$) were estimated using a PSF-matched multi-wavelength catalogue including 10 photometric optical/near-IR passbands (from the U band to 4.5 $\mu$m), through a comparison with a library of galaxy SED templates, spanning a wide range of galaxy types (from elliptical, to star-forming to QSO-dominated) and star formation histories (SFHs). The construction of the multi-wavelength catalogue and the photometric redshift estimates will be described in detail in a paper by Pannella et al. (in prep). 
The photometric redshift catalogue includes 1893 $z_{phot}$ within the GOODS--\her\ area considered here. 
Photometric redshift estimates were available for 671 of the 24 $\mu$m detected sources ($\sim$35\%) without $z_{spec}$ measurements. To verify the reliability of the photometric redshifts we compared the $z_{phot}$ with the spectroscopic redshifts from \citet{barger2008}; in the photometric catalogue by Pannella et al. there are 1030 sources overlapping with the \citet{barger2008} spectroscopic sample. The relative accuracy of $z_{phot}$, defined as the average absolute scatter ($AAS=\rm mean [\mid\Delta z\mid /(1+z_{spec})]$, where $\Delta z=(z_{phot}- z_{spec})$; e.g. \citealt{rafferty2011}) is $\approx$5\%, with $\approx$4\% of outliers ($AAS>0.2$; see also \citealt{mullaney2012}). 
Two further $z_{phot}$ were taken from \citet{pope2006}, yielding a total of 673 photometric redshift estimates for our sample. This gives a final redshift identification completeness of $\sim$98\% (1900/1943 sources, including spectroscopic and photometric redshifts) amongst the 24 $\mu$m detected sources in the GOODS--\her\ field with a redshift range $z=0.02-6.54$. 

For the purposes of our analysis, we want to investigate here only the sources with a significant radio detection, in order to have reliable measurements of the FIR--radio correlation. We therefore only included in our sample sources with a redshift identification amongst the VLA/24 $\mu$m sample (484/489 sources). Due to the limitations dictated by our SED fitting tool (Sect. \ref{sed}), we also imposed a redshift limit of $z\le3.0$ to our sources, yielding a sample of 458 VLA/24 $\mu$m sources with spectroscopic or photometric redshifts of $z\le3.0$. The analysis of the whole 24 $\mu$m detected sample, including radio undetected sources, will be presented in a future paper (Del~Moro et al., in preparation).

\subsection{Stellar masses}\label{mass}
The multi-wavelength optical/near-IR catalogue used to estimate the photometric redshifts (Pannella et al., in prep.; see Sect. \ref{reds}) was also used to calculate the galaxy stellar masses ($M_*$).
The stellar masses have been derived using the SED fitting code detailed in \citet{drory2004, drory2009} to fit our multi-wavelength data. The star formation histories have been parameterised with a linear combination of a main SF event, with SFR exponentially declining with time as $\psi(t)\propto \rm exp(-t/\tau)$ (where the time-scale $\tau=0.1-20$ Gyr), and a secondary burst. The main component has solar metallicity and an age between 0.01 Gyr and the age of the Universe at the source redshift, while the secondary burst is limited to $<$10\% of the galaxy total stellar mass and is modelled as a 100 Myr old constant SFR episode with solar metallicity. We adopted a \citet{salpeter1955} IMF for both components and an extinction law (\citealt{calzetti2000}), allowing ranges of $A_V=0-1.5$ mag and $A_V=0-2.0$ mag to extinguish the main component and the burst, respectively (Pannella et al., in preparation; see also \citealt{mullaney2012}). The uncertainties on the stellar masses are estimated from the dispersion on the mass-to-light ratio ($M/L$) distribution of the entire library of models adopted, as well as from the systematic uncertainties (due to the adopted models, IMF, SFH, metallicity, etc.; see e.g. \citealt{marchesini2009}). The uncertainties are typically larger at low stellar masses and range from $\approx$0.4 dex at $\rm log\ M_*=9.0\ \rm M_{\odot}$ to $\approx$0.2 dex at $\rm log\ M_*=11.0\ \rm M_{\odot}$ (Pannella et al., in prep.). We note that also AGN emission in the UV/optical band can cause uncertainties on the stellar mass estimates. However, the contamination from the AGN affects the stellar mass only when the AGN is very luminous ($L_{\rm X}>10^{44}$ erg~s$^{-1}$; e.g. \citealt{rovilos2007,xue2010,mullaney2012}).

Stellar masses were measured for 1894 of the 1900 24 $\mu$m detected sources with a redshift identification within the GOODS--\her\ field (Sect. \ref{reds}), which include 456 of the 458 VLA/24 $\mu$m sources with spectroscopic or photometric redshift $z\le3.0$ that constitute our final sample. The galaxy stellar mass values obtained range between log $M_*=7.0-12.5\ \rm M_{\odot}$ (log~$M_*=8.7-11.9\ \rm M_{\odot}$ for the VLA/24 $\mu$m sample), with the large majority of the sources ($\approx$97\%) having log $M_*=9.0-11.5\ \rm M_{\odot}$ (Fig. \ref{fig.mass}).\footnote{We note that the fraction of sources with very low (log $M_*<8.0\ \rm M_{\odot}$) or very high (log $M_*>12.0\ \rm M_{\odot}$) mass values amongst the 24 $\mu$m detected sample (3\%) is consistent with the estimated fraction of photometric redshift outliers (see Sect. \ref{reds}).} The median stellar mass for the 24 $\mu$m sample is log $M_*\approx10.4\ \rm M_{\odot}$, and for the VLA/24 $\mu$m sample the median is log $M_*\approx10.8\ \rm M_{\odot}$. We note that the stellar mass distribution of the VLA/24 $\mu$m sample sources is consistent with that of the X-ray detected sources (median stellar mass: log $M_*\approx10.9\ \rm M_{\odot}$; see Fig. \ref{fig.mass}).

\section{FIR--radio correlation}
\label{rexc}
\subsection{SED fitting approach}
\label{sed}

The emission observed in the MIR and FIR bands is produced by dust heated by the radiation emitted through star formation and/or accretion onto a SMBH. Star-formation, which occurs on large scales in galaxies, heats the dust to a wide range of temperatures: the hot dust produces emission at near-IR (NIR; $\lambda\approx2.0-5.0\ \mu$m) and MIR wavelengths and gives rise to the characteristic PAH features (e.g. \citealt{chary2001,smith2007}), while a large amount of colder dust ($T_{dust}\approx20-50$ K; see also Sect. \ref{intro}) produces a typical SED that peaks at FIR wavelengths ($\lambda\sim$100 $\mu$m). AGN activity yields on average hotter dust temperatures ($T_{dust}\approx200-1000$ K) than star formation, so that the bulk of the AGN emission is produced in the MIR band with a peak at shorter wavelengths ($\lambda\sim20-30\ \mu$m; e.g. \citealt{netzer2007,mullaney2011}). The lack of colder dust ($T_{dust}<200$ K) causes a fast decline of the SED at wavelengths longer than $\lambda\gtrsim30\ \mu$m (e.g. \citealt{netzer2007,mullaney2011}). AGN activity and star formation are often coupled in a galaxy and it is not trivial to separate the emission due to these two processes. To disentangle the two components we therefore performed a detailed analysis of the IR SEDs of the 458 VLA/24 $\mu$m detected sources, with spectroscopic or photometric redshifts out to $z=3$ (Sect. \ref{reds}). 

To represent the galaxy emission we used five star-forming galaxy (SFG) templates defined by \citet{mullaney2011}, covering the wavelength range 6--1000 $\mu$m. These five templates are defined as composites of a sample of local star-forming galaxies with $L_{\rm IR}\lesssim10^{12}\ L_{\odot}$ (\citealt{brandl2006}) and are designed to sample the full range of {\it IRAS} colours observed for these galaxies (see \citealt{mullaney2011} for details).
We have extended the five SFG templates to shorter wavelengths (3 $\mu$m) using the average starburst SED derived by \citet{dale2001}. To verify whether the \citet{dale2001} template was suitable for extending the \citet{mullaney2011} templates, we obtained publicly available NIR and MIR data (from the NASA/IPAC Extragalactic Database, NED\footnote{http://ned.ipac.caltech.edu/}) for the \citet{brandl2006} sample of local star-forming galaxies and we plotted these data points over our extended templates as a check; we thus verified that data and models agreed with reasonable scatter. 

\begin{figure}[!t]
\centering{
\includegraphics[scale=0.6]{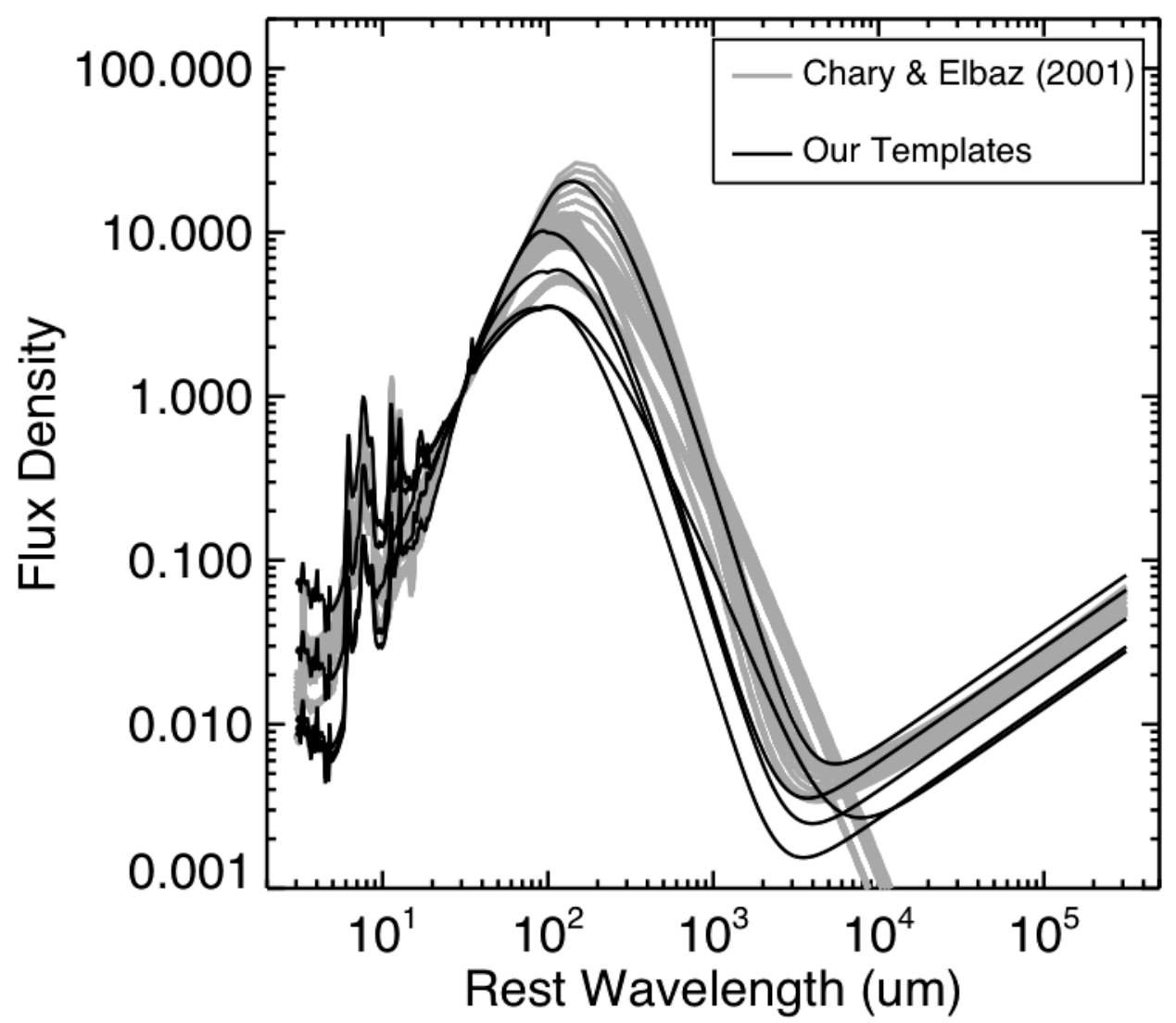}}
\caption{Comparison of the five star-forming galaxy templates used in our approach (black lines) with the 105 \protect\citet{chary2001} galaxy templates (grey lines); all of the templates have been normalised to 1 at 30 $\mu$m. With our 5 SFG templates we cover a broad range of star-forming galaxy properties, even broader than those of the \protect\citet{chary2001} templates.}
\label{fig.sb}
\end{figure}

We also extended the five SFG templates to the radio band with a power-law slope $S_{\nu}\propto\nu^{-\alpha}$, with $\alpha=0.7$ (e.g., \citealt{ibar2009,ibar2010}; see Fig. \ref{fig.sb}). The normalisation of the radio power-law component was fixed according to the typical FIR--radio relation found for local star-forming galaxies \citep{helou1985, condon1992}. We stress that even though we limited our SFG template library to only 5 templates, they represent a wide range of IR color-color properties of star-forming galaxies \citep{mullaney2011}, even broader than those reproduced by, e.g., the \citet{chary2001} galaxy template library (105 templates; see Fig. \ref{fig.sb}).

To reproduce the emission from the AGN we used the empirically defined AGN template by \citet{mullaney2011}, composed by a broken power-law, declining at wavelengths longward of $\lambda\gtrsim30\ \mu$m as a modified black-body. We note that this template is in agreement with the typical SEDs produced by clumpy torus models (e.g. \citealt{nenkova2008a,nenkova2008b}). We fixed the power-law indices at the average values $\Gamma_1=1.7$ and $\Gamma_2=0.7$, with a break point at $\lambda_{Brk}=19\ \mu$m (see \citealt{mullaney2011}). We have allowed the AGN component to be modified due to dust extinction, using the extinction law of \citet{draine2003}, which mainly affects the template at $\lambda\lesssim30\ \mu$m and also produces the typical silicate absorption feature at 9.7 $\mu$m, often observed in AGN (e.g. \citealt{roche1991,shi2006,roche2007,alejo2008}).

\begin{figure*}[!t]
\centerline{
\includegraphics[scale=1]{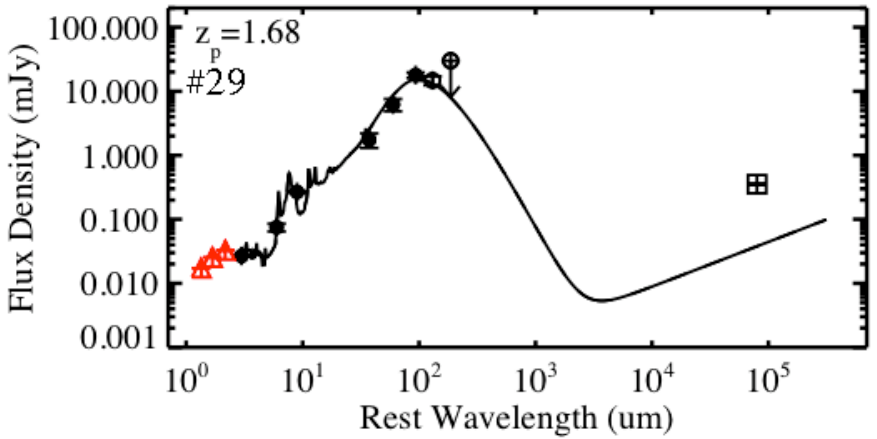}
\hspace{-0.1 cm}
\includegraphics[scale=1]{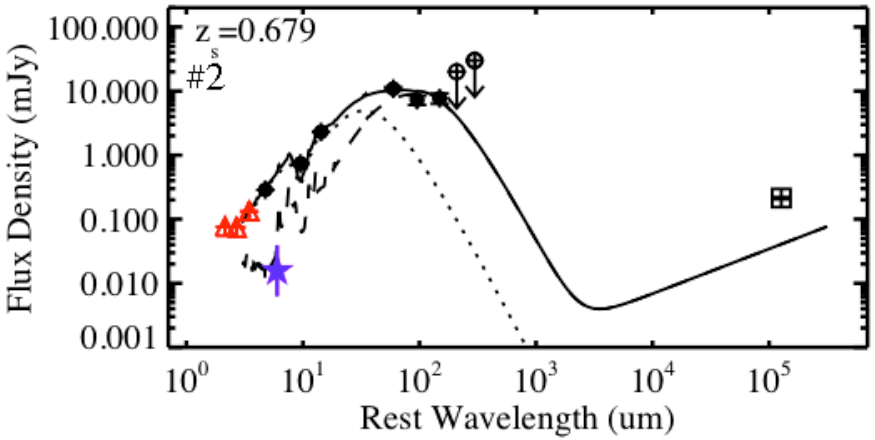}
}
\centerline{
\hspace{-0.1 cm}
\includegraphics[scale=1]{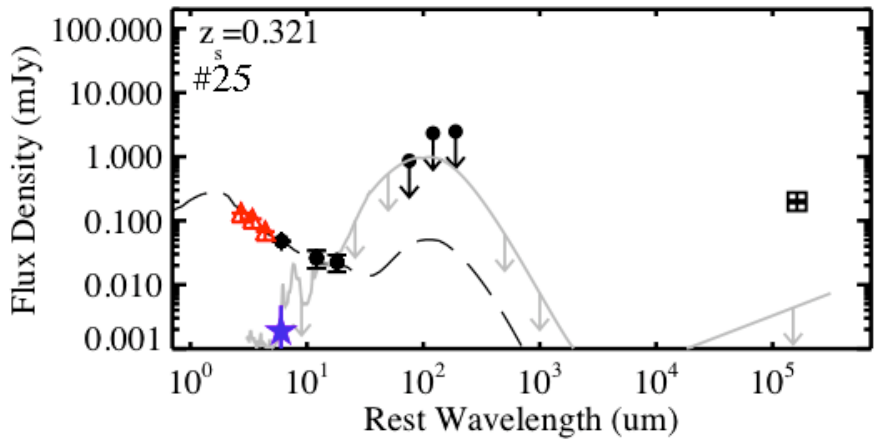}
\hspace{-0.1 cm}
\includegraphics[scale=1]{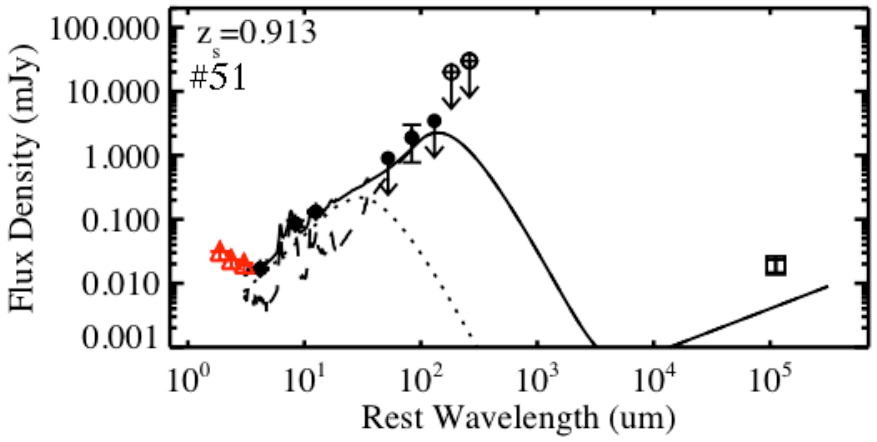}}
\caption{Examples of spectral energy distributions (SEDs) to demonstrate the variety of SEDs found for the radio-excess sources (Sect. \ref{sample}). The SEDs on the left are consistent with a simple galaxy template: SFG template (``IR SFG'', top) or elliptical galaxy template (``passive'', bottom; see Sect. \ref{seds}); the SFG template upper limit (grey line) is also shown. We note that the elliptical template (long dashed line) is not fitted to the data, but it is only shown to demonstrate that it can well represent the data. The SEDs on the right are fitted with a star-forming galaxy (dashed line) $+$ AGN (dotted line) model (``IR AGN''): on the top there is an X-ray detected AGN, on the bottom an X-ray undetected AGN. The total SEDs are shown as black solid lines. The filled circles represent the {\it Spitzer} 8, 16, 24 $\mu$m and the {\it Herschel} 100, 160, 250 $\mu$m flux densities (in mJy), used to constrain the SEDs. The open symbols indicates the data that were not included in the SED fitting process: red triangles are {\it Spitzer}-IRAC 3.6, 4.5, 5.8 $\mu$m flux densities, black circles are SPIRE 350 and 500 $\mu$m and black squares are VLA 1.4 GHz flux densities; the radio data do not match the SEDs in these cases, because the sources have excess radio emission compared to that expected from pure star formation (Sect. \ref{sample}). The blue star represents the 6 $\mu$m luminosity of the AGN predicted from the X-ray luminosity in the rest-frame 2-10 keV, using the \protect\citet{lutz2004} relation for local unobscured AGN; we note that this point does not match the IR AGN component because the X-ray luminosity tends to underestimate the intrinsic AGN power if the AGN emission is heavily absorbed (see Sect. \ref{iragn}). On the top left corner of each plot the source redshifts are indicated as well as the corresponding source number in Table 1 (col.1).}\label{fig.sed}
\end{figure*}

The flux densities at 8 $\mu$m, 16 $\mu$m, 24 $\mu$m from \spz\ and 100 $\mu$m, 160 $\mu$m and 250 $\mu$m from \her\ have been used in the SED fitting process to constrain the SEDs of our sources (Fig. \ref{fig.sed}). In the case of non-detections (S/N$<$3), the measured fluxes at each source positions, with the large associated uncertainties, were used in the SED fits. The fluxes in the shorter \spz--IRAC bands (3.6, 4.5, 5.8 $\mu$m) were not included in the SED fitting process as they fall out of the wavelength range covered by our templates at relatively low redshifts $z>0.2$. Moreover, at these wavelengths the observed SED is often dominated by the old stellar population emission, which is not accounted for in our SED templates. Data at longer wavelengths from \her--SPIRE 350 $\mu$m and 500 $\mu$m (\citealt{elbaz2011}) were also not included in the fits, because of the larger uncertainties on the measured fluxes due to strong blending (see Sect. \ref{fir}) and the low number of significantly detected sources. However, these fluxes (or upper limits) have been included when plotting the SEDs, as a visual check of the best-fitting SED solutions (see Fig. \ref{fig.sed} and Fig. \ref{fig.sedall}). 

We note that only using photometry at $\lambda<250\ \mu$m (SPIRE-250 $\mu$m band) means that at high redshifts ($z\gtrsim1.3$) we are not able to fully constrain the FIR SED peak at wavelengths longer than $\lambda\approx100\ \mu$m (rest-frame) and therefore we cannot exclude the presence of a colder dust component, which would yield higher FIR fluxes than that predicted from our model. However we can anticipate (see Sect. \ref{sample}) that we find very good agreement between our average FIR--radio relation and that found in several previous works.

The SED fitting process was developed as follows: 
\begin{enumerate}
\item Initially, only the star-forming galaxy templates were used in the fit; we fitted the data of each source with each of the five SFG templates (Fig. \ref{fig.sb}) using $\chi^2$ minimisation to evaluate the goodness of the fit; \smallskip 

\item As a second step, we performed new fits, again using $\chi^2$ minimisation, by adding the AGN component, also including extinction of the AGN component as a free parameter (varying between $A_V\approx0-30$ mag, corresponding to $N_{\rm H}\approx 0-5\times10^{22}$ cm$^{-2}$ assuming the average galactic dust-to-gas ratio $A_V=N_{\rm H}/(1.8\times10^{21})$; e.g. \citealt{predehl1995}), to each SFG template (SFG $+$ AGN); \smallskip 

\item Finally, an $f$-test was performed using the $\chi^2$ values and the degrees of freedom ($d.o.f.$) for all of the five pairs of solutions to evaluate the improvement of the fit due to the addition of the AGN component. We accepted the SFG $+$ AGN model as the best-fit if the AGN component significantly (i.e. $>90$\% confidence level, according to the $f$-test probability) improved the resulting $\chi^2$ for the majority of the solutions, i.e. in at least three out of five fitting solutions.
\end{enumerate}
The criteria we adopted to define the best-fitting model were established after performing several tests on our SED fitting approach; in particular, we tested these criteria on a sub-sample of our sources for which detailed \spz$-IRS$ MIR spectra are available (see Appendix A, for details). 
\begin{figure*}[!t]
\centering{
\includegraphics[scale=0.7]{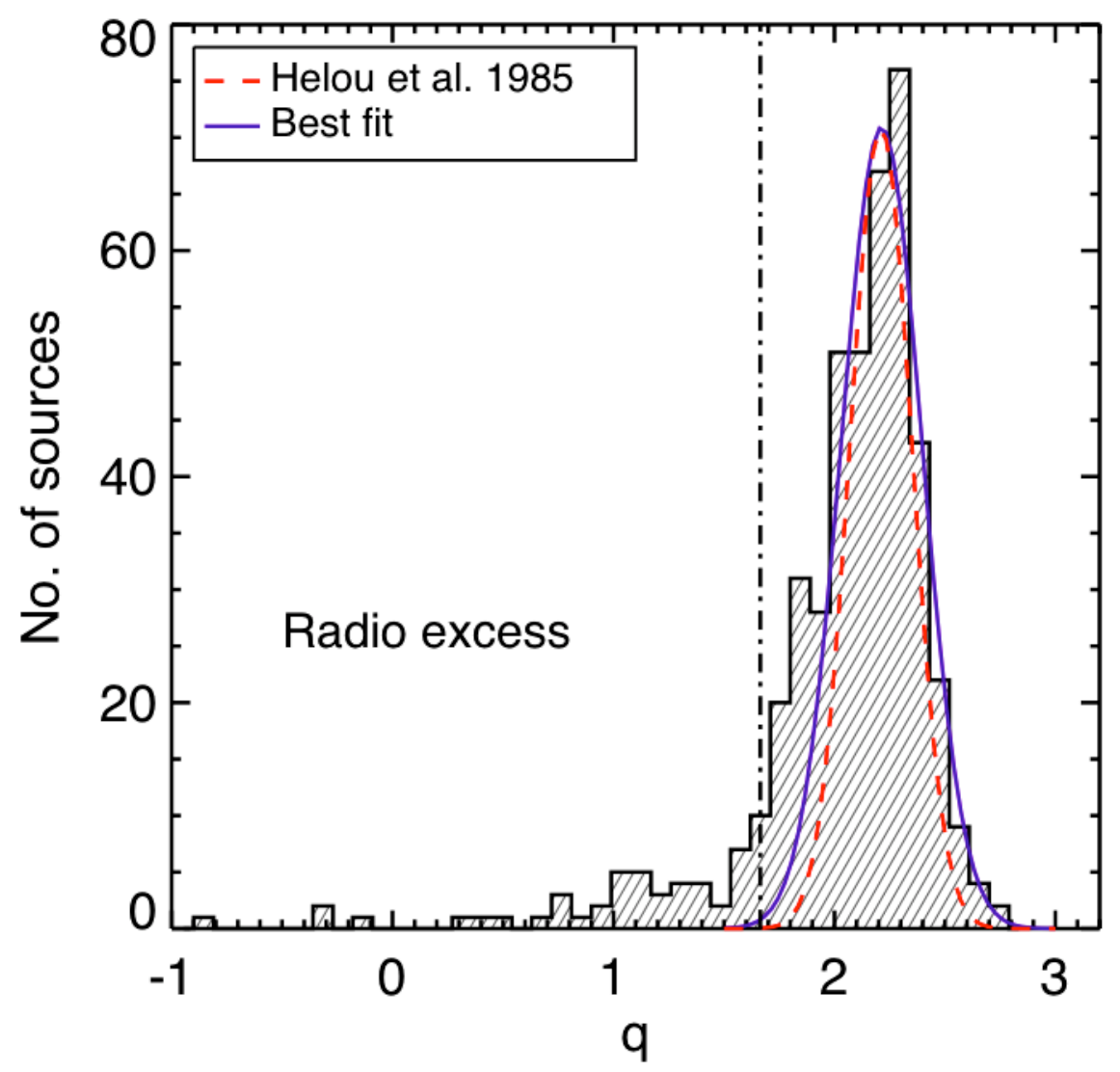}
\hspace{0.3cm}
\includegraphics[scale=0.7]{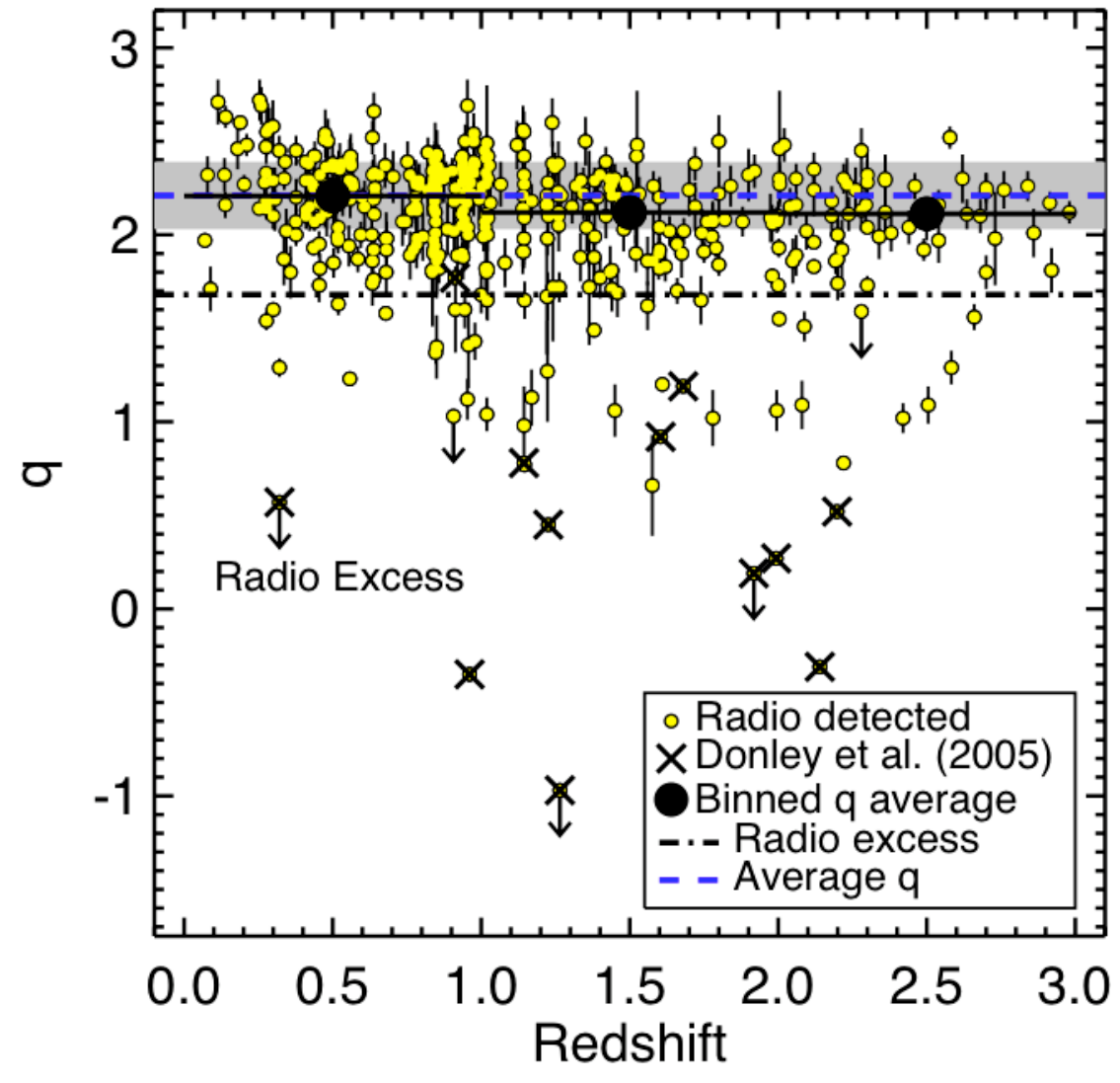}
}
\caption{{\it Left:} Distribution of FIR--radio flux ratio ($q$) for all of the sources in the VLA/24 $\mu$m sample (458 sources, excluding the upper limits; see Sect. \ref{sample}); the blue Gaussian profile represents the best fit to the peak ($q_{mean}=2.21\pm0.18$), while the red dashed Gaussian is the $q$ distribution found by \protect\citet{helou1985} for local starburst galaxies. The vertical line indicates the separation we assumed to select our radio-excess sources, $q=1.68$, corresponding to 3$\sigma$ from the peak (2$\sigma$ when include radio upper limits).
{\it Right:} FIR--radio flux ratio ($q$) as a function of redshift for all of the VLA/24 $\mu$m sources; the radio-excess sources have $q<1.68$ (dot-dashed line); the solid horizontal line represents the average $q$ value for the ``radio-normal'' sources ($q=2.21$) and the shaded region indicates the $\pm1\sigma$ error from this average. The radio-excess sources identified by \protect\citet{donley2005} are plotted as crosses (see Sect. \ref{q24}). The binned q average for the ``radio-normal'' sources, in three redshift bins, is also shown as black circles; we found no significant evidence of evolution of the FIR--radio correlation out to $z=3$.}
\label{fig.sel}
\end{figure*}


Once the best-fit model was defined, the final measurements of the properties of the sources (i.e., FIR flux, relative AGN/SFG contribution, etc.) and their errors were derived as weighted averages of the values obtained from the five best-fit model solutions (see Appendix A for details). For the AGN $+$ SFG models we included in the average only the solutions where the AGN component was significant ($>90$\% confidence level). This is because, due to the sparse data used to constrain the SEDs, in some cases the five solutions obtained from the different SFG templates were similar (small difference in $\chi^2$ values) and did not allow us to unambiguously determine a unique solution that best characterised the data (see Appendix A, for details). 

\subsection{Radio-excess sample selection}\label{sample}
The FIR--radio correlation is typically defined as the ratio between the flux in the FIR band ($\sim40-120\ \mu$m; rest-frame) and the flux density in the rest-frame radio band (1.4 GHz; e.g. \citealt{helou1985,sargent2010a}). Many studies performed so far at high redshift relied on either MIR fluxes as a ``proxy'' for the FIR (or bolometric IR) flux (e.g. from the $S_{\rm 24\ \mu m}$; \citealt{appleton2004,donley2005}), 70 $\mu$m flux density (e.g. \citealt{appleton2004,seymour2009,sargent2010a,bourne2011}), or on SED fitting spanning only the rest-frame MIR band (e.g. \citealt{sargent2010a,padovani2011}). These methods are often inaccurate because they require assumptions about the source SED over the whole IR band and/or on the bolometric corrections. Through our detailed SED analysis of the 458 VLA/24 $\mu$m sources with $z\le3.0$, using the approach described in the previous section (Sect. \ref{sed}) and the \her\ data to constrain the FIR SED peak, we can overcome these issues by directly measuring the galaxy emission over the whole IR band.

From the best-fit models, we calculated the FIR flux ($f_{\rm FIR}$) by directly integrating the total SEDs over the rest-frame wavelength range $\lambda=42.5-122.5\ \mu$m (e.g. \citealt{helou1985}). We used the total SED, which in many cases includes contributions from both SFG and AGN components, to calculate the FIR flux because we aim at a conservative selection of radio-excess AGN, since radio quiet AGN often follow the typical FIR--radio relation of star-forming galaxies.

In only $\approx$3\% of the 458 analysed sources (15 sources) was the SED fitting analysis unable to provide a good representation of the mid- and far-IR data.\footnote{The poor fits were flagged by visual inspection of the resulting SEDs; these are the cases where even the best-fitting SEDs deviate significantly from the data.} In most of the cases (6/15), this was due to large uncertainties on the photometric redshift, or possibly to spurious counterpart associations between the different catalogues (see Sect. \ref{reds}). In a smaller number of cases (4/15), the poor fitting results seemed to be due to large uncertainties on the flux density measurements, especially at wavelengths where the sources are not significantly detected ($\rm S/N<3$). We flagged these ``problematic'' cases in Table 1 (column ``Fit''), being aware that the measurements obtained from their SEDs are not fully reliable. The remainder of these sources (5/15) have strong emission in the \spz\ IRAC bands, even stronger than the flux detected at 24 $\mu$m ($S_8/S_{24}>1$), and are not detected at longer wavelengths (FIR) by \her. This suggests that the emission from star formation (or AGN) in these sources is weak, while the emission from old stellar population, usually dominating the rest-frame NIR band ($\lambda<5\ \mu$m, rest-frame), is strong; their infrared SEDs are therefore more consistent with that of galaxies dominated by passive stellar populations rather than by active star formation. For these sources FIR flux upper limits were estimated by normalising the SFG templates to the 24 $\mu$m datapoint (Fig. \ref{fig.sed} and Fig. \ref{fig.sedall}), which is thus used as a proxy of star formation, and integrating the SED between $42.5-122.5\ \mu$m. Of the five measurements obtained from the different SFG templates, the maximum has been taken as the $f_{\rm FIR}$ upper limit.

Using the definition given by \citet{helou1985}, we calculated the ratio between the far-infrared and radio emission ($q$) as: 
\begin{equation}
q=log\ [f_{\rm FIR}/(3.75\times10^{12}{\rm\ Hz})]-log\ [S_{\nu} (1.4\rm\ GHz)]
\end{equation}
where $f_{\rm FIR}$ is in units of W m$^{-2}$, $3.75\times10^{12}$ Hz is the frequency at the centre of the FIR band ($\lambda=80\ \mu$m) and $S_{\nu} (1.4\rm\ GHz)$ is the radio flux density (in units of W m$^{-2}$ Hz$^{-1}$) at rest-frame 1.4~GHz, extrapolated from the VLA data using the power-law slope $S_{\nu}\propto\nu^{-\alpha}$, with $\alpha=0.7$, typical for star-forming galaxies (e.g., \citealt{ibar2009,ibar2010}).

The $q$ distribution obtained for the entire sample is shown in Fig. \ref{fig.sel} (left); the sources where $f_{\rm FIR}$ (and therefore also $q$) is an upper limit are not included in the histogram. The peak of the distribution is at $q\approx2.2$, in excellent agreement with the values typically obtained for star-forming/starburst galaxies (e.g. \citealt{helou1985, condon1992}). However, the distribution is not symmetrical around the peak and shows a broad tail at low $q$ values, indicating a relatively large number of sources with excess radio emission over that expected from star formation processes. 

In order to identify the radio-excess sources, we fitted the peak of the $q$ distribution, considering only sources with $q>2.0$, with a Gaussian profile and estimated the spread of the FIR--radio correlation expected for star-forming galaxies: from the best-fit we obtained $q=2.21\pm0.18$ (see Fig. \ref{fig.sel}). If we also include the radio undetected sources ($\rm S/N<3$) in the $q$ distribution, we obtain a Gaussian profile with a very similar peak, but larger scatter ($q=2.24\pm0.29$), due to the larger uncertainties on the $q$ values. The $q$ values for the radio undetected sources were calculated using the radio 1.4 GHz flux measurements at the 24 $\mu$m source positions (Sect. \ref{rad}). We set the separation between ``radio-normal'' and ``radio-excess'' sources at $q=1.68$, corresponding to a 3$\sigma$ deviation from the peak of the distribution for the VLA/24 $\mu$m sample ($\sim$2$\sigma$ from the peak for the whole 24 $\mu$m sample). We defined ``radio-normal'' as the sources with $q>1.68$, that follow the typical FIR--radio relation ($q\approx2.2$); we note that this population includes star-forming galaxies, but also typical radio-quiet AGN (e.g. \citealt{moric2010,sargent2010a,padovani2011}), while the large majority of the radio-excess sources are most likely to host AGN activity. 

\begin{figure}[!t]
\centerline{\hspace{-0.2 cm}
\includegraphics[scale=0.6]{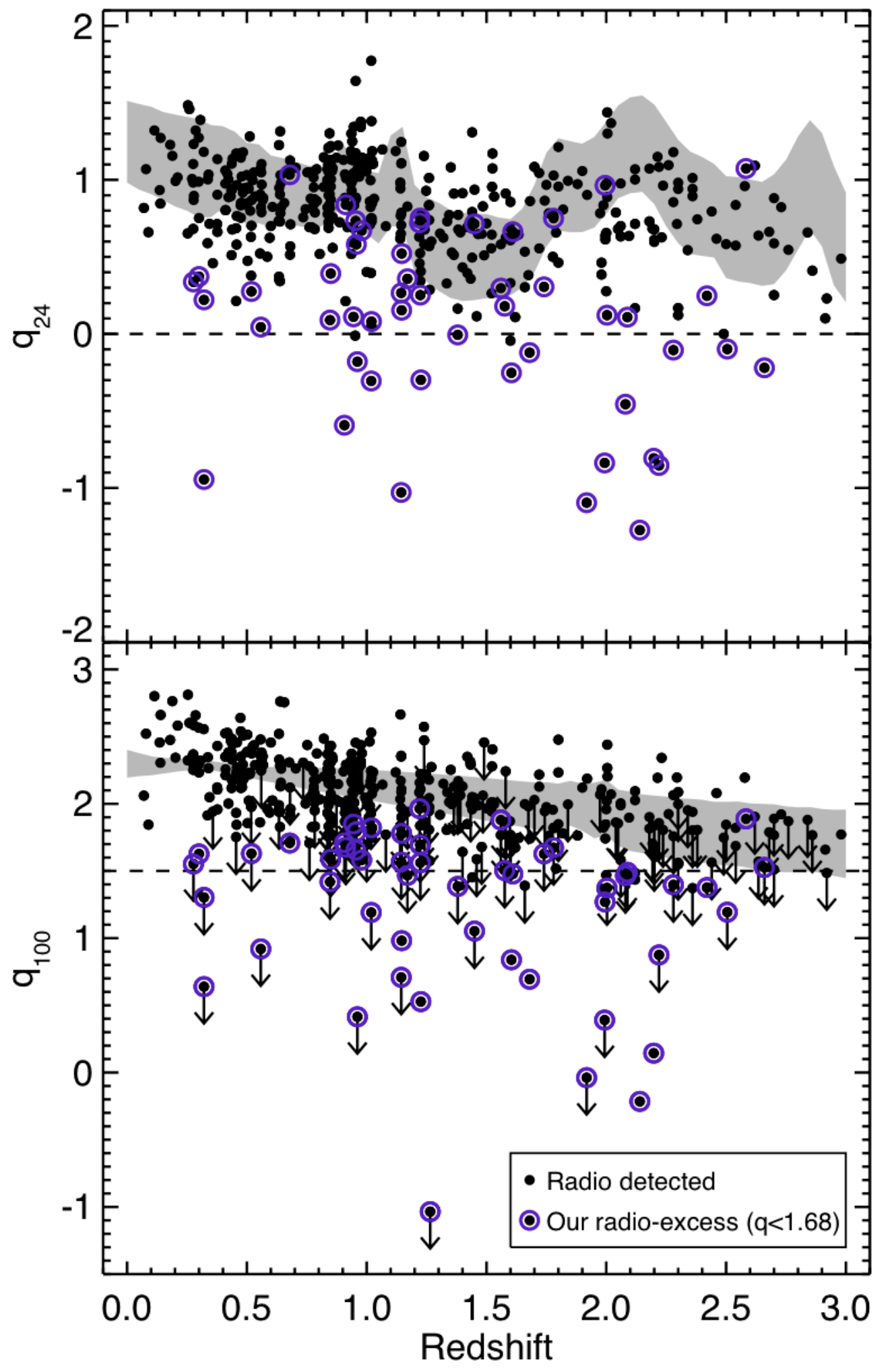}}
\caption{Infrared-to-radio flux ratios, using the observed 24 $\mu$m ($q_{24}$; top) and 100 $\mu$m flux densities ($q_{100}$; bottom); the radio-excess sources in our sample ($q<1.68$) are highlighted by open circles. The shaded regions in the two plots represent the $q_{24}$ and $q_{100}$ ratios predicted for our range of star-forming galaxies as a function of redshift. The horizontal line in the top plot is the $q_{24}=0$ selection used by \protect\citet{donley2005} to define their radio-excess sample. In the bottom panel we identified $q_{100}=1.5$ as an easy selection of radio-excess sources (dashed line), with reasonable completeness ($\approx$60-80\%) at least out to $z\approx2$.}
\label{fig.qz}
\end{figure}

With our selection we obtained a sample of 51 radio-excess AGN candidates, $\sim11$\% of the whole VLA/24 $\mu$m detected sample at $z\le3.0$. 
In Fig. \ref{fig.sel} (right) we show the FIR--radio ratio ($q$) for the entire sample as a function of redshift. The horizontal dashed line indicates the $q$ value at the peak of the distribution ($q=2.21\pm0.18$) with 1$\sigma$ uncertainty (shaded region), while the dot-dashed line at $q=1.68$ represents our threshold for radio-excess sources (3$\sigma$ deviation from the peak). In the plot we also marked 12 of the \citet{donley2005} radio-excess sources that are found within the GOODS-\her\ (North) field and overlap with our sample\footnote{We note that for 6 sources the redshifts listed in \citet{donley2005} are in disagreement with ours (see Table \ref{tab1}): in five cases we have new spectroscopic redshifts where there were only photometric redshift estimates (or no estimates at all) in \citet{donley2005}; in the remaining case, we have a photometric redshift where there was no redshift measurement for this source in \citet{donley2005}.} (see Sect. \ref{q24}). 
In Fig. \ref{fig.sel} (right) we also show that the average $q$ for all of the radio-normal sources, calculated in three different redshift bins ($z_1=0.0-1.0$, $z_2=1.0-2.0$, $z_3=2.0-3.0$, black circles), remains fairly constant, within the errors, over the whole redshift range, and therefore, the apparent decrease of the $q$ values at high redshift is not significant (e.g. \citealt{elbaz2002,sargent2010b,mao2011,bourne2011}).


\begin{longtab}
\begin{footnotesize}
\begin{landscape}
\begin{longtable}{c c c c c c c c c c c c c c c}
\caption{\label{tab1} Radio-excess sample}\tabularnewline
\hline\hline
\rule[-2mm]{0.pt}{4ex} No. & $z$ &  IR Coord. & $S_{\rm 1.4\ GHz}$ & $q$  & log $L_{\rm 1.4\ GHz}$  & VLBI & log $L_{\rm FIR}$ & sSFR & M$^{*}$ & log $\nu L_{\rm 6\ \mu m,\ AGN}$
& Fit & log $L_{\rm 2-10\ keV}$  & IR class &  $z$ Ref. \tabularnewline
  &  &  $h:m:s\ d:m:s$   &  $\mu$Jy  &              & W/Hz      &         & L$_{\odot}$ & Gyr$^{-1}$ & M$_{\odot}$ & erg/s &         &  erg/s &  & \tabularnewline
\rule[-1.7mm]{0.pt}{3ex}  (1) & (2) & (3) & (4)   & (5)  &     (6)    & (7)   & (8) & (9) &  (10)       &  (11) & (12) & (13) & (14) & (15)\tabularnewline
\hline	
\endfirsthead
\caption{Continued} \tabularnewline	
\hline\hline
\rule[-2mm]{0.pt}{4ex} No. & $z$ &  IR Coord. & $S_{\rm 1.4\ GHz}$ & $q$  & log $L_{\rm 1.4\ GHz}$  & VLBI & log $L_{\rm FIR}$ & sSFR & M$^{*}$ & log $\nu L_{\rm 6\ \mu m,\ AGN}$
& Fit & log $L_{\rm 2-10\ keV}$  & IR class &  $z$ Ref. \tabularnewline
  &  &  $h:m:s\ d:m:s$   &  $\mu$Jy  &              & W/Hz      &         & L$_{\odot}$ & Gyr$^{-1}$ & M$_{\odot}$ & erg/s &         &  erg/s &  & \tabularnewline
\rule[-1.7mm]{0.pt}{3ex}  (1) & (2) & (3) & (4)   & (5)  &     (6)    & (7)   & (8) & (9) &  (10)       &  (11) & (12) & (13) & (14) & (15)\tabularnewline
\hline	
\endhead
\hline
\endfoot             
\rule[-0.7mm]{0.pt}{4ex}1 & 2.505 & 12:36:06.84 +62:10:21.4 & 88.7$\pm$9.0 & 1.09$\pm$0.10   & 24.51 &           & 11.59    & 0.311  & 11.60   & 44.41 &    0 & 43.44    & IR AGN  & 1 \tabularnewline
\rule[-1mm]{0.pt}{3ex}  2 & 0.679 & 12:36:08.13 +62:10:35.9 & 213.1$\pm$7.9 & 1.58$\pm$0.04  & 23.58 &\checkmark & 11.15    & 0.456  & 10.90   & 44.47 &    0 & 42.30    & IR AGN  & 1 \tabularnewline
\rule[-1mm]{0.pt}{3ex}  3 & 0.850 & 12:36:08.88 +62:14:30.8 & 29.9$\pm$5.0 & 1.40$\pm$0.17   & 22.95 &           & 10.34    & 0.400  & 10.20   & 43.14 &    0 & $<$41.80 & IR AGN  & 1 \tabularnewline
\rule[-1mm]{0.pt}{3ex}  4 & 1.78  & 12:36:16.06 +62:11:07.7 & 30.9$\pm$4.6 & 1.02$\pm$0.15   & 23.72 &           & 10.73    & 0.030  & 11.20   & 44.61 &    0 & 44.41    & IR AGN  & 3 \tabularnewline
\rule[-1mm]{0.pt}{3ex}  5 & 0.847 & 12:36:17.09 +62:10:11.4 & 65.3$\pm$8.3 & 1.37$\pm$0.06   & 23.29 &           & 10.65    & 0.096  & 11.20   & $<$42.77 & 0 & 43.00    & IR SFG   & 1 \tabularnewline
\rule[-1mm]{0.pt}{3ex}  6 & 1.993 & 12:36:17.55 +62:15:40.5 & 214.7$\pm$8.2 & 0.27$\pm$0.02  & 24.67 &           & 10.93    & 0.0    & 0.00    & $<$43.19 & 0 & $<$42.77 & IR SFG  & 1 \tabularnewline
\rule[-1mm]{0.pt}{3ex}  7 & 2.22  & 12:36:21.27 +62:17:08.2 & 162.9$\pm$7.0 & 0.78$\pm$0.02  & 24.66 &           & 11.43    & 0.354  & 11.40   & $<$43.41 & 2 & $<$42.81 & IR SFG   & 3 \tabularnewline
\rule[-1mm]{0.pt}{3ex}  8 & 2.583 & 12:36:22.94 +62:15:26.5 & 46.3$\pm$4.3 & 1.29$\pm$0.09   & 24.26 &           & 11.54    & $-$    & 11.20   & 45.57 &    0 & 44.77    & IR AGN  & 1 \tabularnewline
\rule[-1mm]{0.pt}{3ex}  9 & 1.918 & 12:36:23.53 +62:16:42.6 & 446.5$\pm$14.4 & $<$0.19       & 24.95 &\checkmark & $<$11.13 & $<$0.082 & 11.70 & $<$42.93 & 0 & $<$42.64 & Passive & 6 \tabularnewline
\rule[-1mm]{0.pt}{3ex} 10 & 0.954 & 12:36:30.06 +62:09:23.9 & 41.1$\pm$4.7 & 1.12$\pm$0.11   & 23.21 &           & 10.32    & 0.109  & 10.60   & 44.36 &    0 & 42.79    & IR AGN  & 1 \tabularnewline
\rule[-1mm]{0.pt}{3ex} 11 & 0.957 & 12:36:32.02 +62:09:17.1 & 18.8$\pm$4.3 & 1.41$\pm$0.23   & 22.87 &           & 10.27    & 0.089  & 10.80   & 43.39 &    0 & $<$42.08 & IR AGN  & 4 \tabularnewline
\rule[-1mm]{0.pt}{3ex} 12 & 0.519 & 12:36:32.48 +62:11:05.4 & 34.6$\pm$4.6 & 1.63$\pm$0.06   & 22.52 &           & 10.13    & 0.024  & 11.30   & $<$42.37 & 0 & $<$41.27 & IR SFG   & 1 \tabularnewline
\rule[-1mm]{0.pt}{3ex} 13 & 1.995 & 12:36:32.55 +62:07:59.8 & 89.1$\pm$10.2 & 1.06$\pm$0.11  & 24.29 &           & 11.34    & $-$    & 11.20   & 45.37 &    0 & 43.54    & IR AGN  & 1 \tabularnewline
\rule[-1mm]{0.pt}{3ex} 14 & 2.66  & 12:36:36.92 +62:13:20.4 & 49.3$\pm$7.7 & 1.56$\pm$0.07   & 24.31 &           & 11.86    & 3.947  & 10.80   & $<$43.83 & 0 & 42.99    & IR SFG   & 3 \tabularnewline
\rule[-1mm]{0.pt}{3ex} 15 & 1.02  & 12:36:38.80 +62:15:10.4 & 19.0$\pm$4.3 & 1.65$\pm$0.10   & 22.94 &           & 10.58    & 0.165  & 10.90   & $<$42.67 & 0 & $<$42.00 & IR SFG   & 3 \tabularnewline
\rule[-1mm]{0.pt}{3ex} 16 & 1.144 & 12:36:40.55 +62:18:32.9 & 349.9$\pm$11.7 & 0.78$\pm$0.01 & 24.32 &           & 11.10    & 0.213  & 11.30   & $<$43.23 & 0 & 41.56    & IR SFG   & 2 \tabularnewline
\rule[-1mm]{0.pt}{3ex} 17 & 1.019 & 12:36:44.51 +62:11:42.0 & 45.1$\pm$9.1 & 1.04$\pm$0.09   & 23.32 &           & 10.35    & 0.759  & 10.00   & $<$42.51 & 0 & $<$41.86 & IR SFG   & 1 \tabularnewline
\rule[-1mm]{0.pt}{3ex} 18 & 0.277 & 12:36:45.61 +62:19:39.3 & 43.7$\pm$5.3 & 1.54$\pm$0.05   & 21.99 &           & 9.52     & 0.071  & 10.20   & $<$41.69 & 0 & $<$40.89 & IR SFG  & 1 \tabularnewline
\rule[-1mm]{0.pt}{3ex} 19 & 2.003 & 12:36:46.07 +62:14:48.8 & 109.6$\pm$5.3 & 1.55$\pm$0.02  & 24.38 &           & 11.92    & 4.347  & 10.80   & $<$43.88 & 0 & 42.08    & IR SFG & 1 \tabularnewline
\rule[-1mm]{0.pt}{3ex} 20 & 0.961 & 12:36:46.34 +62:14:04.6 & 322.4$\pm$12.2 & -0.35$\pm$0.04& 24.11 &\checkmark & 9.75     & $-$    & 11.20   & 44.24 &    0 & 43.92    & IR AGN  & 1 \tabularnewline
\rule[-1mm]{0.pt}{3ex} 21 & 1.223 & 12:36:47.89 +62:10:45.5 & 17.6$\pm$4.7 & 1.27$\pm$0.27   & 23.09 &           & 10.35    & $-$    & 11.00   & 44.25 &    0 & $<$42.15 & IR AGN  & 1 \tabularnewline
\rule[-1mm]{0.pt}{3ex} 22 & 1.610 & 12:36:49.65 +62:07:38.1 & 312.9$\pm$10.9 & 1.20$\pm$0.03 & 24.62 &           & 11.81    & 1.149  & 11.10   & 45.37 &    0 & 44.51    & IR AGN  & 1 \tabularnewline
\rule[-1mm]{0.pt}{3ex} 23 & 0.300 & 12:36:51.73 +62:12:20.4 & 56.0$\pm$4.5 & 1.60$\pm$0.03   & 22.18 &           & 9.77     & 0.427  & 9.70    & $<$42.17 & 0 & 41.78    & IR SFG   & 1 \tabularnewline
\rule[-1mm]{0.pt}{3ex} 24 & 1.224 & 12:36:52.08 +62:12:26.3 & 17.1$\pm$5.0 & 1.67$\pm$0.13   & 23.08 &           & 10.74    & 0.238  & 10.90   & $<$42.92 & 0 & $<$41.92 & IR SFG   & 1 \tabularnewline
\rule[-1mm]{0.pt}{3ex} 25 & 0.321 & 12:36:52.89 +62:14:44.0 & 198.3$\pm$9.6 & $<$0.57        & 22.80 &\checkmark & $<$9.36  & $<$0.005 & 11.10 & $<$41.15 & 0 & 40.98    & Passive & 1 \tabularnewline
\rule[-1mm]{0.pt}{3ex} 26 & 0.321 & 12:36:56.50 +62:12:08.4 & 38.4$\pm$4.1 & 1.29$\pm$0.05   & 22.08 &           & 9.36     & 1.546  & 8.70    & $<$41.42 & 0 & $<$40.74 & IR SFG   & 1 \tabularnewline
\rule[-1mm]{0.pt}{3ex} 27 & 2.14  & 12:36:59.32 +62:18:32.4 & 5505.0$\pm$165.3 & -0.31$\pm$0.03 & 26.15 &        & 11.83    & 2.502  & 10.90   & 44.96 &    0 & 43.38    & IR AGN  & 3 \tabularnewline
\rule[-1mm]{0.pt}{3ex} 28 & 1.146 & 12:36:59.82 +62:19:34.2 & 19.9$\pm$4.5 & 1.65$\pm$0.10   & 23.08 &           & 10.72    & 0.179  & 11.00   & $<$42.85 & 0 & 42.89    & IR SFG   & 1 \tabularnewline
\rule[-1mm]{0.pt}{3ex} 29 & 1.68  & 12:37:00.26 +62:09:09.9 & 353.0$\pm$11.5 & 1.19$\pm$0.01 & 24.72 &\checkmark & 11.89    & 1.286  & 11.30   & $<$44.07 & 0 & $<$42.58 & IR SFG   & 5 \tabularnewline
\rule[-1mm]{0.pt}{3ex} 30 & 1.576 & 12:37:02.84 +62:16:01.2 & 16.0$\pm$4.3 & 0.66$\pm$0.27   & 23.31 &           & 9.95     & 0.091  & 10.30   & 43.60 &    0 & 43.33    & IR AGN  & 1 \tabularnewline
\rule[-1mm]{0.pt}{3ex} 31 & 1.170 & 12:37:04.87 +62:16:01.6 & 28.5$\pm$4.4 & 1.13$\pm$0.15   & 23.26 &           & 10.38    & 0.085  & 10.80   & 43.85 &    0 & 43.60    & IR AGN  & 1 \tabularnewline
\rule[-1mm]{0.pt}{3ex} 32 & 1.450 & 12:37:07.49 +62:21:47.8 & 46.3$\pm$6.4 & 1.06$\pm$0.14   & 23.69 &           & 10.73    & 0.038  & 11.30   & 44.54 &    1 & 44.27    & IR AGN  & 1 \tabularnewline
\rule[-1mm]{0.pt}{3ex} 33 & 0.907 & 12:37:09.43 +62:08:37.6 & 146.8$\pm$6.4 & $<$1.03        & 23.71 &           & $<$10.73 & $<$0.032 & 11.70 & $<$42.52 & 0 & 42.19    & Passive & 1 \tabularnewline
\rule[-1mm]{0.pt}{3ex} 34 & 2.08  & 12:37:09.57 +62:22:02.2 & 36.9$\pm$11.3 & 1.09$\pm$0.13  & 23.95 &           & 11.02    & 1.397  & 10.40   & $<$42.98 & 2 & $<$42.14 & IR SFG   & 3 \tabularnewline
\rule[-1mm]{0.pt}{3ex} 35 & 1.226 & 12:37:09.95 +62:22:58.9 & 703.9$\pm$22.1 & 0.45$\pm$0.03 & 24.70 &           & 11.14    & 0.160  & 11.40   & 44.71 &    0 & 43.51    & IR AGN  & 4 \tabularnewline
\rule[-1mm]{0.pt}{3ex} 36 & 2.28  & 12:37:13.38 +62:10:44.7 & 28.4$\pm$4.6 &  $<$1.59        & 23.92 &           & $<$11.50 & $<$0.152 & 11.80 & $<$43.30 & 0 & $<$42.92 & Passive & 3 \tabularnewline
\rule[-1mm]{0.pt}{3ex} 37 & 0.558 & 12:37:16.38 +62:15:12.3 & 146.3$\pm$8.5 & 1.23$\pm$0.03  & 23.21 &\checkmark & 10.44    & 0.048  & 11.30   & $<$42.66 & 0 & 41.24    & IR SFG   & 1 \tabularnewline
\rule[-1mm]{0.pt}{3ex} 38 & 1.146 & 12:37:16.67 +62:17:33.3 & 373.0$\pm$12.0 & 0.77$\pm$0.03 & 24.35 &\checkmark & 11.12    & 0.133  & 11.10   & 44.95 &    0 & 43.99    & IR AGN  & 1 \tabularnewline
\rule[-1mm]{0.pt}{3ex} 39 & 1.221 & 12:37:17.16 +62:13:52.2 & 13.8$\pm$4.3 & 1.67$\pm$0.31   & 22.99 &           & 10.64    & 0.340  & 10.50   & 43.97 &    0 & $<$42.26 & IR AGN  & 1 \tabularnewline
\rule[-1mm]{0.pt}{3ex} 40 & 0.945 & 12:37:20.26 +62:22:00.0 & 20.3$\pm$4.6 & 1.60$\pm$0.10   & 22.89 &           & 10.49    & 0.065  & 11.20   & $<$42.59 & 0 & $<$42.54 & IR SFG   & 1 \tabularnewline
\rule[-1mm]{0.pt}{3ex} 41 & 2.088 & 12:37:20.70 +62:10:40.7 & 23.0$\pm$4.3 & 1.51$\pm$0.08   & 23.75 &           & 11.24    & 5.784  & 10.00   & $<$43.21 & 0 & $<$42.89 & IR SFG   & 4 \tabularnewline
\rule[-1mm]{0.pt}{3ex} 42 & 1.604 & 12:37:21.36 +62:11:30.9 & 375.6$\pm$12.1 & 0.92$\pm$0.03 & 24.70 &\checkmark & 11.61    & 0.667  & 11.30   & 44.32 &    0 & 42.94    & IR AGN  & 1 \tabularnewline
\rule[-1mm]{0.pt}{3ex} 43 & 0.979 & 12:37:21.86 +62:10:35.7 & 54.5$\pm$5.7 & 1.43$\pm$0.10   & 23.36 &           & 10.78    & 0.074  & 11.30   & 44.10 &    0 & 42.97    & IR AGN  & 6 \tabularnewline
\rule[-1mm]{0.pt}{3ex} 44 & 1.144 & 12:37:24.01 +62:13:04.3 & 23.2$\pm$4.8 & 0.98$\pm$0.21   & 23.15 &           & 10.11    & 0.033  & 11.00   & 43.50 &    0 & 43.53    & IR AGN  & 1 \tabularnewline
\rule[-1mm]{0.pt}{3ex} 45 & 1.56  & 12:37:24.92 +62:13:12.4 & 17.1$\pm$5.0 & 1.62$\pm$0.13   & 23.33 &           & 10.94    & 0.757  & 10.60   & $<$43.09 & 0 & $<$42.58 & IR SFG   & 3 \tabularnewline
\rule[-1mm]{0.pt}{3ex} 46 & 1.74  & 12:37:25.27 +62:10:55.4 & 15.8$\pm$4.6 & 1.65$\pm$0.13   & 23.40 &           & 11.04    & 0.932  & 10.60   & $<$43.13 & 0 & $<$42.73 & IR SFG   & 3 \tabularnewline
\rule[-1mm]{0.pt}{3ex} 47 & 1.265 & 12:37:25.96 +62:11:28.7 &5650.0$\pm$170.1& $<$-0.97      & 25.63 &           & $<$10.66 & $<$0.044 & 11.50 & $<$42.45 & 0 & $<$42.56 & Passive & 1 \tabularnewline
\rule[-1mm]{0.pt}{3ex} 48 & 2.42  & 12:37:36.87 +62:14:29.0 & 67.3$\pm$5.2 & 1.02$\pm$0.08   & 24.36 &           & 11.37    & 0.144  & 11.60   & 44.84 &    0 & 44.32    & IR AGN  & 3 \tabularnewline
\rule[-1mm]{0.pt}{3ex} 49 & 2.198 & 12:37:46.64 +62:17:38.6 & 1200.0$\pm$36.4 & 0.52$\pm$0.01& 25.51 &           & 12.02    & 0.875  & 11.60   & $<$44.07 & 0 & $<$43.30 & IR SFG   & 4 \tabularnewline
\rule[-1mm]{0.pt}{3ex} 50 & 1.38  & 12:37:50.24 +62:13:59.0 & 75.4$\pm$5.8 & 1.49$\pm$0.03   & 23.85 &           & 11.32    & 0.927  & 10.90   & $<$43.61 & 0 & $<$42.85 & IR SFG   & 3 \tabularnewline
\rule[-1mm]{0.pt}{3ex} 51 & 0.913 & 12:37:58.12 +62:15:16.8 & 19.0$\pm$4.3 & 1.60$\pm$0.23   & 22.83 &           & 10.42    & 0.129  & 10.80   & 43.48 &    0 & $<$42.39 & IR AGN  & 1 \tabularnewline
\end{longtable} 
\tablefoot{												  
Col.1: Source ID number; 
Col.2: redshift; 
Col.3: \spz\ MIPS 24 $\mu$m coordinates; 
Col.4:  measured total radio flux density at 1.4 GHz (Daddi et al., in prep.); 
Col.5: $q$ calculated from the FIR flux (rest-frame 42.5-122.5 $\mu$m) and the rest-frame radio 1.4 GHz flux (calculated assuming $\alpha=0.7$; see Sect.\ref{sample}); 
Col.6: logarithm of the rest-frame radio luminosity at 1.4 GHz; 
Col.7: sources with a compact radio core detected in deep VLBI observations (\citealt{chi2009}; Chi et al. 2012, in prep; see Sect. \ref{xray})
Col.8: logarithm of the total (AGN $+$ SFG) rest-frame FIR luminosity (42.5-122.5 $\mu$m) in units of solar luminosities; 
Col.9: specific star-formation rate (sSFR), calculated from the SFRs and the stellar masses, in Gyr$^{-1}$;
Col 10: logarithm of the stellar mass in solar masses; 
Col 11: logarithm of the 6 $\mu$m luminosity of the AGN component obtained from the SED fit (Sect. \ref{rexc}); where the AGN component was not significantly detected an upper-limit of 
30\% of the total 6 $\mu$m luminosity has been reported (see Appendix A);
Col.12: quality flag for the SED fit; $0=$ good fit; $1=$ problems with \her\ photometry affected the fit; $2=$ potentially incorrect photometric redshift; 
Col.13: logarithm of the rest-frame X-ray luminosity in the 2-10 keV energy band, not corrected for absorption; 
Col.14: IR classification (Sect. \ref{discus});
Col.15: Redshift reference: $1=$ \citet{barger2008}, $2=$ \citet{cowie2004}; $3=$ photometric redshift from Panella et al. (in prep.); $4=$ spectroscopic redshift from
Dickinson et al. (in prep); $5=$ photometric redshift from \citet{pope2006}; $6=$ spectroscopic redshift from \citet{chapman2005}.}
\end{landscape}
\end{footnotesize}
\end{longtab}

\subsection{IR--radio relation: $q_{24}$ and $q_{100}$}\label{q24}

Several previous works that have investigated the FIR--radio correlation and radio-excess sources have used different methods to define the FIR--radio flux ratio. In particular, in many studies the FIR flux has been replaced by the monochromatic flux density at 24 $\mu$m (e.g. \citealt{appleton2004, donley2005, ibar2008, sargent2010a}), 60 $\mu$m (e.g. \citealt{vlakis2007}), or 70 $\mu$m \citep{seymour2009, sargent2010a,mao2011}, as a proxy for the FIR emission of the galaxy. 
It is therefore interesting to see how these definitions of the FIR--radio ratios,calculated from monochromatic flux ratios, compare to $q$ estimated by us (Eq. 1) across the full FIR waveband.
In particular we performed a direct comparison of our sample selection with that used by \citet{donley2005}, where the radio-excess sources were selected using $q_{24}<0$, with $q_{24}={\rm log}\ (S_{24\ \mu \rm m}/S_{1.4\ \rm GHz})$, calculated using the observed $24\ \mu$m and 1.4 GHz flux densities.

As shown in Fig. \ref{fig.sel} (right), the majority of the \citet{donley2005} sources (11/12 that overlap with our sample) are also radio-excess sources based on our definition. However $q_{24}$ does not provide a complete selection, since the majority of our radio-excess sources ($\approx$78\%, 40 objects) are missing from the \citet{donley2005} sample. To verify that this is not only due to the improved sensitivity of our radio data (Sect. \ref{rad}), we compared the number of radio-excess sources we would identify with our $q$ definition and selection ($q<1.68$) using the radio catalogue by \citet{richards2000}, the same as that used by \citet{donley2005}: we obtained 24 radio-excess sources in the GOODS-\her\ field, twice the number of those identified by \citet{donley2005} using $q_{24}$ within the same sky area (12 sources). Moreover, in one case the $q_{24}$ selection by \citet{donley2005} disagrees with our FIR--radio ratio, underestimating the FIR power of the sources (Fig. \ref{fig.sel}, right). This effect is shown more clearly in Fig. \ref{fig.qz}, where we calculated the observed $q_{24}$ and $q_{100}$ for all of the VLA/24 $\mu$m detected sources\footnote{$q_{100}={\rm log}\ (S_{100\ \mu \rm m}/S_{1.4\ \rm GHz}$), where $S_{100\ \mu \rm m}$ and $S_{1.4\ \rm GHz}$ are observed monochromatic fluxes. Since we do not have photometric measurements at 60 $\mu$m or 70 $\mu$m to directly compare our results with those obtained in previous works (e.g. \citealt{vlakis2007,seymour2009, sargent2010a,mao2011}), we used the observed 100 $\mu$m flux density, which is the closest available data point to allow any comparison with $q_{60}$ or $q_{70}$.}. The $q_{24}$ and $q_{100}$ ratios predicted for our range of star-forming galaxies are shown as shaded regions in Figure \ref{fig.qz}, where our radio-excess sources ($q<1.68$) are marked with open circles; the $q_{24}=0$ line adopted by \citet{donley2005} to select radio-excess sources is also shown (Fig. \ref{fig.qz}, top panel). Although the $q_{24}<0$ criterion would select some of our radio-excess sources, more than 60\% of them would be missed, testifying the large incompleteness of this method when compared to the FIR--radio relation $q$ (Eq. 1). Moreover, using $q_{24}<0$ to define the radio-excess sample would include some ``radio-normal'' galaxies (according to our definition; Sect \ref{sample}), introducing some contamination to the sample ($\sim$10\%). 

On the other hand, the $q_{100}$ ratio (Fig. \ref{fig.qz}, bottom) seems to be in better agreement with our $q$ selection than the $q_{24}$ ratio, as few of our radio-excess sources fall into the shaded area. This is expected since the flux observed at 100 $\mu$m is closer to the FIR SED peak than that observed at 24 $\mu$m which is also likely to be affected by the PAH emission features at $z\gtrsim1$ (as illustrated by the shaded region in Fig. \ref{fig.qz}, top) and/or by AGN contribution. By defining a separation line below the shaded area to identify radio-excess sources (e.g. $q_{100}=1.5$; Fig. \ref{fig.qz}) we would recover the majority of our radio-excess sources ($\sim$60\%; $\sim$80\% including the upper limits that lie right above the separation line), a much larger fraction than those identified using $q_{24}$. We stress, however, that the $q_{100}$ (observed) selection is still not complete and much less accurate than our selection method using the rest-frame FIR flux to calculate $q$. 
\begin{figure}[!t]
\centering{
\includegraphics[scale=0.6]{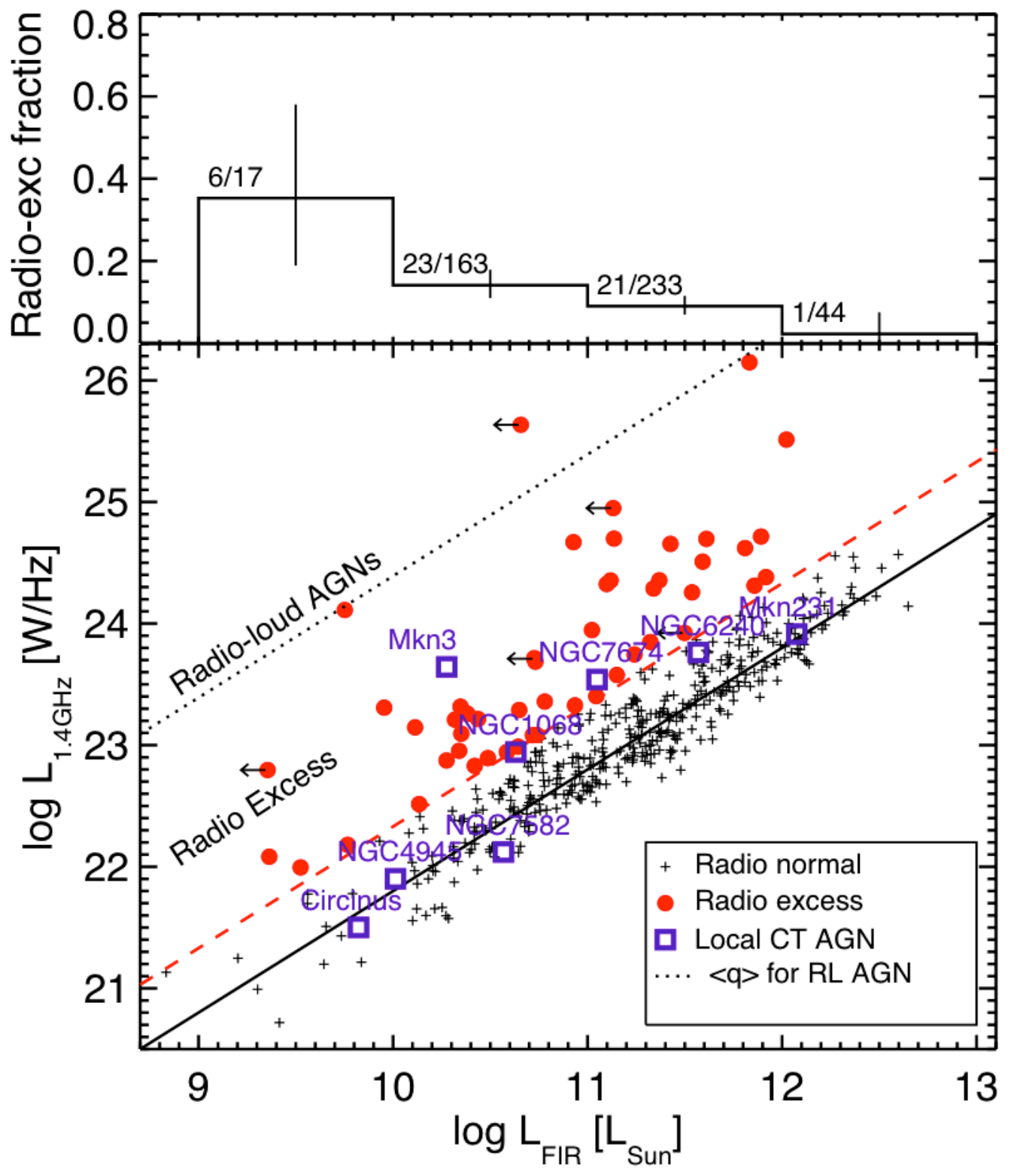}}
\caption{{\it Top panel:} Fraction of radio-excess sources as a function of the rest-frame FIR luminosity (42.5--122.5 $\mu$m, in units of L$_{\odot}$); the fraction decreases with increasing $L_{\rm FIR}$, as expected since sources with high $L_{\rm FIR}$ are more likely to be luminous star-forming galaxies. {\it Bottom panel:} Rest-frame FIR vs. radio 1.4 GHz luminosity (rest-frame); the radio-excess sources are plotted as red circles and some local well known Compton-thick AGNs (open squares) are shown for comparison (data taken from literature). The black dotted line correspond to $\langle q_{\rm RL}\rangle=-0.38$ (average for a sample of RL AGN from \protect\citealt{evans2005}); the black solid line correspond to $q=2.21$ (average for ``radio-normal" sources) and the red dashed line correspond to $q=1.68$, our selection limit for radio-excess sources.}\label{fig.lfir}
\end{figure}

\section{Results}\label{res}
\subsection{Sample properties}\label{res1}
We established in Section \ref{rexc} that the direct measurement of the FIR flux from the best-fit SED provides a reliable and unambiguous way to identify sources that have radio emission in excess to that expected purely from star formation. As described in the previous sections, from our analysis we identified 51 radio-excess sources, which are most likely hosting an AGN. 

In Fig. \ref{fig.lfir}, we show the rest-frame FIR and the radio 1.4 GHz luminosities ($L_{\rm FIR}-L_{1.4\rm\ GHz}$; rest-frame) for all of the ``radio-normal'' (black crosses) and the radio-excess sources (red circles) in the GOODS-\her\ field. The diagonal lines represent the mean $q$ value obtained for the ``radio-normal'' sources ($q=2.21$) and the separation between ``radio-normal'' and radio-excess sources ($q=1.68$). For comparison we also indicate in the plot the average $q$ obtained for a sample of low-redshift ($z\lesssim0.15$) radio-loud AGN (RL; $\langle q_{\rm RL}\rangle=-0.38$) from \citet{evans2005}. For these sources the FIR flux was calculated using the photometry at 60 $\mu$m and 100 $\mu$m, following \citet{helou1985}. Some well studied local Compton-thick AGN taken from literature\footnote{The FIR and radio data for these sources are taken from the NASA/IPAC Extragalactic Database (NED).} (\citealt{dellaceca2008}) are also plotted, as a guide; although the majority of these sources have FIR--radio ratios consistent with that of star-forming galaxies and most radio-quiet AGN ($q\approx2.2$), three out of the eight AGN are radio-excess sources (i.e., Mkn 3, NGC 7674 and NGC 1068), suggesting that some Compton-thick AGN might also be included in our radio-excess sample.

It is important to note that, although some of our radio-excess sources have radio luminosities typical of radio-loud AGN ($L_{rad}>10^{24}$ W~Hz$^{-1}$), the majority lies in a region in between those occupied by RL AGN and ``radio-normal'' sources and therefore can also be referred to as ``radio-intermediate'' sources (e.g. \citealt{drake2003}). We therefore stress that the definition of ``radio-excess'' does not necessarily mean ``radio-loud'' (see Sect. \ref{intro}).

In the top panel of Fig. \ref{fig.lfir}, the fraction of radio-excess sources as a function of the rest-frame FIR luminosity is shown (with 1$\sigma$ uncertainties; \citealt{gehrels1986}). The sources in our sample span a wide range of FIR luminosities, $L_{\rm FIR}\approx10^{9}-10^{12}\ L_{\odot}$ (Table \ref{tab1}), typical of normal star-forming galaxies, luminous IR galaxies (LIRG; $L_{\rm FIR}\approx10^{11} \ L_{\odot}$) and ultra-luminous IR galaxies (ULIRGs; $L_{\rm FIR}\approx10^{12}\ L_{\odot}$). However, the fraction of radio-excess sources decreases with increasing FIR luminosity, and only one source out of 51 reaches the high luminosities typical of ULIRGs. This is not unexpected, since at the high FIR luminosities of ULIRGs, the radio core must be very bright ($L_{\rm rad}>2\times10^{24}$ W~Hz$^{-1}$; RL AGN regime) to be identified as radio-excess above the very strong star-formation activity.

\subsection{X-ray and radio emission from the AGN}\label{xray}
\begin{figure}[!t]
\centerline{
\includegraphics[scale=0.6]{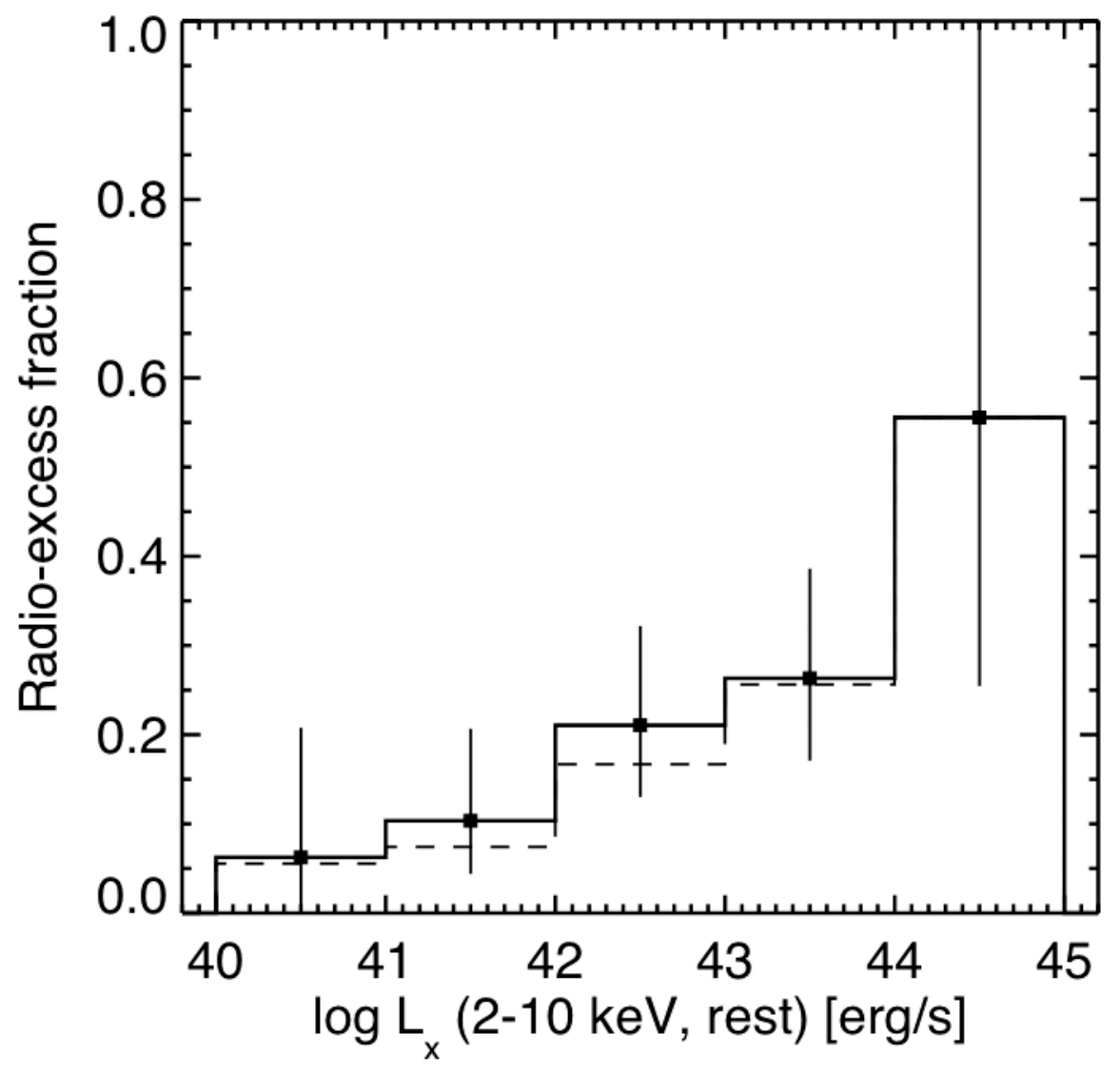}}
\caption{Fraction of radio-excess sources in different X-ray luminosity bins; the fraction of sources with excess radio emission increases with X-ray luminosity, from $\approx$7\% at $L_{\rm X}=10^{40}-10^{41}$ erg~s$^{-1}$ to $\approx$60\% at $L_{\rm X}=10^{44}-10^{45}$ erg~s$^{-1}$. The error bars correspond to 1$\sigma$ uncertainties (\protect\citealt{gehrels1986}). The dashed histogram represents the radio-excess fraction including the X-ray upper limits (Sect. \ref{xx}).}\label{fig.fract}
\end{figure}

Since the excess radio emission is attributed to the presence of an AGN, it is interesting to investigate the X-ray properties of these sources. Amongst our radio-excess AGN sample, only $\sim$53\% (27/51) are detected in the X-ray band (Table 1), hence about half of the sample is X-ray undetected. This suggests that the radio-excess selection is a powerful method to identify AGN that would otherwise be missed by even the deepest X-ray AGN surveys, such as the \ch\ 2 Ms X-ray survey \citep{alexander2003}. 

In Figure \ref{fig.fract}, we show the fraction of radio-excess sources as a function of X-ray luminosity. The sources have been divided into five X-ray luminosity bins from $L_{\rm 2-10\ keV}=10^{40}$ erg s$^{-1}$ to $L_{\rm 2-10\ keV}=10^{45}$ erg s$^{-1}$. We found that the fraction of radio-excess AGN increases from $\approx$7\% at low X-ray luminosities ($L_{\rm 2-10\ keV}<10^{42}$ erg~s$^{-1}$) to $\approx$60\% in the highest luminosity bin ($L_{\rm 2-10\ keV}=10^{44}-10^{45}$ erg~s$^{-1}$). This large increase of radio-excess AGN at high luminosities suggests that the brightest AGN also produce the most powerful radio emission (e.g. \citealt{brinkmann2000, lafranca2010}), presumably the majority of it coming from a radio core. We note that the fact that more luminous AGN are also more radio bright does not mean that high luminosity AGN are typically radio loud. On the contrary, the radio-loudness, measured as $R_{X}=log\ (\nu L_{\rm 1.4\ GHz}/L_{\rm 2-10\ keV})$, decreases with increasing X-ray luminosity (e.g. \citealt{lafranca2010}).

To investigate in detail the radio properties of our radio-excess sources we compared the excess radio emission (i.e. the excess above the radio emission expected from star formation), with the AGN radio core emission detected in deep VLBI 1.4 GHz observations of the Hubble Deep Field North (HDF-N) and the Hubble Flanking Fields (HFF). The details of the VLBI observations are given in \citet{chi2009} and Chi et al. (2012, submitted). Briefly, the sensitivity of the VLBI data varies significantly from the centre of the field (r.m.s. noise level of 7.3 $\mu$Jy/beam within 2$'$ from the phase center) to the outer parts of the field ($2'-8'$; r.m.s. noise level of 14--37 $\mu$Jy/beam), which means that only sources with a radio core brighter than $S_{\rm 1.4\ GHz}\gtrsim100\ \mu$Jy are likely to be detected. Amongst our radio-excess sample, only 13 sources are bright enough (on the basis of the VLA 1.4 GHz flux density; Table \ref{tab1}) to be detected in the VLBI images and for eight of them a compact radio core was indeed detected by VLBI, positively confirming that these sources host an AGN and that the excess radio emission is in fact due to a radio core. The remaining five sources have on average lower radio VLA fluxes compared to the sources detected by VLBI (the only exception is \#47, in Table \ref{tab1}, which is an extended radio-loud, wide-angle-tailed source) and lie in the outer regions of the VLBI field; the non-detection therefore might be due to sensitivity issues, or, as in the case of source \#47, to the radio flux being dominated by extended emission, more than a compact core. The rest of our radio-excess sources lie outside the VLBI region (13 sources) or are typically too faint ($S_{\rm 1.4\ GHz}<100\ \mu$Jy) to be significantly detected above the VLBI sensitivity limit (25 sources). 
\begin{figure}[!t]
\centerline{
\includegraphics[scale=0.6]{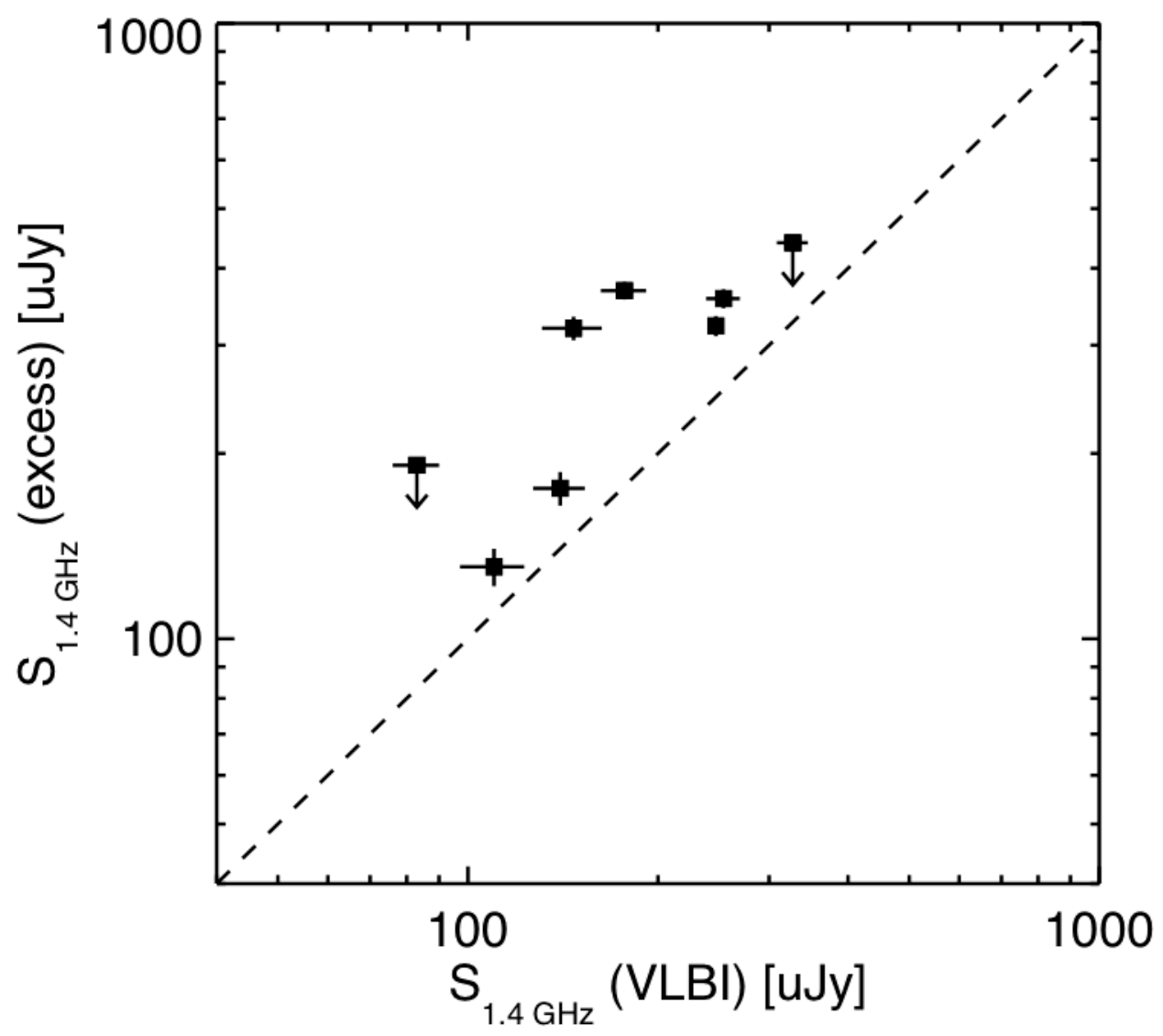}}
\caption{Radio core flux density at 1.4 GHz measured from the VLBI data (Chi et al. 2012, submitted) vs. the excess radio emission (i.e., $S_{\rm1.4\ GHz,\ tot}-S_{\rm1.4\ GHz,\ SFG}$) for our radio-excess sources. Our radio fluxes agree with the VLBI measurements within a factor $\approx2$; our fluxes are typically higher because they might include some extended radio emission from outflows, or ``aborted jets'' \protect\citep[e.g.][]{ghisellini2004}.}\label{fig.vlbi}
\end{figure}

In Figure \ref{fig.vlbi} we show the comparison between the AGN radio core emission measured from the VLBI data and the excess radio emission obtained for our sources. The excess radio emission was estimated by subtracting the 1.4 GHz flux predicted for star formation assuming the average value $q=2.21$ (Sect. \ref{sample}), from the total radio flux density (from the VLA data at 1.4 GHz; Sect \ref{rad}). The obtained radio excess flux densities agree with the VLBI AGN core emission within a factor of $\approx$2. This is a further validation that our SED analysis is reliable and that with it we are able to predict the AGN radio-core luminosities with reasonable accuracy. We note that our estimates tend to slightly over predict the AGN radio-core flux densities because they might include contributions from extended radio emission, such as compact lobes, or ``aborted jets'' (e.g. \citealt{ghisellini2004}), which are invisible for the high resolution VLBI. 

\subsection{SEDs of the radio-excess AGN}\label{seds}

\begin{table*}[!t]
\caption{Summary of the IR and X-ray properties of the radio-excess sources.}
\begin{center}
\begin{tabular}{|c|c c|c c|c c| c |c|}
\hline
   Class   & Total& \% & X-ray detected & \% & X-ray undetected & \% & $\langle q \rangle$ & $\langle \rm SFR\rangle$ (M$_{\odot}$ yr$^{-1}$)\\ 
\hline
Radio Excess & 51  & 100\%  & 27     & 53\% & 24      & 47\%  & 1.10$\pm$0.08  & 58.8$\pm$13.8 \\
IR AGN       & 23  & 45\%   & 18     & 78\% & 5       & 22\%  & 1.02$\pm$0.11  & 45.5$\pm$14.4  \\
IR SFG       & 23  & 45\%   & 7      & 30\% & 16      & 70\%  & 1.33$\pm$0.09  & 69.2$\pm$21.7 \\
Passive      & 5   & 10\%   & 2      & 40\% & 3       & 60\%  & ...        & ... \\
\hline
\end{tabular}
\end{center}
\label{tab.2}
\end{table*}
From the SED fitting analysis we found that the infrared SEDs of the radio-excess sources are rather varied. A significant AGN component, which typically dominates the emission below $\sim$40 $\mu$m (MIR), has been detected in $\sim$45\% of the radio-excess sources (23/51; hereafter, IR AGN; see Fig. \ref{fig.sedall}); this fraction is much higher than that found for the ``radio-normal'' population ($\sim$8\%; Del Moro et al., in preparation), in line with the fact that radio-excess sources are AGNs, as opposed to ``radio-normal'' sources, which are a mix of radio-quiet AGN and star-forming galaxies (Sect. \ref{sample}). The average FIR--radio ratio for the IR AGN amongst the radio-excess sources is $\langle q\rangle_{\rm IR\ AGN}=1.02\pm0.11$, significantly lower than our threshold adopted to separate radio-excess from radio-normal sources ($q<1.68$). The remainder of the sample, however, do not require a significant AGN component in the SED fitting: $\approx$45\% (23 out of 51 sources; Table \ref{tab.2}) have SEDs consistent with those of star-forming galaxies (hereafter, IR SFGs; see Table \ref{tab1} and Fig. \ref{fig.sedall}) and the remaining $\sim$10\% (5 out of 51) have IR SEDs not compatible with either star formation or AGN activity (see Sect. \ref{sample}). For the latter group of sources we verified that the IR data are better represented by a normal elliptical galaxy template (generated with the GRASIL code; \citealt{silva1998}), suggesting that these sources might be ``passive'' (not star-forming) galaxies (see Fig. \ref{fig.sedall}). The mean $q$ value for the IR SFGs is $\langle q\rangle_{\rm IR\ SFG}=1.33\pm0.09$, larger than the average obtained for the IR AGN, but still well below the radio-excess selection limit ($q<1.68$). For the ``passive'' sources, the FIR--radio ratio calculated from the SEDs are all upper limits and therefore we cannot provide a reliable average $q$ value. 

In the far-infrared band, above $\lambda\approx 40\ \mu$m, the emission is dominated by star formation peaking around 100 $\mu$m (rest-frame) for the majority of the radio-excess sources ($\approx$86\%; 44 sources); this fraction also includes the majority of the sources that are dominated by AGN emission at shorter wavelength (i.e., IR AGN). However, we found a number of sources ($\approx$14\%; 7 sources) where the AGN emission outshines the star formation over the whole IR band, with the SFG contributing less than 50\% to the total FIR luminosity; these sources are amongst the brightest AGN in our radio-excess sample and the majority of them (6/7) are also detected in X-rays. 
Constraining the actual contribution from star formation (if there is any) to the total SED in these objects would require deep sub-mm observations, in order to sample the wavelengths beyond the rest-frame 100 $\mu$m.  

Using the broad-band IR luminosities ($\lambda=8-1000\ \mu$m, rest-frame) of the star formation component (i.e. removing the AGN contamination) derived from the best-fit SEDs in the same way as for the FIR luminosities (Sect. \ref{sample}), we estimated the star formation rates (SFRs) for our radio-excess sources. To convert the IR luminosity ($L_{\rm IR}$) into SFR we used the relation from \citet{kennicutt1998}, assuming a \citet{salpeter1955} IMF. The SFRs obtained for our sample span a wide range of values, from $\rm SFR\approx0.8$ M$_{\odot}$ yr$^{-1}$, indicating very low star formation, to $\rm SFR\approx350$ M$_{\odot}$ yr$^{-1}$, typical of starbursting systems. The average star formation rate of our radio-excess sources is $\rm \langle SFR\rangle=58.8\pm13.8$, with the IR SFG having slightly higher values than the IR AGN ($\rm \langle SFR\rangle_{IR\ SFG}=69.2\pm21.7$ and $\rm \langle SFR\rangle_{IR\ AGN}=45.5\pm14.4$; Table \ref{tab.2}). We note that the SFR could not be calculated for four of the IR AGN where the AGN component significantly dominates the total IR SED ($>$90\% contribution; i.e. \#8, \#13, \#20 and \#21, see Fig. \ref{fig.sedall}) and therefore the star formation component was not constrained from the SED fitting. Moreover, the ``passive'' sources have not been included in the SFR average for the radio-excess, since the SFRs estimated for these sources are upper limits. 

We stress that in our calculation of the SFRs we removed any contribution from the AGN to the IR luminosity, thanks to our SED decomposition. It is important to note that if we did not acknowledge the fact that AGN can contribute significantly also in the FIR band, and therefore assume that the total IR luminosity is only due to star formation (see e.g., \citealt{elbaz2010,shao2010,mullaney2012,santini2012}), we would over predict the SFR up to a factor of $\sim$2 when the AGN emission contributes $<$50\% to the total SED, and up to a factor of $\sim$4, when the AGN dominates ($>$50\% contribution) the SED at FIR wavelengths.   

\section{Discussion}
\label{discus}

We have shown in the previous sections that through detailed SED analysis in the IR band using deep \spz\ and \her\ data we were able to identify a relatively large fraction of radio-excess AGN from an initial radio-IR selected sample in the GOODS-North field. Although the excess radio emission is attributed to an AGN, we found that the multi-wavelength properties of these sources are rather heterogeneous; from the results obtained for the radio-excess source SEDs (Sect. \ref{seds}), we can distinguish three main types of objects amongst our sample: i) infrared AGN, ii) infrared star-forming galaxies (IR SFGs) and iii) ``passive'' systems. We aim here to explore in detail the variety of sources that compose the radio-excess population and whether the radio-excess AGN are different from the general X-ray selected AGN population.

\subsection{Infrared AGN}\label{iragn}
About 45\% of our radio-excess sample is made of sources that we can define as ``typical AGN'': these sources show a clear AGN component at IR wavelengths, which contributes more than 50\% to the total emission in the MIR band, and are generally also detected in the X-rays ($\approx$78\%; Table \ref{tab.2}): they are typically X-ray bright AGN, with luminosity above $L_{\rm X}\gtrsim10^{43}$ erg s$^{-1}$.

\begin{figure*}
\begin{center}
\includegraphics[scale=0.7]{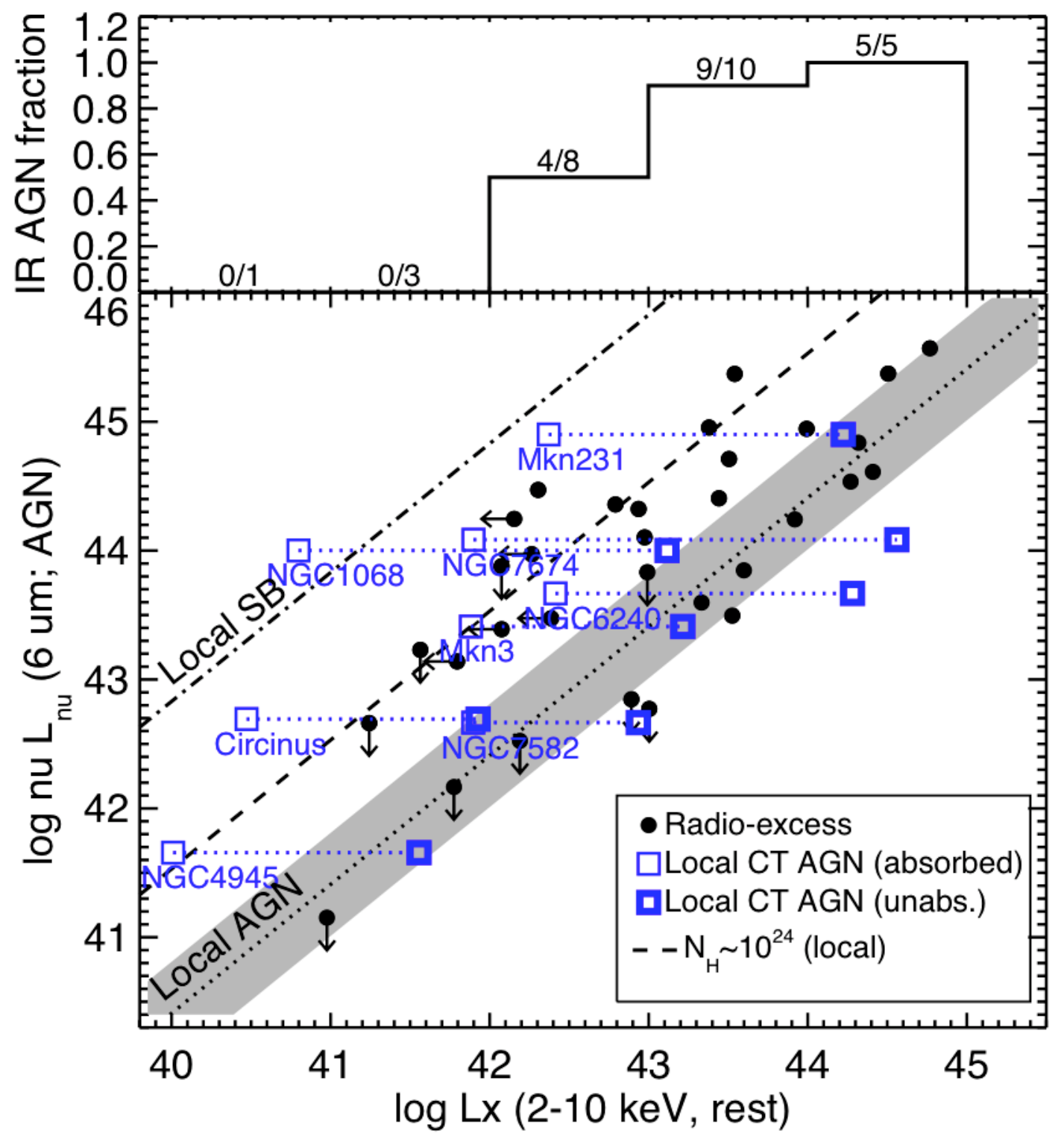}
\end{center}
\caption{Fraction of IR detected AGN as a function of the X-ray luminosity (top panel); the sources have been divided in five luminosity bins, between $L_{\rm X}=10^{40}-10^{45}$ erg~s$^{-1}$, although no AGN is expected to be found below $L_{\rm X}=10^{41}$ erg~s$^{-1}$. The bottom panel shows the X-ray luminosity at $2-10$ keV (rest-frame; not corrected for absorption) vs. the monochromatic 6 $\mu$m AGN luminosity ($\nu$L$_{\nu}$) for the radio-excess sources. Some local Compton-thick AGN are also plotted as open squares (thin marks: observed L$_{\rm X}$; thick marks: absorption corrected L$_{\rm X}$). The shaded region represents the typical $L_{\rm 6\ \mu m}-L_{\rm X}$ relation for local AGN \protect\citep{lutz2004}, while the dashed line is the relation expected for heavily obscured AGN ($N_{\rm H}\approx10^{24}$ cm$^{-2}$) and the dash-dotted line is the typical relation for starburst galaxies (taken from \protect\citealt{alexander2008b}).}
\label{fig.lxlr}
\end{figure*}

In Fig. \ref{fig.lxlr} (top panel) we show the fraction of IR detected AGN at different X-ray luminosity bins. Although the number of sources in each bin is relatively small, it is evident that the fraction of AGN detected at IR wavelengths decreases with decreasing luminosity of the sources. While at $L_{\rm X}>10^{43}$ erg s$^{-1}$ we are able to identify almost all of the X-ray detected AGN ($>$90\%) through the IR SED analysis, at lower X-ray luminosities ($L_{\rm X}=10^{42}-10^{43}$ erg s$^{-1}$) this fraction drops to $\approx$50\%. This means that when the AGN is intrinsically bright its emission is strong enough to outshine that of the host galaxy and it can be clearly detected at IR wavelengths (at least in the  MIR band); for the brightest sources the AGN dominates the whole IR band, even at far-infrared wavelengths (e.g. \#8 and \#13, see Fig. \ref{fig.sedall}; Sect. \ref{seds}). 

From the best-fit SEDs we can calculate the MIR luminosity at $\lambda=6\ \mu$m for the AGN (extracted from the detected AGN component only), which is often used as a diagnostic for the intrinsic power of the AGN (e.g. \citealt{lutz2004, alexander2008b, gandi2009, georgantopulos2011}), since the IR band is only lightly affected by extinction. In Fig. \ref{fig.lxlr} (bottom panel) we plot the monochromatic 6 $\mu$m luminosity (L$_{\rm 6\ \mu m}$) of the AGN, derived from the SED fits, as a function of the X-ray luminosity ($2-10$ keV, rest-frame; not corrected for absorption). Where the AGN component was not significantly detected in the IR band, an upper limit was calculated for L$_{\rm 6\ \mu m}$ to be 30\% of the total (AGN $+$ SFG) 6 $\mu$m luminosity estimated from the best-fit SEDs (Table \ref{tab1}; see Appendix A for justification). The shaded area in the figure represents the typical X-ray/IR luminosity correlation found for local, unobscured AGN by \citet{lutz2004}, while the dashed line represents the L$_{\rm 6\ \mu m}$--L$_{\rm X}$ relation predicted for sources obscured by large column densities ($N_{\rm H}\approx10^{24}$ cm$^{-2}$; \citealt{alexander2008b}). When sources are heavily obscured (e.g., Compton thick), their X-ray luminosity can be strongly suppressed, as opposed to the IR luminosity, and they should lie on the left hand side of the local AGN relation (or even on the left of the dashed line). Indeed, for comparison, we also show in the figure some local Compton-thick AGN, with observed (thin open squares) and intrinsic (i.e., corrected for absorption; thick open squares) X-ray luminosities, taken from literature (NED$^3$). Since for these local CT AGN we did not perform SED decomposition, we plotted the total 6 $\mu$m luminosity\footnote{For sources like Mkn 3 and Mkn 231, this is a good estimate of the AGN 6 $\mu$m luminosity since their MIR spectra are dominated by the AGN \citep[e.g.][]{weedman2005,goulding2012}; however, for other sources, e.g. NGC 4945, where the MIR emission is dominated by star formation (e.g. \citealt{peeters2004,goulding2012}) the plotted 6 $\mu$m luminosity is likely to be an overestimate of the intrinsic AGN emission at 6 $\mu$m.} for these sources (obtained from \spz\ or ISO data, taken from literature). 

About 59\% of our radio-excess AGN (19 out of 32 sources\footnote{The 32 sources plotted in Fig. \ref{fig.lxlr} include all of the radio-excess AGN with a measured AGN luminosity at 6 $\mu$m and/or in the X-ray band; namely, they are 23 IR AGN, 7 X-ray detected IR SFGs and 2 X-ray detected passive systems (see Table \ref{tab.2}).} plotted in Fig. \ref{fig.lxlr}, bottom) follow the local L$_{\rm 6\ \mu m}$-L$_{\rm X}$ relation from \citet{lutz2004}, indicating good agreement between the X-ray luminosity and the MIR luminosity estimated from our SEDs; 
the remaining X-ray detected AGN (8 sources) lie closer to the region expected for heavily obscured AGN ($N_{\rm H}\approx10^{24}$ cm$^{-2}$; see \citealt{alexander2008b}), together with all the X-ray undetected sources (5 sources), for which we have X-ray luminosity upper limits. The MIR luminosity of these objects indicates that the intrinsic power of the AGN is higher than that detected in the X-rays, suggesting they may be heavily obscured, possibly by Compton-thick material surrounding the black hole ($N_{\rm H}>10^{24}$ cm$^{-2}$). Also for the X-ray undetected IR AGN ($\sim$22\%, 5/23 sources; see Table \ref{tab.2}) the IR luminosity measured from the SEDs ($\nu L_{\rm 6\ \mu m}>10^{43}$ erg~s$^{-1}$) implies that these sources should be bright enough in the X-rays to be detected by the deep \ch\ data available in GOODS-N. 
Since our capability to detect the AGN emission in the IR decreases with decreasing AGN power (Fig. \ref{fig.lxlr}, top), this suggests that the five IR AGN that are X-ray undetected are likely to be intrinsically luminous sources, and therefore they are candidate Compton-thick AGN. 

On the basis of these analyses, amongst the IR detected AGN we can therefore estimate the fraction of candidate Compton-thick AGN to be up to $\sim$43\% (10 sources that lie close to the $N_{\rm H}\approx10^{24}$ cm$^{-2}$ line in Fig. \ref{fig.lxlr}, bottom), considering both X-ray detected and undetected AGN. However, to confirm the classification of these AGN as Compton-thick and their fraction, other diagnostics are necessary \citep[e.g.][]{georgantopulos2011,goulding2011}. The most efficient way to unambiguously classify these sources as Compton-thick AGN, would be the detection of a strong iron line (Fe K$\alpha$, $E=6.4$ keV) and a reflection dominated spectrum from detailed X-ray spectral analysis. Unfortunately, since our sources are at high redshifts, they are typically weakly detected (or undetected) in the deep \ch\ data, preventing the good quality spectra necessary for this kind of analysis. 


\subsection{Infrared star-forming galaxies}\label{irsb}

Together with the IR AGN, the other main group amongst our radio-excess sources ($\approx$45\%) have IR SEDs consistent with star formation emission, with no significant evidence at IR wavelengths for the presence of an AGN. However, the excess radio emission still suggests that these sources are not simple star-forming galaxies. 

\citet{miller2001} found a significant number of star-forming galaxies not hosting AGN ($\sim20$\%) with enhanced radio emission compared to the typical FIR--radio relation. However this phenomenon has been observed in the centre of galaxy clusters and the radio excess for these galaxies has been estimated to be not larger than a factor of three from the field galaxy FIR--radio relation (corresponding to $q\approx1.73$; \citealt{gavazzi1986}). \citet{bressan2002} interpret this effect as a post starburst phase, when the star formation has stopped, while the radio emission produced by supernova remnants (SNRs) has still not dimmed, as it fades more slowly. This phenomenon can not explain the origin of the radio excess in all of our IR SFGs, since the selection criteria adopted in our analysis ($q<1.68$), which correspond to more than a factor three enhancement of the radio emission, should already exclude the majority of these radio-excess post-starburst galaxies; moreover, it is unlikely that about half of our radio-excess sources reside in the centre of galaxy clusters. In fact, the average $q$ calculated for our IR SFG ($\langle q\rangle_{\rm IR\ SFG}=1.33$; see Sect. \ref{seds}) suggests that the radio emission in these sources is much higher than that observed in post-starburst galaxies. Moreover, we verified that the SFR for the IR SFGs is consistent with that of star-forming/starburst galaxies (Sect. \ref{seds}), meaning that the star formation is still ongoing in these systems. Therefore, the excess radio emission in these IR SFG can only be due to the presence on an AGN.

A number of the IR SFGs (7/23) are detected in the X-ray band, although they are typically less luminous than the X-ray detected IR AGN ($L_{\rm X}\lesssim10^{43}$ erg s$^{-1}$). For these sources we estimated the IR luminosity expected for the AGN at 6 $\mu$m from the X-ray luminosity in the $2-10$ keV band (rest-frame) using the \citet{lutz2004} relation for local unobscured AGN. In four cases, the estimated $L_{\rm 6\ \mu m}$ of the AGN is consistent with the total 6 $\mu$m luminosity calculated from the SEDs (see Fig. \ref{fig.lxlr}), indicating that the MIR emission of these sources is unlikely to be produced by star formation only, despite our SED fits suggest that this is the case. The AGN in these sources might be obscured at MIR wavelengths (e.g. \citealt{deo2009,goulding2009,goulding2012}) and therefore the emission appears dominated by star formation. These IR SFGs are the brightest sources amongst this group\footnote{We stress that the X-ray luminosities are not corrected for absorption (Sect. \ref{xx}), and therefore if large amounts of dust and gas are present in these sources, the intrinsic X-ray luminosity is likely to be higher.} ($L_{\rm X}\approx10^{43}$ erg s$^{-1}$) and indeed their SEDs could be consistent with a combination of SFG $+$ AGN emission; however, given our conservative criteria to identify the best-fitting SED model, the AGN component in these cases did not pass the $f-$test probability threshold (see Appendix A). At the luminosities of these sources, the SED fitting procedure struggles to disentangle the two components, especially if the AGN contribution is not significantly larger than that from star formation (Fig. \ref{fig.lxlr}, top, and Appendix A). For the other three sources, both the X-ray and the predicted IR luminosities are below $L_{\rm X}\lesssim10^{42}$ erg s$^{-1}$ (see Table \ref{tab1}), which are consistent both with a faint AGN or with star-forming galaxy emission. The remaining $\approx$70\% of the IR SFGs (16/23) are not detected in the X-ray band. 

The most likely interpretation for these IR SFGs amongst our radio-excess sources is that they are low--moderate luminosity AGN, typically with $L_{\rm X}\lesssim10^{43}$ erg s$^{-1}$ (see Fig. \ref{fig.lxlr}), or possibly luminous AGN heavily obscured at MIR wavelengths, and hosted in dusty star-forming galaxies, whose IR emission is thus overwhelmed by star formation and eludes our detection capability (Sect. \ref{iragn}). If a large amount of dust is present in these galaxies, the nuclear emission from the AGN could be heavily obscured and remain undetected even in the X-rays. Supporting this interpretation is the fact that one of our X-ray undetected IR SFGs (\#29, in Table \ref{tab1}) was found to have a compact radio core in the VLBI observations by \citet{chi2009}, clearly revealing the presence of an AGN in this star-forming galaxy (see Sect. \ref{xray}); this source also shows extremely red optical/NIR colours, indicating significant dust reddening \citep{chi2009}. Although the inability to unambiguously identify the presence of an AGN in the IR and/or X-ray bands for the majority of these IR SFGs represents an evident limitation in finding AGN, the radio excess provides a clear signal, often the only signal, of nuclear activity in these sources, bringing us closer to completing the census of the AGN population.

\begin{figure}[!t]
\centerline{
\includegraphics[scale=0.67]{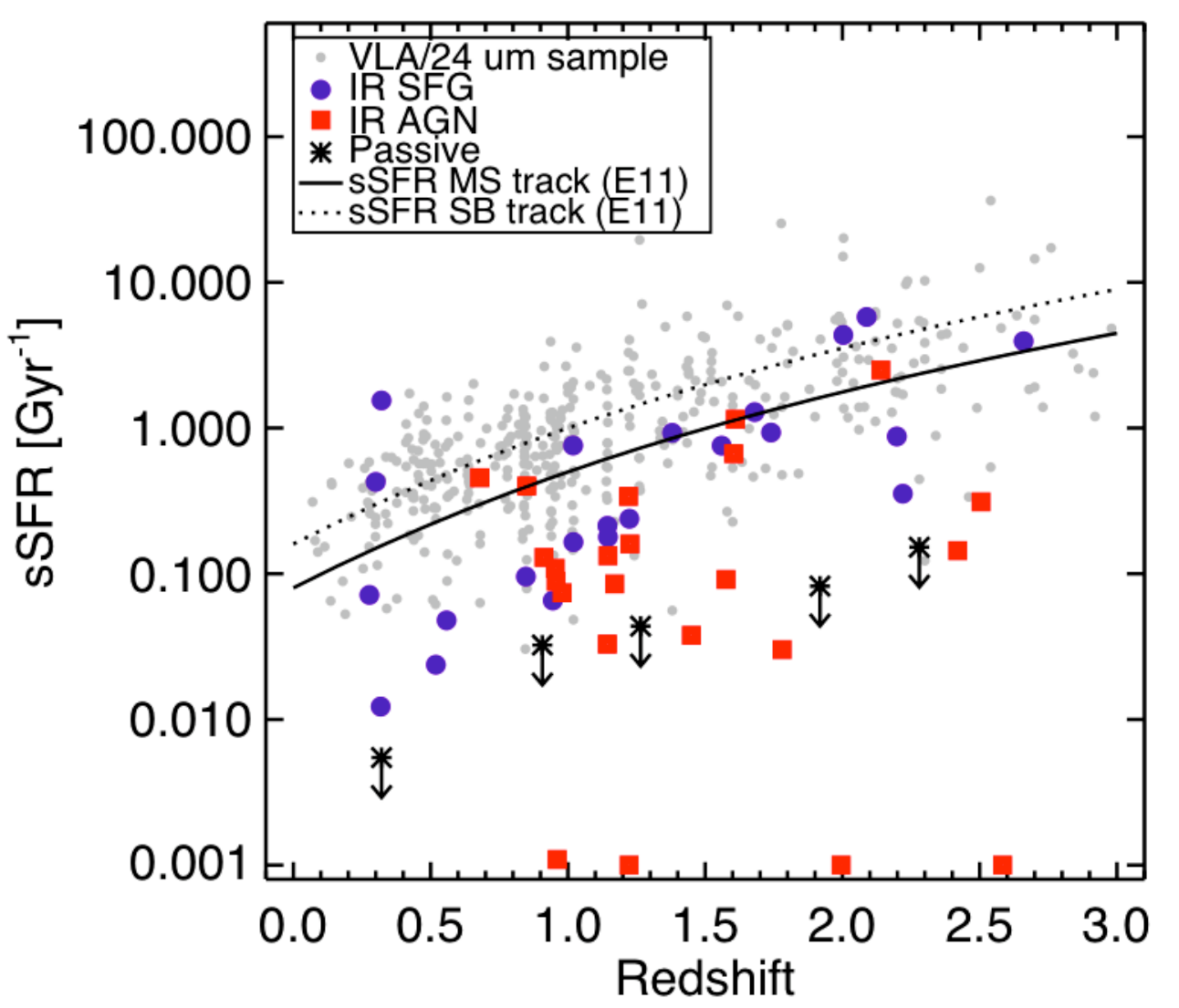}}
\caption{Specific star formation rate (sSFR) vs. redshift for the entire VLA/24 $\mu$m sample (grey dots) and for the radio-excess sources, divided into their three IR categories: IR AGN (red squares), IR SFG (blue circles) and passive systems (asterisks). The four sources plotted at the bottom (with a symbolic value of sSFR$=0.001$ Gy$^{-1}$) are the four IR AGN for which the SFR could not be estimated (see Sect. \ref{seds}). The tracks in the plot show the sSFR evolution with redshift for main sequence star-forming galaxies (MS; solid line) and for star-forming galaxies (SFG; dotted line) from \protect\citet{elbaz2011}. The radio-excess sources have typically lower star formation than the star-forming galaxy population detected at 24 $\mu$m.}\label{fig.sfr}
\end{figure}

\subsection{``Passive'' systems}

A small fraction ($\approx$10\%, 5/51 objects) of the radio-excess sources seem to have weak contributions from AGN activity and star formation, as their IR SEDs are in fact consistent with those of relatively quiescent elliptical galaxies (Sect. \ref{seds}). These sources have lower redshifts ($z\le1$, except for one with $z\sim1.5$) than the average of the radio-excess sample and some of them are amongst the most powerful radio emitters (Table 1). This suggests that they might be radio galaxies, such as Fanaroff--Riley type I radio galaxies (FR Is), which are powered by low accretion rate AGNs \citep{chiaberge1999, ghisellini2001}. At low accretion rates ($L/L_{\rm Edd}<0.01$) AGNs are predicted to have radiatively inefficient accretion flows (RIAFs; \citealt{narayan1995,narayan2008}), which are typically associated with strong radio outflows \citep{narayan1995,maccarone2003}. Indeed, one of the passive sources in our sample (\#47 in Table \ref{tab1}) is a well known wide-angle-tail (WAT) radio source at $z=1.265$ and shows strong extended radio jets (e.g., \citealt{bauer2002}). 
These galaxies may have already passed the active phase when the SMBH is rapidly growing and the host galaxy experiences strong star formation and they are left with a low accreting SMBH in a passive red galaxy (e.g. \citealt{hopkins2006a,hickox2009}). 

\subsection{Radio excess or dimmed star formation?}\label{ssfr}

In this subsection we aim to investigate the properties of the galaxies hosting the radio-excess AGN, comparing them to the properties of X-ray selected AGN hosts. In Section \ref{seds} we have shown that the radio-excess sources have on average star formation rates typical of normal star-forming/starburst galaxies. However, to have better insights into the host galaxy properties one should explore the specific star formation rate (sSFR), i.e. the SFR per unit stellar mass. In fact, recent studies have shown that a correlation exists between the SFR and the galaxy stellar mass (e.g. \citealt{noeske2007,daddi2007a,elbaz2007,pannella2009a}), meaning that more massive galaxies have typically higher SFRs. This correlation also evolves with redshift, producing the so called ``main sequence'' of star-forming galaxies (MS; e.g., \citealt{noeske2007,elbaz2011}). Therefore, removing the mass dependence of the SFRs gives a better estimate on the relative star formation activity of galaxies.   

We calculated the sSFR by simply dividing the SFRs (Sect. \ref{seds}) by the galaxy stellar masses (see Sect. \ref{mass} and Fig. \ref{fig.mass}). In Figure \ref{fig.sfr} we show the sSFR as a function of redshift for the entire VLA/24 $\mu$m sample (grey dots) and for the radio-excess sample, distinguishing between the different types of sources according to their IR SEDs (i.e., IR SFG, IR AGN and passive). In the plot we also show the sSFR$-z$ evolution track for MS star-forming galaxies and for starburst (SB) galaxies defined by \citet{elbaz2011}. Our radio-excess sources typically lie below the sSFR tracks for star-forming galaxies, indicating that they have, on average, lower star formation contribution when compared to radio-normal sources of the same stellar mass. However, if we consider the average sSFR for the IR SFGs, IR AGN and passive systems amongst our radio-excess sample, we find that for the IR SFGs $\rm \langle sSFR \rangle_{IR\ SFG}=1.15\pm0.26$ Gy$^{-1}$, consistent with the MS star-forming galaxies (assuming an average redshift of $z\approx1.5$), while for the IR AGN, the average sSFR is significantly lower: $\rm \langle sSFR \rangle_{IR\ AGN}=0.38\pm0.14$ Gy$^{-1}$.\footnote{The four IR AGN where the SFR could not be calculated (see Sect. \ref{seds}) were not included in the average sSFR calculations, and are plotted in Fig. \ref{fig.sfr} with a symbolic value of sSFR$=0.001$ Gy$^{-1}$.} The sources identified as passive from our SED fitting analysis have, in fact, sSFR$<10$\% of the average MS galaxies at a given redshift (Fig. \ref{fig.sfr}) and are more massive ($\langle M_*\rangle \approx$11.5 $\rm M_\odot$; see Table \ref{tab1}) than the rest of the sample ($\langle M_*\rangle\approx11.0$ $\rm M_\odot$ and $\langle M_*\rangle\approx10.7$ $\rm M_\odot$ for IR AGN and IR SFGs, respectively).\footnote{We note that the range of stellar mass values for the radio-excess sources is consistent with that of the entire VLA/24 $\mu$m sample (see Sect. \ref{mass}); however, on average, the radio-excess sources have larger stellar masses (median $M_*\approx11.1\ \rm M_\odot$) than the ``radio-normal'' sources.}

Previous studies have found that galaxies hosting X-ray selected AGN typically have star formation rates in agreement with the MS star-forming galaxies (e.g. \citealt{alonso-herrero2008, silverman2009, xue2010, mullaney2012, santini2012}) and do not show different properties from galaxies not hosting AGN. This suggests that the radio-excess technique tends to select a different AGN population from those selected in X-rays, with the host galaxies of radio-excess AGN having on average less ongoing star formation when compared to the hosts of X-ray selected AGN. We note that thanks to our SED decomposition, we were able to calculate the SFRs (and sSFRs) more accurately than in many of the previous works mentioned above by removing any AGN contribution to the IR luminosity (see Sect. \ref{seds}). This gives us smaller sSFR values than those resulting assuming a pure SFG dominated IR luminosity. However, we found that the discrepancies between the sSFR of radio-excess AGN and X-ray AGN hosts are real and are not due to the technique adopted to calculate the $L_{IR}$ (and therefore sSFR). In fact, if we estimate the sSFRs for all of the X-ray detected sources in the VLA/24 $\mu$m sample (see Sect. \ref{xx}), we obtain good agreement with the MS sSFR (Del Moro et al., in prep.). We therefore conclude that the host galaxies of radio-excess AGN are growing at a slower rate than the typical X-ray selected AGN hosts. Since many of our radio-excess sources are not detected in X-rays, it seems that the excess radio emission can reveal a complementary AGN population to that selected in the X-ray band.  

Since in the local Universe more radio-loud AGN are preferentially found in galaxy groups and clusters (e.g. \citealt{best2004, kauffmann2008}), as opposed to typical radio-quiet AGN that tend to reside in less dense regions (e.g. \citealt{kauffmann2004}), we want to verify whether the radio-excess AGN also reside in different environments than the typical X-ray selected AGNs. We therefore cross-matched the positions and redshifts of our radio-excess sources with those of extended X-ray sources, corresponding to known clusters, in the GOODS-N field (e.g. \citealt{bauer2002}). As search regions we used the ellipses defining the extent of the X-ray emission, reported in Table 1 of \citet{bauer2002}. We found that only two out of the 51 radio-excess AGN in our sample (specifically sources \#6 and \#17 from Table \ref{tab1}) lie within the X-ray extended emission regions from \citet{bauer2002}; only one of these two sources (\#17; i.e. only $\approx$2\% of the radio-excess sample) has a redshift matching that of the galaxy cluster ($z=1.01$; \citealt{bauer2002}) and therefore is likely to be associated with the cluster. The large majority of our radio-excess sources lie in less dense environments, similarly to the X-ray selected AGN population.

The lower sSFRs observed for our sources compared to MS galaxies and X-ray AGN hosts suggest that the radio-excess characteristic of our sources is not only due to a stronger radio core emission from the AGN than in ``radio-normal'' sources, but also to a dimming of the emission from star formation, which enhances, by contrast, the radio excess. The decreasing average sSFR for the different types of radio-excess sources identified through our SED analysis (i.e. IR SFG, IR AGN, passive; Sect. \ref{seds}) indicates that we might be looking at different stages of BH-galaxy evolution: from the IR SFG that are the more active and possibly the most dust-obscured sources (see Sect. \ref{irsb}), where the presence of the AGN is testified only by the radio-excess, to the IR AGN, where the dust obscuration is progressively lower (although still very high in some cases; see Sect. \ref{iragn}), and the AGN is detected at radio, IR and often X-ray bands, while the star formation is quenching. At the final stage there are the passive systems, where the star formation has stopped and the SMBH is slowly accreting in a massive, passive galaxy (e.g. \citealt{hopkins2006a,hickox2009}).   



\section{Summary and conclusions}
\label{conc}
Using deep IR \spz\ and \her\ data, as part of the GOODS--\her\ program, and radio VLA data in the GOODS-N field, we performed a detailed SED analysis for 458 sources with spectroscopic or photometric redshift identification ($z\le3.0$). From the best-fit SEDs we calculated the FIR flux ($f_{\rm FIR}$; $\lambda=42.5-122.5\ \mu$m) and the FIR--radio relation ($q$) to identify a sample of sources with excess radio 1.4 GHz emission over that expected for star-forming galaxies ($q\approx2.2$; \citealt{helou1985}). We obtained a sample of 51 ($\approx$11\% of the initial sample) radio-excess AGN ($q<1.68$) and investigated their radio, IR and X-ray properties. The main results of our analysis can be summarised as follows: 
\begin{itemize}

\item{} We found that the fraction of radio-excess sources increases with X-ray luminosity (Sect. \ref{xray}), suggesting that the more luminous AGN are also more powerful in the radio band, although not necessarily radio-loud. This suggests that there is a wide distribution of radio power amongst the AGN population and that the radio-excess sources (i.e. sources with intermediate radio power) constitute a significant part of the total AGN population. \smallskip 

\item{} The radio-excess sample seems to be composed of a heterogeneous mix of sources: i) IR AGN ($\approx$45\%, 23/51), which are the ``classical'' highly accreting AGN hosted in star-forming galaxies; ii) IR SFG ($\approx$45\%, 23/51), which are likely to host moderate luminosity ($L_{\rm X}<10^{43}$ erg s$^{-1}$) or dust-obscured luminous AGN, whose emission eludes our detection capabilities in the IR and often the X-ray bands, but they are identifiable through their radio excess; iii) low accretion rate AGNs (e.g. RIAFs, FR Is) hosted in passive, non-star forming galaxies ($\approx$10\%, 5/51), where the excess radio emission is likely to be due to the presence of strong radio jets and lobes, in addition to a radio core (Sect. \ref{discus}). \smallskip


\item{} Only 27 of the radio-excess AGN ($\sim$53\%) are detected in the X-ray band; the large number of X-ray undetected sources suggests that many radio-excess sources might be heavily obscured (possibly Compton-thick) AGN (Sect. \ref{iragn}). Indeed, amongst the 24 X-ray undetected sources, five sources (i.e. $\sim$20\%) show a strong AGN component in the IR band. The majority of the X-ray undetected sources are IR SFGs, which are likely to be low--moderate luminosity AGN, or luminous dust-obscured AGN, as demonstrated by the detection of a compact radio core in deep VLBI observations for one of these X-ray undetected IR SFGs (Sect. \ref{irsb}). Assuming that the local correlation between the AGN $L_{\rm 6 \mu m}$ and the X-ray (2-10 keV) luminosity holds out to high redshift, we estimate the fraction of Compton-thick candidates amongst our radio-excess AGN to be $\sim$20\% ($\sim$43\% amongst the IR detected AGN 10/23); however the unambiguous classification of these sources requires other diagnostics, which are not feasible with the current data (Sect. \ref{iragn}).\smallskip

\item{} The specific star formation rates (sSFR) estimated for the radio-excess AGN are on average lower than those observed for main sequence star-forming galaxies, or X-ray AGN host galaxies, indicating that the radio-excess technique tends to select a different AGN population than the X-rays (Sect. \ref{ssfr}). This also suggests that the excess radio emission measured for our sources is not only due to strong radio emission from the AGN, but to a combination of two effects: i) radio core emission from the AGN and ii) dimmed star formation emission. Moreover, the progressively lower sSFR observed for IR SFGs, IR AGN and passive systems, suggests that through our radio-excess sources we might look at different stages of BH-galaxy co-evolution.

\end{itemize}

\begin{acknowledgements}
We thank the anonymous referee for the useful comments on our paper. We gratefully acknowledge support from the STFC Rolling Grant (ADM; DMA). This work is based on observations made with \her, a European Space Agency Cornerstone Mission with significant participation by NASA. FEB acknowledges support from Programa de Financiamiento Basal, CONICYT-Chile (under grants FONDECYT 1101024 and FONDAP-CATA 15010003), and {\it Chandra} X-ray Center grant SAO SP1-12007B. WNB, BL, YQX acknowledge the {\it Chandra} X-ray Center grant SP1-12007A and NASA ADP grant NNX10AC99G. RG acknowledges support from the Italian Space Agency (ASI) under the contract ASI-INAF  I/009/10/0. YQX acknowledges support of the Youth 1000 Plan (QingNianQianRen) program and the USTC startup funding. We are grateful to N. Drory for sharing the SED fitting code used to estimate galaxy stellar masses.
\end{acknowledgements}

\bibliography{radio_xcs_paper3_r1final.bbl}

\begin{appendix}
\section{Testing the SED fitting technique}

As described in Section \ref{sed}, our fitting approach consists of using 5 SFG templates, derived from a sample of local star-forming galaxies and an empirically derived AGN template \citep{mullaney2011} modified by an extinction law (with $A_V\approx0-30$ mag). We applied these templates to decompose the total SEDs of high redshift sources ($z\le3$) into AGN and galaxy components. To determine the best-fitting solutions, we firstly fit to the \spz\ and \her\ photometric data a simple model including only the SFG template, using $\chi^2$ minimization, yielding 5 SED solutions (one for each SFG template). We then fit the data with a more complex SFG $+$ AGN (plus extinction on the AGN component only) model, using an $f-$test to determine the best-fitting solutions. The criteria we used to choose whether the AGN component significantly improves the fit are: i) an $f-$test probability $>$90\% confidence when adding the AGN template to the fit; ii) this first condition must be met in the majority of the SED fitting solutions (i.e. at least 3 out of 5).

 
This SED fitting approach (Sect. \ref{sed}) and the criteria chosen to identify the best-fit models were selected based on the results of careful tests performed on a sample of 30 sources in the VLA/24 $\mu$m detected sample (Sect. \ref{data}) for which \spz\ IRS low-resolution spectroscopy is available (covering the wavelength range $\lambda\approx3-20\ \mu$m, rest-frame; \citealt{pope2008b, murphy2009, kirkpatrick2012}). These sources are all detected in the X-ray band \citep{alexander2003} and the majority of them are also detected in at least one of the \her\ bands at 100, 160 and/or 250 $\mu$m (29/30 sources; Table \ref{tab3}). Our tests consist of two main steps: 
\begin{enumerate}
\item constraining the AGN and star formation components using the IRS spectra (plus \her\ data), which provide more detailed information (e.g. PAH features or featureless power-law continuum) than photometry alone on the source emission at MIR wavelengths, where the AGN contribution affects the SED the most ($\lambda\lesssim 30-40\ \mu$m); 
\item performing the SED fits using only \spz\ and \her\ photometric points (not including the IRS spectra), testing various criteria to select, amongst the different solutions, the best-fitting models that more closely matched the IRS spectral fitting results.
\end{enumerate}
In analysing the IRS spectra, we also included the \her\ 100, 160 and 250 $\mu$m flux densities to help constrain the SEDs in the FIR band, as the spectra alone cover too small a wavelength range to allow a reliable extrapolation of the SEDs in the broad IR band. In fact, fitting just the IRS spectra (not including the \her\ data) tend to underestimate the amount of star formation contribution to the total emission, favoring higher contribution from the AGN component, instead.

\begin{figure}[!t]
\centering{
\includegraphics[scale=0.6]{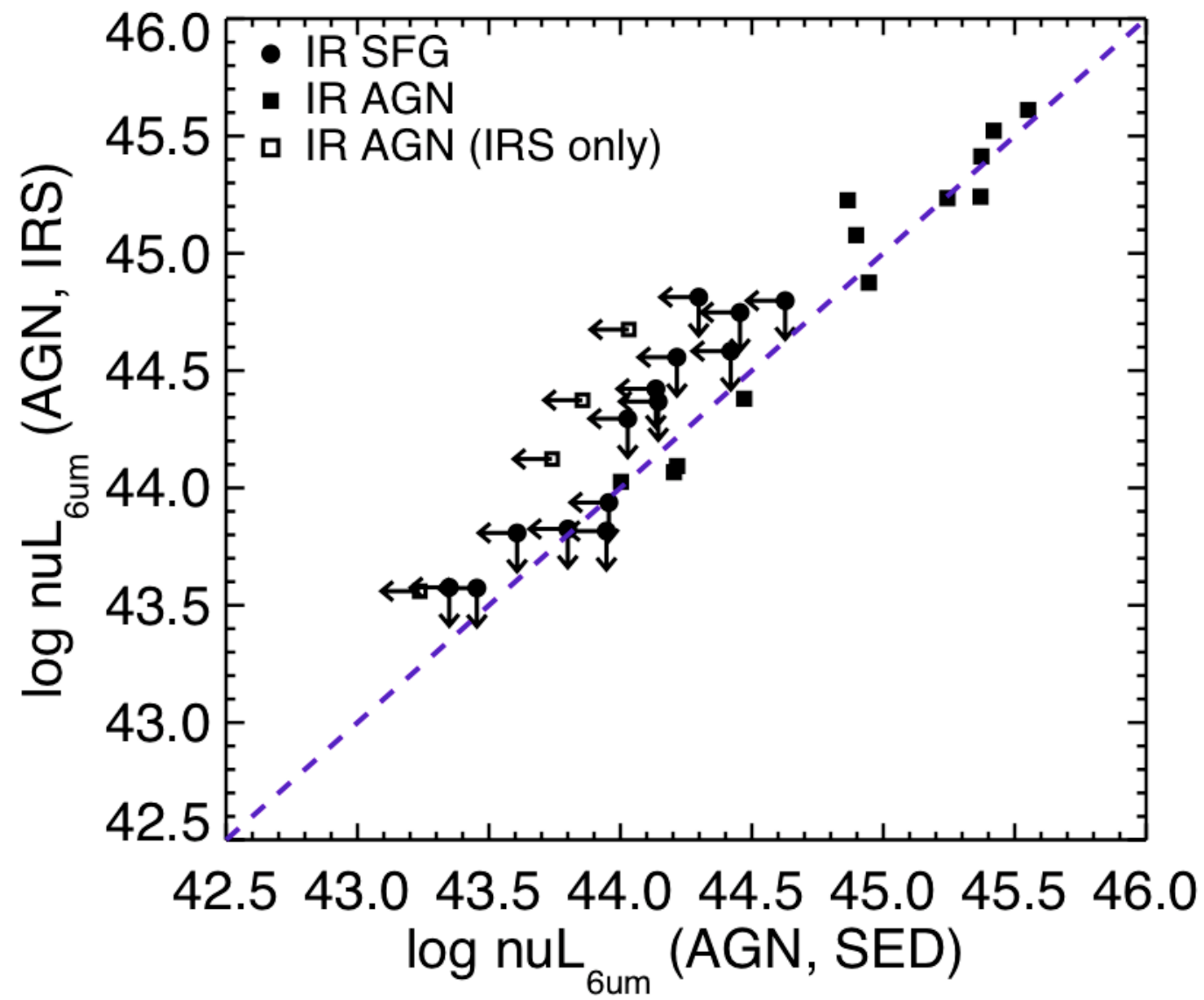}}
\caption{Comparison of the AGN 6 $\mu$m luminosity obtained from the SED fitting of the photometric \spz\ and \her\ data points ($\nu L_{\rm 6\ \mu m}$, SED) versus the $L_{\rm 6\ \mu m}$ estimated from the IRS spectral ($+$ \her\ data points) fitting ($\nu L_{\rm 6\ \mu m}$, IRS). For the IR SFGs, where the AGN component was not significantly detected, we plotted an upper limit of the AGN $L_{\rm 6\ \mu m}$ (i.e., 30\% of the total 6 $\mu$m luminosity; see Appendix A). Sources where the classification from the two analyses agree are shown as filled symbols: IR AGN (filled squares) and IR SFG (filled circles). The AGN identified only from the IRS spectral fit are shown as open squares.}
\label{fig.l6c}
\end{figure}

Since the \spz\ IRS spectra provide a large amount of data for fitting our SED templates, we fitted the spectra (and \her\ data) directly using a model including both SFG and AGN (plus extinction) templates described in Sect. \ref{sed} to measure galaxy and AGN contributions to the total IR emission. We note that because the IRS data have much smaller uncertainties compared to the \her\ data, we increased the errors on the IRS spectra by a constant factor to be of the same order of magnitude of the \her\ data errors\footnote{For the \her\ undetected source, i.e. where S/N$<3$ in the \her\ bands, the errors on the 100, 160, 250 $\mu$m flux density measurements are still much larger than the \spz\ IRS spectra ``increased'' uncertainties.} (when S/N$>3$); this is to avoid the SED fitting results being dominated by the \spz\ IRS data (which would have much more ``weight'' on the resulting $\chi^2$ than the \her\ data), and also to account for the intrinsic scatter between our 5 discrete SFG templates. We adopted $\chi^2$ minimisation to evaluate the best fit of the 5 SFG templates. We note that due to the high signal-to-noise ratio of the IRS spectra (despite the enhancement applied to the errors), the resulting $\chi^2$ values for all the fits are very high ($\chi^2/d.o.f.>>1$) and therefore cannot be used as an absolute measure of the goodness of the fit in the usual way (see also \citealt{mullaney2011}). 
However, since the $\chi^2$ calculation is affected by this issue in the same way for each solution, a comparison between $\chi^2$ values can still be used to identify the best-fit amongst the five different solutions: the spectral fit yielding the minimum $\chi^2$ was chosen as the best-fitting solution. It is important to note however, that in some cases the difference in $\chi^2$ between different solutions is small, and therefore, the chosen best-fit model is not the only acceptable solution to reproduce the data. To overcome this issue, instead of choosing one solution as the best-fit, we used a weighted mean between the five different fitting solutions to retrieved the 6 $\mu$m luminosity of the AGN (from the fitted AGN component), the FIR fluxes and the $q$ values for the sources. The $\chi^2$ values were used to calculate appropriate weights in order to give the maximum weight to the solution with the minimum $\chi^2$ (i.e. $w_i=\chi^2_{min}/\chi_i^2$). 
A dominant AGN component (i.e. $L_{\rm 6\ \mu m, AGN}/L_{\rm 6\ \mu m, tot}>0.5$) was measured in 16 cases. From these IRS spectra ($+$\her) fits, we also determined that when the AGN is found to contribute less than $\sim$30\% to the total (AGN $+$ SFG) emission at 6 $\mu$m, the total SED is not significantly affected by the presence of the AGN component compared to that of a pure SFG. Therefore, when the AGN is not significantly detected with our SED fitting approach (i.e. when using photometric \spz\ and \her\ data only), we chose the upper limit for the AGN luminosity at 6 $\mu$m to be 30\% of the total 6 $\mu$m luminosity (see Sect. \ref{iragn}).

After using the IRS spectra to obtain good constraints on the AGN and star formation components, we then performed several sets of SED fits to these sources using \spz\ 8, 16, 24 $\mu$m and \her\ 100, 160, 250 $\mu$m flux densities only (not including the IRS spectra), testing different criteria to select the best-fitting solutions, in particular to establish the significance of the AGN component in the fits. These tests were performed to find the best ``recipe'' to recover, with these sparse photometric data points (which is the approach adopted in this paper; see Sect. \ref{sed}), the results obtained from the IRS spectral (plus \her\ data) fits (Fig. \ref{fig.sed0} and Table \ref{tab3}). As for the IRS spectra, the \spz\ photometric data have much smaller uncertainties than the \her\ data; we therefore increased the errors on the 8, 16 and 24 $\mu$m flux densities (by constant factors for all of the sources) during the SED fitting process in order for each data point to have a similar ``weight'' in the fit and to allow for the intrinsic scatter on the templates. We note, however, that in these SED fits, as for the IRS $+$ \her\ data fits, the reduced $\chi^2$ alone (i.e. $\chi^2/d.o.f.\approx1$) is not a reliable indicator of the goodness of fit since the $\chi^2$ is very sensitive to the errors on the data points used in the fits, especially when the number of data points is limited, as in our case. Nevertheless, as we pointed out above, the relative $\chi^2$ can still be used to determine the best-fit model between two sets of solutions (SFG only, or SFG $+$ AGN; Figure \ref{fig.sed}), so we performed an $f-$test, which compares the $\chi^2$ values and $d.o.f.$ of the two models (see Sect. \ref{sed}), to measure the improvement of the fit obtained by including the AGN component to the model.

We tested the reliability of the best-fitting solutions by varying the $f-$test probability threshold between the two adopted models (SFG and SFG $+$ AGN) and also varying the number of solutions that needed to pass this threshold for the model to be accepted as the best-fit. A higher confidence level (e.g. 95\% or 99\%) allowed us to recover the AGN component detected in the IRS spectra only in a small number of cases, while missing some very obvious, strong IR AGN, where the AGN emission found from the IRS contributes 100\% to the total 6 $\mu$m luminosity. We therefore lowered the confidence level threshold from the $f-$test to be at least 90\%. The $f-$test was performed between each pair of SED fitting solutions (i.e. for each SFG template) in order to verify whether the need of the AGN component was dependent on the SFG template used in the fits. Indeed, we recognised that the fits performed with one particular SFG template (``SB5'' in \citealt{mullaney2011}) typically required an AGN component with higher confidence than the others. To avoid biases on the results due to the choice of SFG template and thus to remove the dependence of the fitting solutions on any specific templates, we decided to accept the AGN component as significant only if the $f-$test criterion was met in at least 3 of the 5 SED fitting solutions, otherwise we conservatively assumed the simple SFG model as the best-fit. 

As previously seen from the \spz\ IRS spectral fits, in some cases different fitting solutions yielded small differences in $\chi^2$ values meaning that a unique solution could not be defined. Therefore, once we established the best-fit model (SFG or SFG $+$ AGN), we obtained the 6 $\mu$m luminosity of the AGN, the total FIR fluxes and the $q$ values for our sources using a weighted average of the values calculated from all the best-fit model solutions\footnote{For the SFG $+$ AGN model we only included in the weighted average calculation the solutions where the AGN component was significant in the fit ($>$90\% confidence level from the $f-$test).} (as for the fits of the \spz\ IRS spectra $+$ \her\ data); the errors on the estimated averages were calculated as weighted average variances. 

To test the results for our chosen SED fitting approach, we compared the 6 $\mu$m luminosity of the AGN obtained from the IRS spectral fitting ($+$ \her\ data) to that obtained from the SED fitting performed using \spz\ and \her\ photometry (Figure \ref{fig.l6c}). In Fig. \ref{fig.l6c} we distinguish between sources with a dominant AGN component at 6 $\mu$m (IR AGN: $L_{\rm 6\ \mu m,\ AGN}/L_{\rm 6\ \mu m,\ tot}>0.5$) from those where the AGN component is not significantly detected (IR SFG: $L_{\rm 6\ \mu m,\ AGN}/L_{\rm 6\ \mu m,\ tot}<0.5$). In $\approx$87\% of the cases the IR classification of the sources obtained from the IRS spectra and the SED fits are consistent with each other (12 IR AGN and 14 IR SFGs) and the 6 $\mu$m luminosities measured from the two types of analysis are in good agreement (Table \ref{tab3}). In the remaining cases ($\approx$13\%), the two different analyses yielded different results: the IRS spectral analysis identified an AGN component in four further sources, which are missed by our SED fitting analysis; this is because the criteria we adopted in defining the AGN component as significant in our SED fits are very conservative and thus, while we can easily identify the AGN where $L_{\rm 6\ \mu m,\ AGN}/L_{\rm 6\ \mu m,\ tot}>>0.5$, we are likely to miss some cases where the AGN component is not strongly dominant in the MIR band. With this in mind, we can consider the detection of the AGN emission component through our SED fitting approach as robust.  

As a final verification of the reliability of our SED fitting approach, we plot in Figure \ref{fig.sed0} the best-fit SEDs (using \spz\ and \her\ photometric points only) for the 30 sources analysed here in comparison to the \spz\ IRS spectra; we showed only one of the best-fitting solutions for each source. The SEDs were fitted using only \spz\ 8, 16, 24 $\mu$m, and \her\ 100, 160, 250 $\mu$m photometric points (black circles); the \spz-IRAC data points at 3.6, 4.5, and 5.8 $\mu$m (red triangles) and the VLA radio data points (black star) are over-plotted to the SEDs (not used in the fits) to show the agreement with the resulting SEDs. The \spz\ IRS spectra (cyan line) are also shown, but they are not included in the SED fits. The panels on the top right of each plot is a zoom on the IRS spectra ($2.5-25\ \mu$m rest-frame) to better show the comparison between the best-fit SEDs and the spectra. In general, there is very good agreement with the resulting SED and the IRS spectra, even though we stress that here the spectra were not used to constrain the SEDs. This result is a confirmation of the validity of our templates and SED fitting approach. The only few cases (2/30; i.e. CXOJ123555.13$+$620901.7 and CXOJ123726.51$+$622026.8) where the SEDs differ from the MIR IRS spectra are those where deep silicate features are present; in these cases the photometric data points do not provide enough information to predict the amount of extinction needed to reproduce these strong features. 
However, since the purpose of our analysis is not to measure the strength of the spectral features, such as silicate absorption/emission features or PAH emission lines, but to estimate the AGN and SFG contribution to the overall IR SEDs, we can ignore the discrepancies between SEDs and spectra for these sources. 

\begin{sidewaystable*}
\caption{X-ray AGN in GOODS-N with \spz-IRS spectra.\label{tab3}}
\begin{footnotesize}
\centering
\begin{tabular}{c c c c c c c c c c c c}
\hline\hline
\rule[-2mm]{0.pt}{4ex} XID & $z$ & $S_{\rm 8}$  & $S_{\rm 16}$ & $S_{\rm 24}$ & $S_{\rm 100}$ & $S_{\rm 160}$  & $S_{\rm 250}$ &
$S_{\rm 1.4\ GHz}$ & $log\ \nu L_{\rm 6\ \mu m,\ AGN}$& $log\ \nu L_{\rm 6\ \mu m,\ AGN}$ & IRS Ref. \\
  &  &  $\mu$Jy & $\mu$Jy  &  $\mu$Jy  & mJy  & mJy & mJy &   $\mu$Jy & (SED) erg/s &  (IRS) erg/s &  \\
\rule[-1.7mm]{0.pt}{3ex}  (1) & (2) & (3) & (4)   & (5)  &     (6)    & (7)   & (8) & (9) &  (10)       &  (11) & (12) \\
\hline	
\rule[-0.7mm]{0.pt}{4ex}CXOJ123555.13+620901.7 &  1.875 &   97.3$\pm$0.9 &	$-$	 & 373.0$\pm$ 8.7 &  3.3$\pm$0.6 &  5.4$\pm$1.9 &  8.3$\pm$1.9 & 203.9$\pm$8.4  &     44.87 &	 45.23 & 2 \\
\rule[-1mm]{0.pt}{3ex} CXOJ123603.31+621111.0 &  0.638 &  109.2$\pm$0.5 & 700.8$\pm$22.8 &1250.0$\pm$10.1 & 29.8$\pm$1.5 & 33.9$\pm$2.1 & 25.0$\pm$1.3 & 155.3$\pm$6.5  &  $<$43.61 & $<$43.81 & 2 \\
\rule[-1mm]{0.pt}{3ex} CXOJ123608.12+621036.0 &  0.679 &  287.6$\pm$0.5 & 740.8$\pm$24.1 &2300.0$\pm$13.5 & 10.9$\pm$0.6 &  7.2$\pm$1.2 &  7.7$\pm$1.4 & 213.1$\pm$7.9  &     44.47 &	 44.38 & 3 \\
\rule[-1mm]{0.pt}{3ex} CXOJ123616.11+621513.7 &  2.578 &   43.4$\pm$0.5 &  52.4$\pm$10.8 & 326.0$\pm$ 8.0 &  5.6$\pm$0.5 & 14.8$\pm$1.1 & 30.4$\pm$2.3 &  35.8$\pm$4.9  &  $<$44.63 & $<$44.80 & 2 \\
\rule[-1mm]{0.pt}{3ex} CXOJ123619.45+621252.4 &  0.473 &  170.9$\pm$0.5 & 506.8$\pm$16.5 & 971.0$\pm$10.9 & 28.5$\pm$1.4 & 34.2$\pm$2.1 & 29.0$\pm$2.8 &  65.3$\pm$4.8  &  $<$43.35 & $<$43.58 & 3 \\
\rule[-1mm]{0.pt}{3ex} CXOJ123622.53+621545.2 &  0.639 &   72.7$\pm$0.4 & 363.0$\pm$11.9 & 725.0$\pm$ 6.9 & 20.3$\pm$1.0 & 25.7$\pm$1.6 & 23.7$\pm$1.3 &  35.1$\pm$8.3  &  $<$43.45 & $<$43.57 & 2 \\
\rule[-1mm]{0.pt}{3ex} CXOJ123622.66+621629.8 &  1.790 &   27.9$\pm$0.4 &  90.6$\pm$ 9.1 & 417.0$\pm$ 8.2 &  3.0$\pm$0.4 &  9.7$\pm$1.2 & 27.7$\pm$2.5 &  91.6$\pm$9.9  &  $<$44.22 & $<$44.56 & 2 \\
\rule[-1mm]{0.pt}{3ex} CXOJ123629.11+621045.9 &  1.013 &   73.0$\pm$0.5 & 425.0$\pm$13.9 & 730.0$\pm$12.7 &  7.5$\pm$0.5 & 19.4$\pm$1.2 & 33.4$\pm$1.2 & 100.8$\pm$12.9 &  $<$43.95 & $<$43.82 & 3 \\
\rule[-1mm]{0.pt}{3ex} CXOJ123630.00+620542.4 &  0.483 &  311.4$\pm$0.7 &	$-$	 &2480.0$\pm$26.0 & 15.0$\pm$1.0 & 14.0$\pm$1.3 & 10.8$\pm$2.2 &  64.8$\pm$9.9  &     44.00 &	 44.03 & 3  \\
\rule[-1mm]{0.pt}{3ex} CXOJ123632.59+620759.8 &  1.994 &  116.9$\pm$0.5 & 363.0$\pm$11.9 & 820.0$\pm$ 9.6 &  1.7$\pm$0.4 &  0.2$\pm$1.7 &  1.4$\pm$7.1 &  89.1$\pm$10.2 &     45.37 &	 45.24 & 3  \\
\rule[-1mm]{0.pt}{3ex} CXOJ123633.23+620834.8 &  0.934 &  126.5$\pm$0.5 & 497.5$\pm$16.2 & 779.0$\pm$ 7.9 & 10.6$\pm$0.6 & 20.3$\pm$1.5 & 22.0$\pm$1.4 &  50.8$\pm$10.2 &  $<$43.86 &	 44.37 & 2 \\
\rule[-1mm]{0.pt}{3ex} CXOJ123633.67+621005.7 &  1.016 &   52.4$\pm$0.5 & 497.5$\pm$16.2 & 581.0$\pm$ 9.0 &  8.6$\pm$0.6 & 16.2$\pm$1.2 & 21.6$\pm$1.2 &  58.5$\pm$9.1  &  $<$43.80 & $<$43.83 & 3 \\
\rule[-1mm]{0.pt}{3ex} CXOJ123634.50+621241.2 &  1.224 &   75.3$\pm$0.4 & 874.9$\pm$28.5 & 444.0$\pm$ 5.5 & 33.6$\pm$1.7 & 56.0$\pm$10.2& 50.3$\pm$23.9& 201.1$\pm$10.3 &  $<$44.14 & $<$44.37 & 2 \\
\rule[-1mm]{0.pt}{3ex} CXOJ123635.58+621424.1 &  2.005 &  310.5$\pm$0.4 & 662.9$\pm$21.6 &1520.0$\pm$14.4 & 12.1$\pm$0.7 & 23.5$\pm$1.5 & 30.2$\pm$1.4 &  76.0$\pm$7.9  &     45.55 &	 45.61 & 3 \\
\rule[-1mm]{0.pt}{3ex} CXOJ123642.20+621545.5 &  0.858 &  134.9$\pm$0.4 & 571.7$\pm$18.6 & 866.0$\pm$10.3 & 10.9$\pm$0.6 & 20.7$\pm$1.3 & 22.8$\pm$1.3 & 151.3$\pm$9.4  &     44.22 &	 44.09 & 3\\
\rule[-1mm]{0.pt}{3ex} CXOJ123645.81+620753.9 & 1.433  &   47.0$\pm$0.6 & 304.2$\pm$10.0 & 374.0$\pm$ 7.9 & 14.3$\pm$0.8 & 21.2$\pm$1.6 & 22.0$\pm$1.2 &  73.2$\pm$5.6  &  $<$44.13 & $<$44.42 & 3 \\
\rule[-1mm]{0.pt}{3ex} CXOJ123646.73+620833.6 &  0.971 &   68.7$\pm$0.5 & 650.8$\pm$21.2 & 991.0$\pm$ 8.8 & 17.0$\pm$0.9 & 32.3$\pm$1.8 & 34.6$\pm$2.1 &  81.7$\pm$5.1  &  $<$43.96 & $<$43.94 & 3 \\
\rule[-1mm]{0.pt}{3ex} CXOJ123646.72+621445.9 &  2.004 &   27.9$\pm$0.3 & 146.8$\pm$ 9.7 & 422.0$\pm$29.1 &  4.2$\pm$0.5 &  7.5$\pm$1.8 &  6.1$\pm$6.3 &  15.4$\pm$4.8  &     44.90 &	 45.08 & 3 \\
\rule[-1mm]{0.pt}{3ex} CXOJ123649.66+620738.3 &  1.610 &  233.4$\pm$0.6 & 727.2$\pm$23.7 &1440.0$\pm$12.6 &  9.4$\pm$0.7 &  6.5$\pm$1.0 &  8.6$\pm$1.3 & 312.9$\pm$10.9 &     45.37 &	 45.41 & 3 \\
\rule[-1mm]{0.pt}{3ex} CXOJ123653.37+621139.6 &  1.270 &   44.8$\pm$0.4 & 421.1$\pm$13.7 & 336.0$\pm$ 7.5 & 15.8$\pm$0.9 & 23.9$\pm$1.5 & 21.7$\pm$2.0 &  84.2$\pm$8.9  &  $<$44.03 & $<$44.29 & 3 \\
\rule[-1mm]{0.pt}{3ex} CXOJ123655.89+620807.6 &  0.792 &   89.4$\pm$0.6 & 516.3$\pm$16.8 & 859.0$\pm$10.8 & 27.9$\pm$1.4 & 39.6$\pm$2.0 & 33.4$\pm$1.6 & 118.0$\pm$5.6  &  $<$43.74 &	 44.12 & 2 \\
\rule[-1mm]{0.pt}{3ex} CXOJ123701.62+621146.2 &  1.760 &   58.0$\pm$0.4 & 180.4$\pm$ 8.9 & 739.0$\pm$10.7 &  7.0$\pm$0.6 & 17.1$\pm$1.4 & 38.6$\pm$2.7 & 121.7$\pm$10.3 &  $<$44.46 & $<$44.75 & 2 \\
\rule[-1mm]{0.pt}{3ex} CXOJ123703.94+621156.7 &  3.406 &    8.1$\pm$0.4 &  37.1$\pm$ 8.5 & 105.0$\pm$ 9.5 &  1.1$\pm$0.4 &  2.0$\pm$0.9 &  0.0$\pm$1.2 &  36.6$\pm$4.2  &     45.24 &	 45.24 & 3 \\
\rule[-1mm]{0.pt}{3ex} CXOJ123711.38+621330.8 &  1.996 &   37.8$\pm$0.4 &  83.2$\pm$ 8.4 & 537.0$\pm$ 9.3 &  7.2$\pm$0.5 & 22.6$\pm$1.6 & 43.9$\pm$2.0 & 130.8$\pm$9.9  &  $<$44.42 & $<$44.58 & 2 \\
\rule[-1mm]{0.pt}{3ex} CXOJ123712.04+621325.7 &  1.996 &   12.5$\pm$0.4 &  31.3$\pm$ 8.9 & 219.0$\pm$ 6.6 &  3.5$\pm$0.6 & 11.2$\pm$1.8 & 13.9$\pm$1.5 &  50.4$\pm$8.1  &  $<$44.03 &	 44.67 & 1 \\
\rule[-1mm]{0.pt}{3ex} CXOJ123716.57+621643.5 &  1.820 &   55.3$\pm$0.4 & 148.2$\pm$ 7.3 & 508.0$\pm$ 6.3 &  4.8$\pm$0.5 & 14.2$\pm$1.1 & 22.4$\pm$1.2 &  79.1$\pm$5.2  &  $<$44.30 & $<$44.81 & 2 \\
\rule[-1mm]{0.pt}{3ex} CXOJ123716.65+621733.3 &  1.146 &  239.6$\pm$0.5 & 516.3$\pm$16.8 &1240.0$\pm$15.5 &  3.6$\pm$0.5 &  4.0$\pm$1.0 &  4.3$\pm$1.3 & 373.0$\pm$12.0 &     44.94 &	 44.88 & 3 \\
\rule[-1mm]{0.pt}{3ex} CXOJ123726.51+622026.8 &  1.750 &  175.3$\pm$0.5 & 933.4$\pm$30.3 & 931.0$\pm$ 8.6 &  9.5$\pm$0.7 & 14.9$\pm$1.4 & 17.3$\pm$2.9 & 100.3$\pm$6.4  &     45.42 &	 45.52 & 2 \\
\rule[-1mm]{0.pt}{3ex} CXOJ123737.14+621205.3 &  0.410 &  211.8$\pm$0.7 & 556.0$\pm$18.1 & 763.0$\pm$ 8.0 & 15.5$\pm$0.9 & 26.0$\pm$1.9 & 26.6$\pm$1.2 &  55.7$\pm$10.7 &  $<$43.24 &	 43.56 & 3 \\
\rule[-1mm]{0.pt}{3ex} CXOJ123739.19+622059.4 &  0.837 &  153.8$\pm$0.6 & 453.5$\pm$14.8 & 760.0$\pm$ 8.2 &  7.5$\pm$0.6 & 10.5$\pm$1.3 & 15.8$\pm$2.4 &  56.9$\pm$10.9 &     44.20 &	 44.07 & 3 \\
\hline
\end{tabular}
\end{footnotesize}
\tablefoot{
Col.1: X-ray source ID; 
Col.2: redshift; 
Col.3: \spz-IRAC 8 $\mu$m flux density in $\mu$Jy; 
Col.4: \spz\ 16 $\mu$m flux density in $\mu$Jy; 
Col.5: \spz-MIPS 24 $\mu$m flux density in $\mu$Jy; 
Col.6: \her\ 100 $\mu$m flux density in mJy; 
Col.7: \her\ 160 $\mu$m flux density in mJy; 
Col.8: \her\ 250 $\mu$m flux density in mJy; 
Col.9: measured VLA total radio flux density at 1.4 GHz (Daddi et al., in prep.; \citealt{morrison2010});
Col.10: logarithm of the 6 $\mu$m luminosity of the AGN component obtained from the SED fit of the \spz\ and \her\ photometric data only; 
where the AGN component was not significantly detected an upper limit of 30\% of the total 6 $\mu$m luminosity has been reported (see Appendix A);
Col.11: logarithm of the 6 $\mu$m luminosity of the AGN component obtained from the IRS spectral fit (plus \her); where the AGN component 
was not significantly detected an upper-limit of 30\% of the total 6 $\mu$m luminosity has been reported (see Appendix A);
Col.12: Reference for the IRS spectra: $1=$\citet{pope2008b}, $2=$\citet{murphy2009}, $3=$\citet{kirkpatrick2012}.}
\end{sidewaystable*}

\begin{figure*}[!t]
\centering{
\includegraphics[scale=0.9]{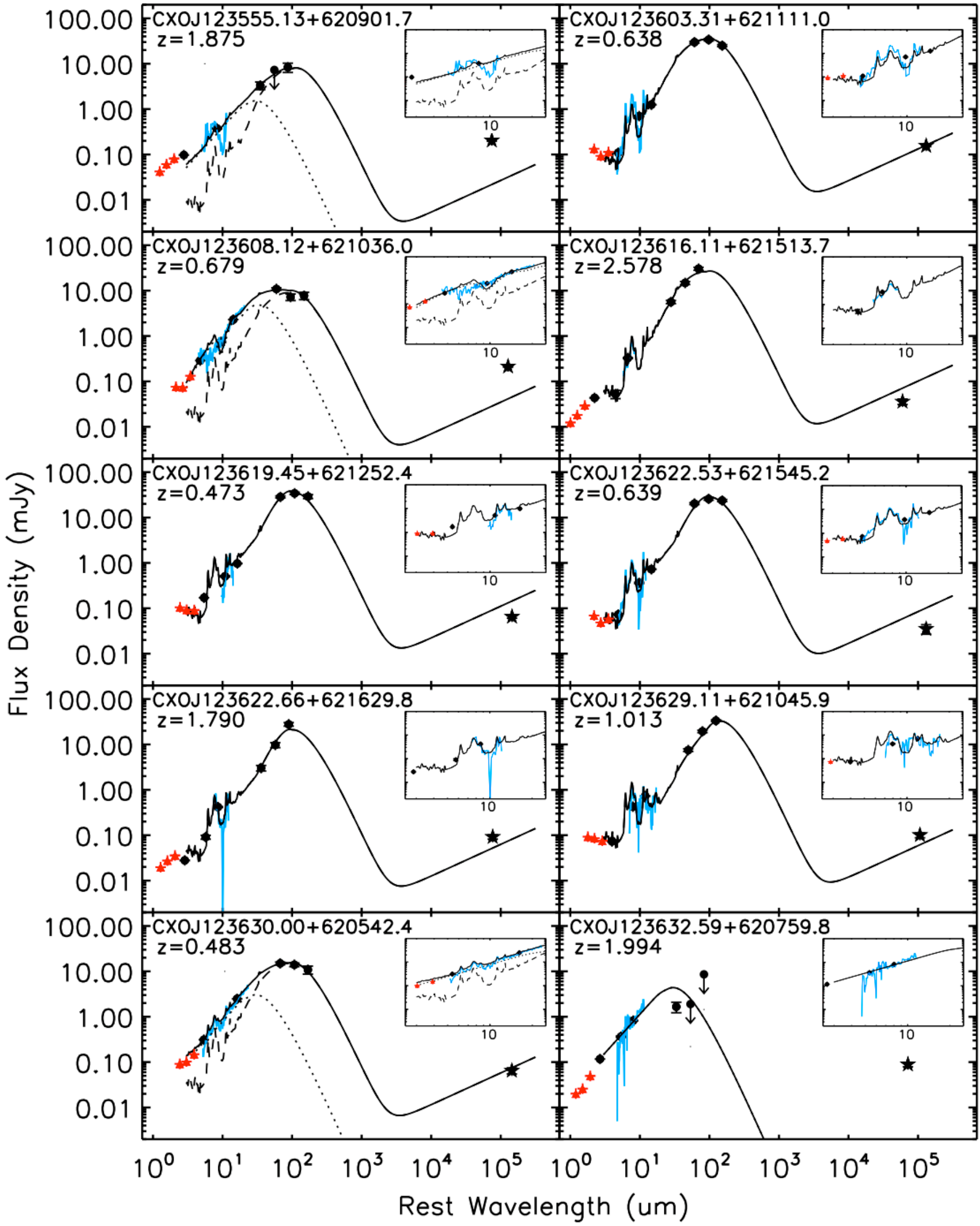}}
\caption{Best-fitting solutions to the spectral energy distributions (SEDs) of the X-ray detected sources in GOODS-N with \spz-IRS spectroscopy (\protect\citealt{pope2008b,murphy2009,kirkpatrick2012}). The dotted lines represent the AGN component and the dashed lines indicate the SFG component; the total SEDs are represented as solid lines. The \spz\ 8, 16, 24 $\mu$m and the \her\ 100, 160, 250 $\mu$m data points (black circles) have been used to constrain the SEDs, while the \spz-IRAC 3.6, 4.5, 5.8 $\mu$m (red stars) and the VLA 1.4 GHz (black star) radio data points are over plotted on the SEDs but they are not included in the fit. The \spz-IRS spectrum is also shown (cyan line), but it is NOT used to constrain the SEDs. On the top right-hand side of each plot, a zoom on the \spz-IRS spectrum is shown (2.5-25 $\mu$m); our best-fit SEDs are typically in very good agreement with the \spz-IRS spectra.}
\label{fig.sed0}
\end{figure*}
\begin{figure*}[!t]
\ContinuedFloat
\centering{
\includegraphics[scale=0.9]{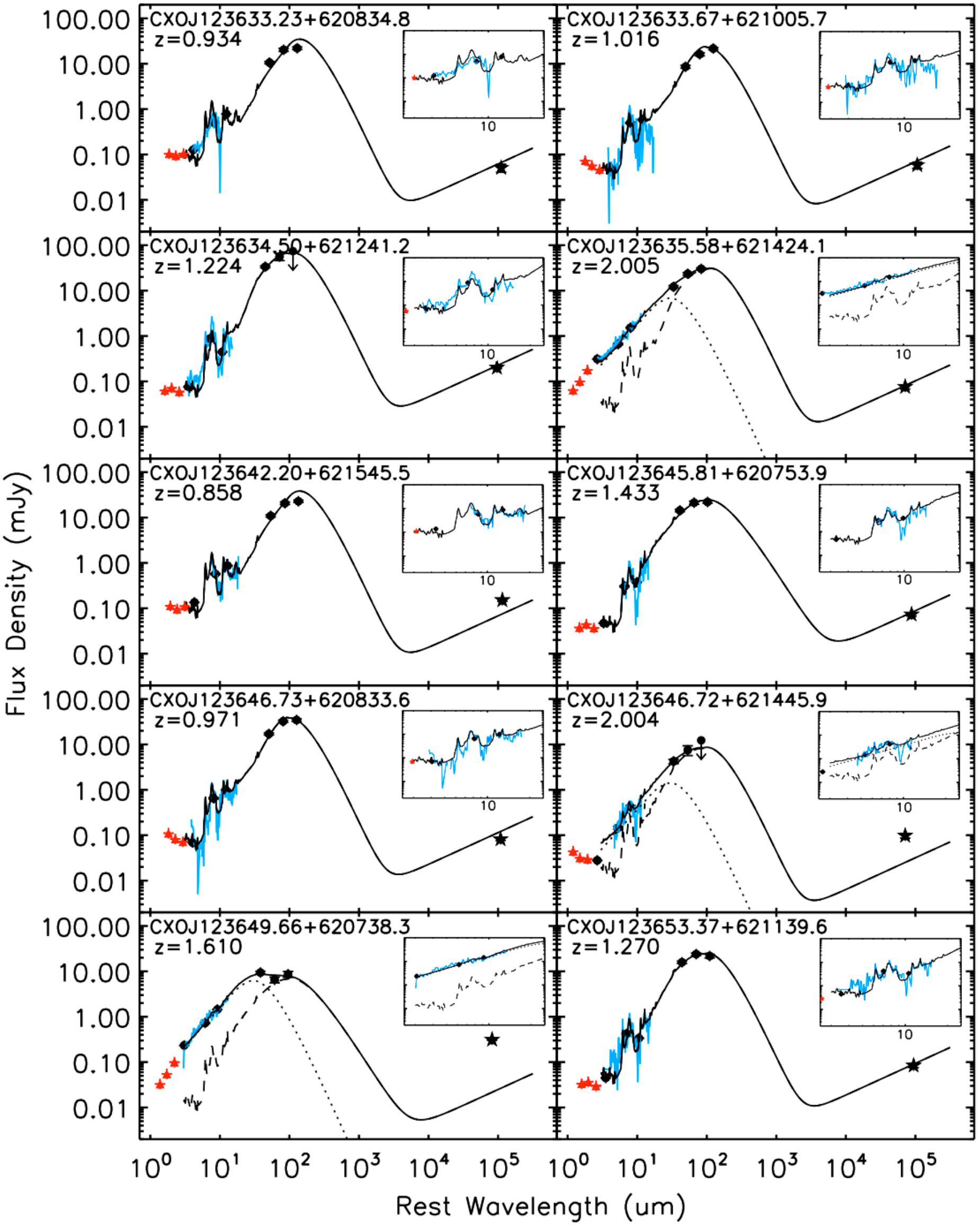}}
\caption{Continued.}
\end{figure*}
\begin{figure*}[!t]
\ContinuedFloat
\centering{
\includegraphics[scale=0.9]{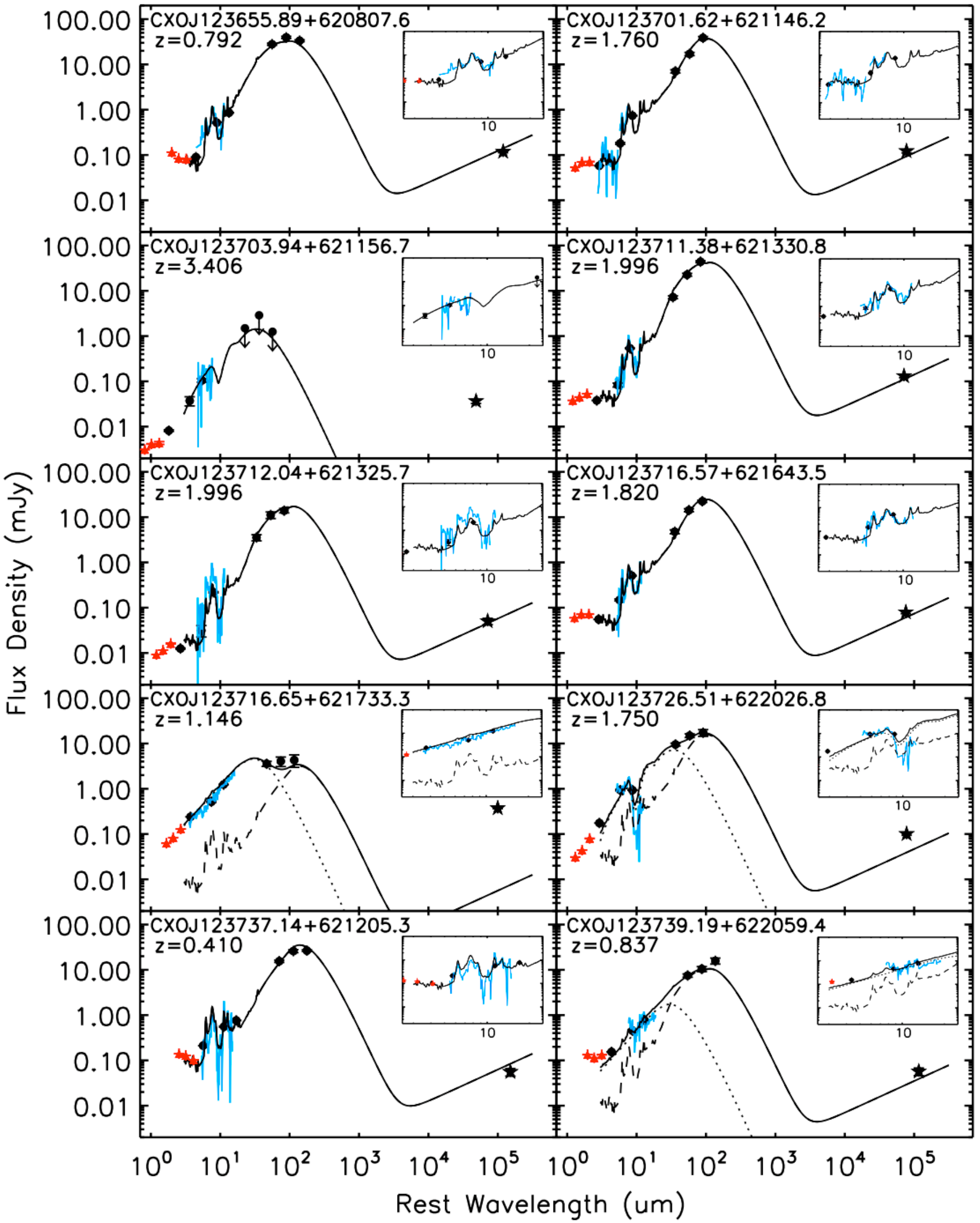}}
\caption{Continued.}
\end{figure*}

\section{SEDs of the radio-excess sources}

In this appendix the best-fit SED plots for the entire radio-excess sample (51 sources) are reported (Figure \ref{fig.sedall}); for each source we only plotted the SED with the lowest $\chi^2$ value amongst the best-fitting model solutions (see Sects. \ref{sed} and Appendix A).
\begin{figure*}[!t]
\centering{
\includegraphics[scale=0.87]{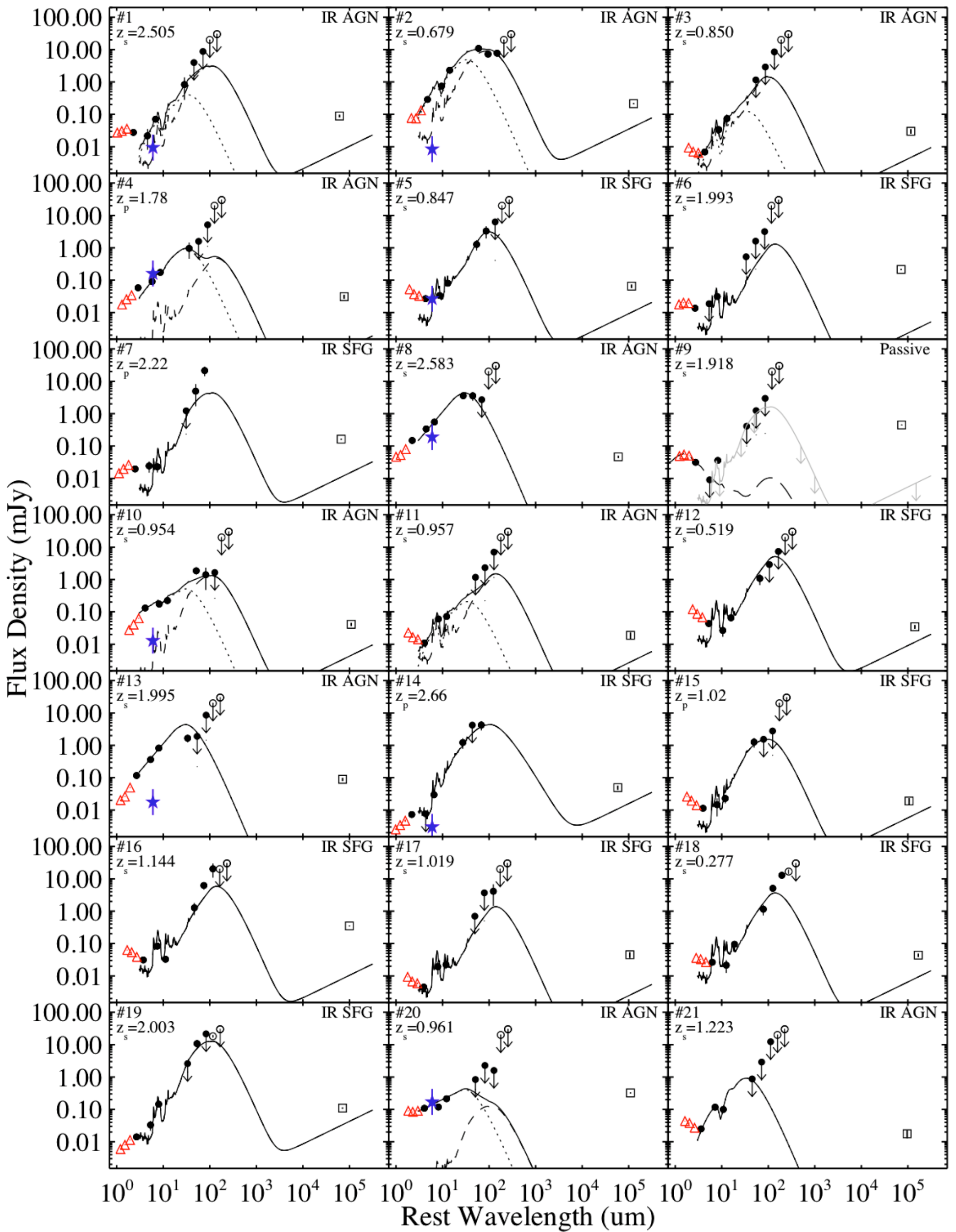}}
\vspace{-0.4cm}
\caption{Best-fit SEDs for the 51 sources in the radio-excess sample. The total SEDs are shown as black solid lines, the AGN templates are shown as dotted lines and the SFG templates as dashed lines. Filled circles represent the {\it Spitzer} 8, 16, 24 $\mu$m and the {\it Herschel} 100, 160, 250 $\mu$m flux densities, which are used to constrain the SEDs. Open symbols indicate the data that were not included in the SED fitting process: red triangles are {\it Spitzer}-IRAC 3.6, 4.5, 5.8 $\mu$m flux densities, black open circles are SPIRE 350 and 500 $\mu$m, and black squares are VLA 1.4 GHz flux densities; the blue stars represent the 6 $\mu$m luminosity of the AGN predicted from the X-ray luminosity (2--10 keV, rest-frame) using the \protect\citet{lutz2004} relation for local unobscured AGN; we note that these points do not always match the IR AGN component because the X-ray luminosity tends to underestimate the intrinsic AGN power if the AGN emission is heavily absorbed. For the passive sources we plotted the SFG template upper limit (grey line) and an elliptical galaxy template (long dashed line; Sect. \ref{seds}) to show that it could well represent the data, although we stress that this latter template is not included in our SED fitting analysis. The rise of the IRAC data point at short wavelengths observed for many sources is due to the emission from the galaxy old stellar population.}
\label{fig.sedall}
\end{figure*}
\begin{figure*}[!t]
\ContinuedFloat
\centering{
\includegraphics[scale=0.87]{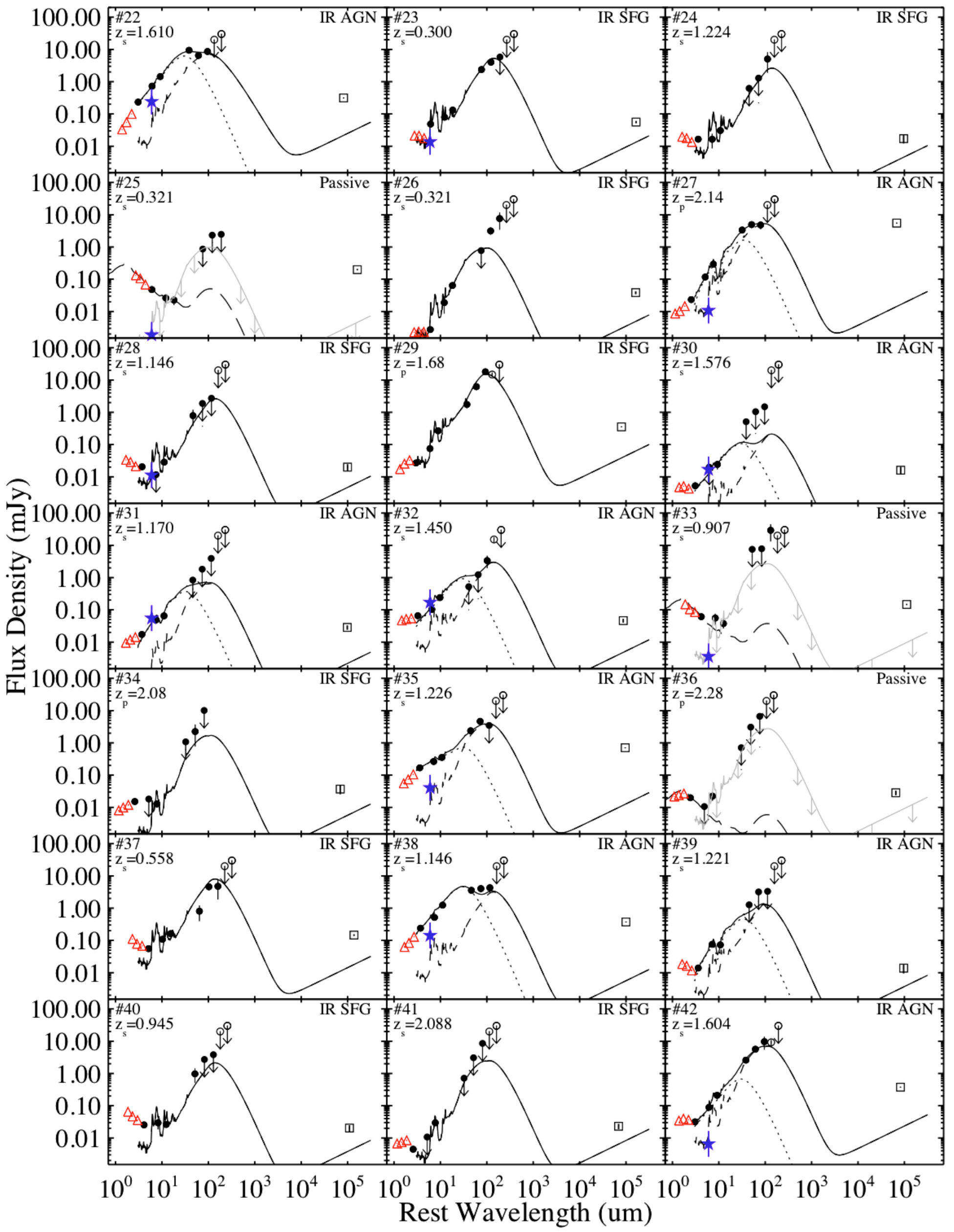}}
\caption{Continued.}
\end{figure*}
\begin{figure*}[!t]
\ContinuedFloat
\centering{
\includegraphics[scale=0.87]{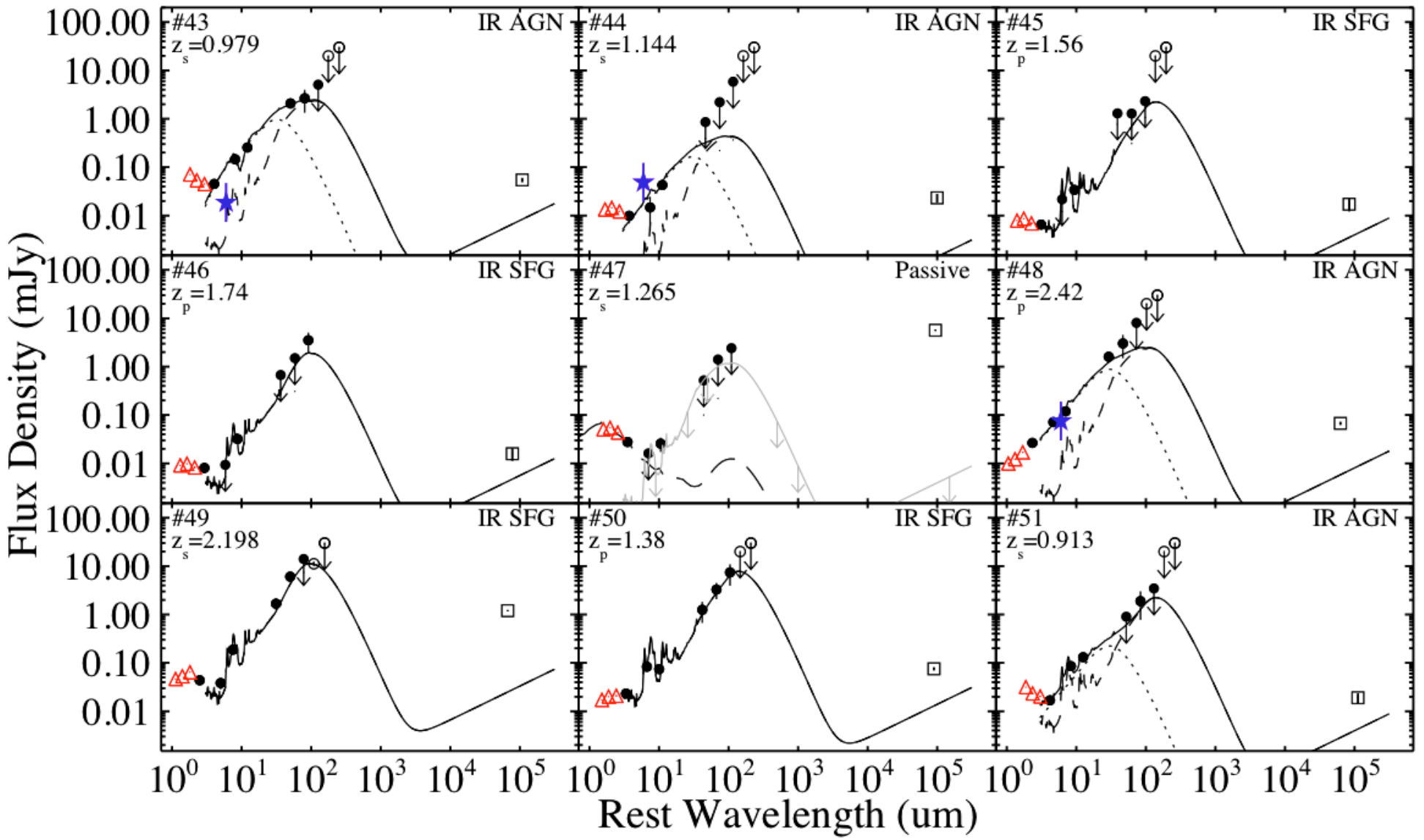}}
\caption{Continued.}
\end{figure*}
\end{appendix}

\end{document}